\documentclass{aastex62}

\usepackage{amsmath}
 
\revised{\today}

\shorttitle{SED curvature of \textit{Fermi} Blazars}
\shortauthors{Anjum, Chen \& Gu}

\begin{document}
\title{On the Origin and Evolution of Curvature of the Spectral Energy Distribution of \emph{Fermi} Bright Blazars}

\author[0000-0001-6820-717X]{Muhammad S. Anjum}
\affil{Key Laboratory for Research in Galaxies and Cosmology, Shanghai Astronomical Observatory, Chinese Academy of Sciences, 80 Nandan Road, Shanghai 200030, China}
\affil{University of Chinese Academy of Science, 19A Yuquan Road, Beijing 100049, China}
\email{mshahzadanjum@yahoo.com}

\author[0000-0002-1908-0536]{Liang Chen}
\affil{Key Laboratory for Research in Galaxies and Cosmology, Shanghai Astronomical Observatory, Chinese Academy of Sciences, 80 Nandan Road, Shanghai 200030, China}
\email{chenliang@shao.ac.cn}

\author[0000-0002-4455-6946]{Minfeng Gu}
\affil{Key Laboratory for Research in Galaxies and Cosmology, Shanghai Astronomical Observatory, Chinese Academy of Sciences, 80 Nandan Road, Shanghai 200030, China}


\begin{abstract}
The origin and evolution of spectral curvature in blazar spectral energy distribution (SED) is still unclear. Since the observed SED curvature is related to an intrinsic curvature in emitting electron energy distribution (EED), we study this question by employing a log-parabolic EED with a curvature parameter and peak energy to model the quasi-simultaneous broadband SEDs of selected blazars in Fermi-LAT Bright AGN Sample (LBAS) using synchrotron and inverse Compton (IC) processes. We find that log-parabolic IC model can successfully explain the emission in all blazars in our sample. On average, FSRQs have higher magnetic field, Doppler factor, and curvature than BL Lac objects. The BL Lac objects show an anticorrelation between the curvature parameter of the EED and its peak energy, which is a signature of stochastic acceleration. FSRQs do not manifest such correlation and rather show a mild positive relationship between these parameters. This suggests that the evolution of spectral curvature in the BL Lac objects is dominated by a strong stochastic acceleration component, whereas the curvature in FSRQs evolves in a cooling dominated regime due to an additional external Compton (EC) component. The strong cooling in FSRQs not only restricts the electron peak energy but also adds extra curvature to the high energy tail of emitting EED. Since the curvature decreases from FSRQs toward high peak BL Lac objects (HBLs), opposite to peak energy, the curvature parameter can be considered a third parameter of the blazar sequence in addition to peak frequency and luminosity.
\end{abstract}

\keywords{galaxies: active --- galaxies: jets --- galaxies: nuclei --- radiation mechanisms: non-thermal}

\section{Introduction} \label{sec:introduction}
Active galactic nuclei (AGNs) are bright centers of massive galaxies powered by accretion matter onto supermassive black holes. AGNs have been categorized based on their observational features into many classes and are unified under an orientation scheme \citep{1995PASP..107..803U} and blazars are the brightest type of AGNs with their relativistic plasma jets pointed within a small angle to our line of sight. Onwards from the 1970s, many studies found that spectra of blazars are significantly curved, even for single band observations \citep{1974ApJ...192L.115R, 1977ApJ...214L.105O, 1983PASP...95..724S}. When combining two or more observing bands, the resulting multiwavelength spectral energy distribution (SED) shows prominent spectral curvature and SEDs of some objects even present significant breaks \citep{1981ApJ...243...47L, 1985ApJ...298..630L, 1986ApJ...310..317G, 1987ApJ...318..175B}. The SED from radio to UV or X-rays revealed a bump in $\log\nu$-$\log\nu f_{\nu}$ representation and this component is believed to be synchrotron emission of accelerated electrons in the relativistic jet \citep[see, e.g.,][]{1978bllo.conf..328B, 1986ApJ...308...78L, 1989MNRAS.241P..43G, 1998MNRAS.301..451G, 1995PASP..107..803U, 1996ApJ...463..444S}. The high energy bump in the SED from hard X-rays up to very high energy (VHE) $\gamma$-rays is attributed to inverse Compton (IC) emission in leptonic models \citep{2007Ap&SS.309...95B}. Based on the synchrotron peak frequency $\nu_p$, blazars are often divided into low synchrotron peak (LSP), intermediate synchrotron peak (ISP), and high synchrotron peak (HSP) sources \citep[e.g.,]{2010ApJ...716...30A}. For FSRQs the $\nu_p$ usually vary within $10^{12.5-14}$ Hz \citep{2010ApJ...716...30A}, whereas for BL Lac objects it usually varies in a larger range ($10^{13-17}$ Hz).

Many high energy detectors with broadband energy coverage (e.g, BeppoSAX provided $0.1-300$ keV and currently NuStar covering $3-79$ keV X-rays) can provide details of SED shape. The Fermi-LAT, working in the $0.1-300$ GeV energy range, discovered that the extragalactic $\gamma$-ray sky is dominated by blazars. Fermi's Third AGN Catalog (3LAC) \citep{2015ApJ...810...14A} consisted of 1591 blazars, while the number of sources increased to 2863 sources in the recently released 8 yr Fourth LAT AGN catalog (4LAC) with increased energy range from 50 MeV to 1 TeV \citep{2020ApJ...892..105A}. Many blazars observed by LAT shows curvature in the $\gamma$-ray spectra \citep[see, e.g.,][]{2011ApJ...743..171A, 2020ApJ...892..105A}. High energy observation of some blazars \citep[e.g., PKS 2155-304, Mrk 501 and Mrk 421, see,][]{1998ApJ...501L..17S, 1999ApJ...511..149K, 2007ApJ...664L..71A, 2008MNRAS.386L..28G, 2012A&A...542A.100A} by Imaging Atmospheric Cerenkov Telescopes (IACT) also showed a substantial curvature at TeV $\gamma$-ray energies. The curved spectra even in a single band suggests that the electron distribution may be intrinsically curved, since a power-law distribution $N(\gamma)\propto \gamma^{-p}$ would emit a power-law spectrum $ F_\nu \propto \nu^{-\alpha} $. Combining X-ray through $\gamma$-ray observations the high energy component of blazar SED, which usually peaks in the $\gamma$-ray band, can be characterized. This high energy bump, in leptonic models, is usually explained as IC emission of the same electron population accounting for the synchrotron component emission \citep[see,][and references therein]{1998MNRAS.301..451G, 2010MNRAS.402..497G}. The high energy emission arises as synchrotron self-Compton (SSC), if the seed photons come from internal synchrotron photons \citep{1992ApJ...397L...5M}, or external Compton (EC) if the dominated seed photons come from the accretion disk \citep{1993ApJ...416..458D}, broad line region \citep[BLR][]{1994ApJ...421..153S}, and infrared (IR) photons from an extended dusty torus \citep{2000ApJ...545..107B}. See \citet{2007Ap&SS.309...95B} for a review on the models for blazar non-thermal emission.

Since the multiwavelength spectra are significantly curved, the characterization of a broadband bump demands at least three parameters: peak frequency, peak flux, and a "curvature" parameter to measure the "width" of the bump. The relationship between the former two parameters has been extensively explored in blazars, the so-called blazar sequence, i.e., the anticorrelation between peak luminosity against synchrotron peak frequency, which may arise due to radiative cooling of energetic particles in the jet \citep{1998MNRAS.299..433F, 1998MNRAS.301..451G, 2011ApJ...735..108C, 2012MNRAS.420.2899G}. To measure the curvature of a bump, there are at least two methods \citep[][referred as C14 hereafter]{2014ApJ...788..179C}: (i) if a log-parabolic law is employed to fit the bump, i.e., $\log\nu f_{\nu}=-b\left(\log\nu-\log\nu_{p}\right)^{2}+\log\nu_{p}f_{\nu_{p}}$, the coefficient of second order term $b$ measures the curvature; and (ii) if a broken power law is adopted to fit the SED, the difference of spectral indices $\left|\alpha_{2}-\alpha_{1}\right|$ roughly measures the curvature of the SED bump. The curved SED may not be just the result of purely radiative cooling, because it predicts that the spectral index always steepens as $\left|\alpha_{2}-\alpha_{1}\right|=0.5$, while the observations show a large range of curvature (C14). The intrinsic curvature in the electron energy distribution (EED) arises due to the combined effect of particle acceleration and radiative cooling. \citet{2004A&A...413..489M} showed that if the acceleration gain decreases with particle energy, the spectrum of accelerated electrons can be described by a log-parabola. Since the broadband blazar spectra are always curved, it is important to understand how the curvature of electron distribution evolves in physical conditions of blazar jets.

Multiple X-ray observations of Mrk 421 by Swift, XMM-Newton, and BeppoSAX can be fitted by a log-parabolic law model and the spectral curvature is found to be inversely correlated with the peak energy $E_p$ \citep[see,][]{2004A&A...413..489M, 2007A&A...466..521T, 2009A&A...501..879T}. Furthermore, many other BL Lac objects (including, e.g., Mrk 501, PKS 2155-304, and PKS 0548-322) have been extensively studied by \citet{2008A&A...478..395M}, who found an inverse correlation between the X-ray component peak frequency and its curvature for a small sample of BL Lac objects. Besides the high energy studies, the low energy part of SED from radio to optical band also presents a similar feature \citep[][]{1986ApJ...308...78L}. In the past three decades, despite the fact that many large blazar samples were used to study the properties of the peak frequency and flux \citep[see,][]{1996ApJ...463..444S, 1998MNRAS.299..433F, 1998MNRAS.301..451G, 2006A&A...445..441N, 2009MNRAS.397.1713C, 2009RAA.....9..168W, 2010ApJ...716...30A, 2010MNRAS.402..497G, 2011ApJ...735..108C}, only few works explored the properties of spectral curvature by using broadband SEDs. \citet{2011MNRAS.417.1881R} found an anticorrelation between the curvature and peak frequency by fitting the radio to optical band synchrotron SEDs of 10 BL Lac objects. In 2014, C14 mathematically fitted the quasi-simultaneous broadband SEDs spanning from radio to $\gamma$-rays, of both synchrotron and IC component of 48 Fermi bright blazars, by a log-parabolic law. C14 found a significant anticorrelation between synchrotron peak frequency and its curvature with chance probability $P$ down to $1.35\times10^{-17}$. Even employing $\left|\alpha_{2}-\alpha_{1}\right|$ as a "surrogate" of curvature, the correlation remained still significant ($P=5.35\times10^{-5}$, see C14). By exploring a larger blazar sample, \citet{2016MNRAS.463.3038X} confirmed the above findings, with FSRQs showing slight departure from the BL Lac objects. These studies suggested that there might be an intrinsic inverse relationship between curvature and the peak energy of EED of log-parabolic shape, produced by a statistical or stochastic particle acceleration mechanism \citep{2004A&A...413..489M, 2007A&A...466..521T, 2009A&A...501..879T, 2011ApJ...739...66T}. However, all these previous studies investigated the relationship between SED peak frequency and its curvature by fitting the SEDs through a log-parabolic function, rather than any physically motivated model.

Although an anticorrelation between synchrotron peak frequency and the SED curvature has already been found, it may not correspond to an intrinsic anticorrelation between electron peak energy and EED curvature since synchrotron peak shifts can also be caused by magnetic field $B$ and Doppler factor $\delta$ changes other than electron peak energy $\gamma_p$ (i.e., $\nu_p \propto B \gamma_p^2 \delta$). Physical modeling of simultaneous SEDs is important to check if the expected anti-correlation between curvature and peak energy of underlying EED truly holds, which means studying the evolution of intrinsic electron spectral curvature in blazar jets. Since blazars are extremely variable especially at high energy, i.e., from X-rays to TeV $\gamma$-rays, the simultaneous SEDs of a sample are important to study the evolution of curvature against the peak energy. Until now, although some works employed a log-parabolic EED to fit the blazar broadband SEDs \citep[e.g.,][]{2017MNRAS.464..599D, 2018ApJS..235...39C, 2020ApJS..248...27T}, only \citet{2016MNRAS.456.2173Y} explored the relationship between intrinsic curvature and peaked electron energy of underlying EED  by SED modeling of a single source 3C 279 at various epochs.

In this work, we collect the quasi-simultaneous broadband SEDs of a complete sample of blazars \citep{2010ApJ...716...30A} including both FSRQs and BL Lac objects and use a log-parabolic IC model to fit these spectra. The model parameters are used to study the properties of electron spectral curvature and its relationship with electron peak energy, i.e., how the curvature evolves in different underlying conditions in blazars. \citet{2018ApJS..235...39C} modeled a large sample of blazars from the 3LAC by an approximation method and estimated jet physical parameters (size, bulk velocity, magnetic field strength of emission region and peak energy and curvature of EED). We compare our results based on our exact SED modeling with the 3LAC sample. Section \ref{sec:sample} presents the blazar samples used in our study. We define the physical model in Section \ref{sec:model} and the results are discussed in Section \ref{sec:results}. Finally, we summarize our results and conclusions in Section \ref{sec:summary}. In this work, a $\Lambda$CDM cosmology is assumed and values within 1$\sigma$ of the Wilkinson Microwave Anisotropy Probe results \citep{2011ApJS..192...18K} are used; in particular, $H_{0}=70$ km s$^{-1}$ Mpc$^{-1}$, $\Omega_{\Lambda}=0.73$, and $\Omega_{\rm M}=0.27$.

\section{The Sample} \label{sec:sample}
We choose the quasi-simultaneous broadband SEDs of selected blazars from the LAT Bright AGN Sample \citep[LBAS;][]{2009ApJ...700..597A} for IC modeling. \citet{2010ApJ...716...30A} compiled the quasi-simultaneous SEDs of LBAS blazars, from radio through $\gamma$-rays, including the first 3 months Fermi-LAT integrated data, Swift-XRT and UVOT observations coincident with the Fermi observational run and other space- and ground-based telescopes observations at longer wavelengths. Our sample includes those 48 LBAS sources with quasi-simultaneous SEDs. We divide the sample into two main classes as FSRQs and BL Lac objects, while the BL Lac objects are further characterized into LBLs, IBLs, and HBLs, based on the observed synchrotron peak frequency, \citep{2010ApJ...716...30A}. Thus, in total, the whole sample consists of 23 FSRQs, 9 LBLs, 8 IBLs and 8 HBLs. Since the observed trends between SED parameters in a blazar sequence must be related to physical jet parameters, such a categorization gives us a complete sample to study the curvature and its relationship with other source parameters, mainly the electron peak energy. The LSP blazars (FSRQs and LBLs) make up 32 sources ($>65\%$ of the sample) and the curvature properties of these sources are investigated for the first time in this study. C14 fitted the same sample with a log-parabolic mathematical function; however, those parameters cannot be related to physical parameters of the IC model. Since simultaneous observations are not available at many frequencies around the synchrotron peak in some LSP sources, we include additional archival data from the ASI Space Science Data Center (SSDC)\footnote{www.ssdc.asi.it} in such cases. In some sources, time integrated observations at $2-10$ keV X-rays from Swift 1SWXRT catalog \citep{2013A&A...551A.142D} are included to better constrain the SSC component. For few sources with recently measured redshifts, we got their redshift values from NASA/IPAC Extragalactic Database (NED)\footnote{www.ned.ipac.caltech.edu} to determine the rest frame SEDs. Only in case where the redshift is not available, we used a value of $z=0.4$ similarly as in \cite{2010ApJ...716...30A}. In many FSRQs the optical-UV data may be dominated by the thermal big blue bump, so we use them as an upper limit for synchrotron emission. Since the jet model is self-absorbed from low to high radio frequencies, we do not fit these data.
 
We also compiled the curvature and peak energy of EED for 1392 blazars from the 3LAC sample, for which the physical jet parameters of BL Lac objects and FSRQs have been obtained using an IC model \citep{2018ApJS..235...39C}. It is important to note that \citet{2018ApJS..235...39C} derived the jet parameters of blazars from approximated IC peak frequency and luminosity, rather than fitting the broadband SEDs. Furthermore, \citet{2018ApJS..235...39C} employed an SSC model for all BL Lac objects, however, $\gamma$-ray emission of LBLs is usually needs an EC component \citep{2014MNRAS.439.2933Y}. However, although the physical parameters for the 3LAC sample are not tightly constrained for individual sources, it may not significantly affect the statistical trends between the parameters due to large sample size.

\section{The Model} \label{sec:model}
We employ a one-zone synchrotron and IC model to fit the blazar SEDs \citep[see][for details]{2017ApJ...842..129C}, assuming a homogeneous spherical source of radius $R$ having uniform magnetic field $B$, filled with an isotropic emitting EED. The emission blob moves with a relativistic velocity $v$ and Lorentz factor $\Gamma=1/\sqrt{1-\left(v/c\right)^{2}}$. For viewing angle $\theta$, one has a Doppler beaming factor $\delta=1/\left[\Gamma\left(1-\left(v/c\right)\cos\theta\right)\right]$, which transforms frequency and luminosity from the jet frame to the AGN frame as $\nu=\delta\nu'$ and $\nu L(\nu)=\delta^{4}\nu' L'(\nu')$, respectively. Following the discussion in \citet{2018ApJS..235...39C} we adopted $\delta\lesssim\Gamma$ in our calculations. The reasons for this choice is (in case of jet opening angle $\theta_{j}\sim\theta$) that the causality argument requires $\theta_{j}\Gamma\lesssim1$ \citep{2013A&A...558A.144C}, which is supported by simulations of axis-symmetric magnetically driven outflows \citep{2009MNRAS.394.1182K}. The radio observation \citep[e.g.,][]{2005AJ....130.1418J, 2009A&A...507L..33P} and SSC process \citep{2014ApJ...789..161N} also indicate $\theta_{j}\Gamma\gtrsim0.1-0.7$. We employ a log-parabolic EED because (1) particle acceleration mechanism can easily produce a log-parabolic EED \citep[see e.g.,][and C14]{1962SvA.....6..317K, 2004A&A...413..489M, 2006A&A...448..861M, 2011ApJ...739...66T} and (2) the coefficient of the second-order term can be easily taken to measure curvature. Within an energy range $\gamma_{min}-\gamma_{max}$, the energy distribution follows,
\begin{equation}
N\left(\gamma\right)=N_{0}\left(\frac{\gamma}{\gamma_{p}}\right)^{-3} 10^{-b\log\left(\gamma/\gamma_{p}\right)^{2}},
\label{EED}
\end{equation}
where $b$ is the curvature parameter, $N_0$ is particle energy density, and $\gamma_p$ is the electron peak energy. With this form, electrons peaked at energy $\gamma_{p}$ emit photons at SED peak. We employ the EED ranging from $\gamma_{min} = 2$ through peak energy $\gamma_p$ up to a maximum energy $\gamma_{max} = 10^4 \gamma_p$. The exact determination of $\gamma_{max}$ may be practically irrelevant as the IC losses are limited by the Klein-Nishina cross section. The total luminosity from radiating electrons is \citep[see][for more details]{2017ApJ...842..129C},
\begin{equation}
L'(\nu')=2\pi^{2}R^{3}j(\nu')\frac{2\tau^{2}-1+\left(2\tau+1\right) e^{-2\tau}}{\tau^{3}},
\end{equation}
with optical depth $\tau=k(\nu')R$, the absorption coefficient $k(\nu')$ and the emitting coefficient $j(\nu')$ \citep{1970RvMP...42..237B, 1979rpa..book.....R}. The emission coefficient of synchrotron/IC component radiation becomes,
\begin{equation}
j(\nu')= \frac{1}{4\pi} \int N(\gamma) P(\nu',\gamma) d\gamma
\end{equation}
where $P(\nu',\gamma)$ is the power of a single electron. For SSC, we use the average seed photon energy density $u_s=(9/4) L_s'/(4 \pi R^2 c)$ in our calculation \citep[for the case of the optical thin region, see][for details]{2017ApJ...842..129C}, leading to a SSC dominance $u_s/u_B$. In the case of EC, the external seed photon density in jet frame is enhanced as $u_{ext}'=\frac{17}{12} \Gamma^2 u_{ext}$, leading to EC dominance $u_{ext}/u_B$. We use the full Klein-Nishina cross section \citep{1970RvMP...42..237B} to calculate the IC losses in SSC and EC components. For FSRQs and LBLs, EC becomes important and may dominate their $\gamma$-ray emissions as compared to SSC. Recent works show that the emission region may be outside of BLR \citep[see, e.g.,][]{2014ApJ...789..161N, 2018A&A...616A..63A}, and therefore the seed photons for EC dominantly come from dusty torus \citep[see, e.g.,][]{2013MNRAS.436.2170C, 2014ApJS..215....5K, 2016MNRAS.456.2173Y}. In our calculation, we assume seed photons arising from a cold dusty torus emitting nearly as a blackbody with peak frequency $\nu_{ext}=3\times 10^{13}$ Hz. Only in a few cases where the dust seed photons fail to fit the high energy $\gamma$-ray emission, we include the contribution from H$\alpha$ seed photons from a thermal BLR centered at $\nu_{ext}= 2\times10^{15}$ Hz. We leave external seed photon energy density $u_{ext}$ as a free parameter constrained by fitting the Fermi-LAT GeV observations \citep{2017ApJ...837...38K}.\\

\begin{figure}[t]
\centering
\includegraphics[height=4cm, width=3cm, angle=-90]{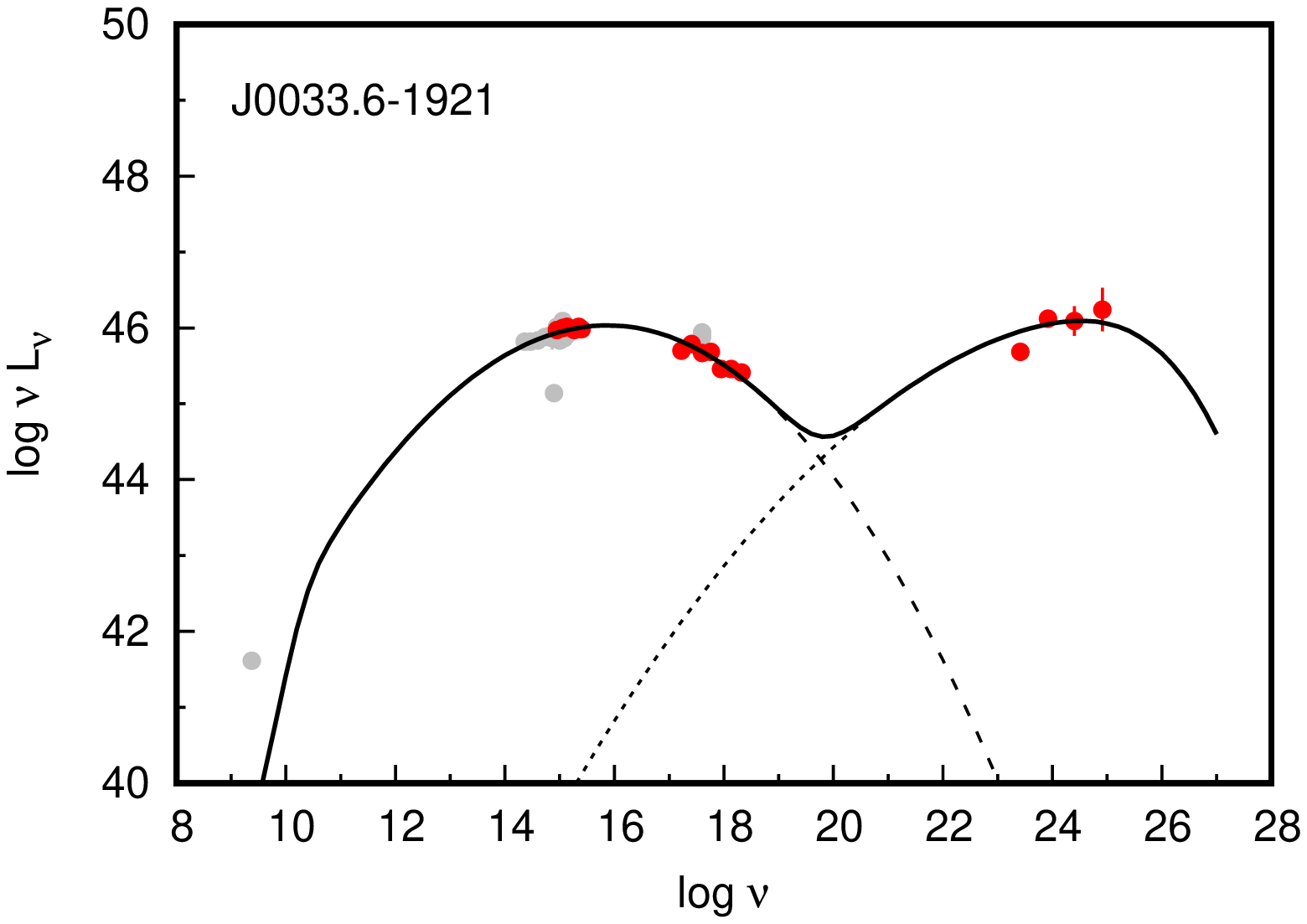}
\includegraphics[height=4cm, width=3cm, angle=-90]{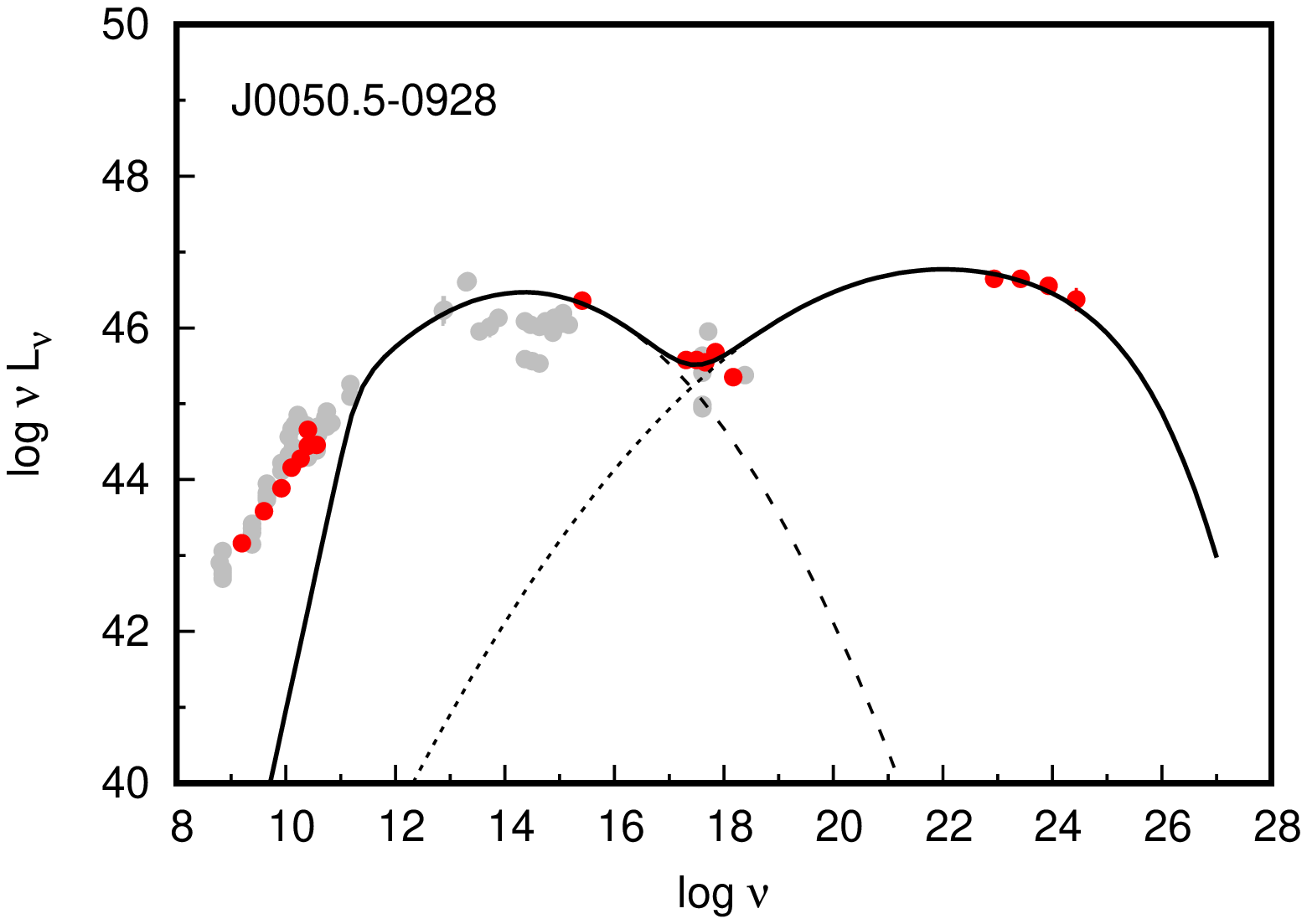}
\includegraphics[height=4cm, width=3cm, angle=-90]{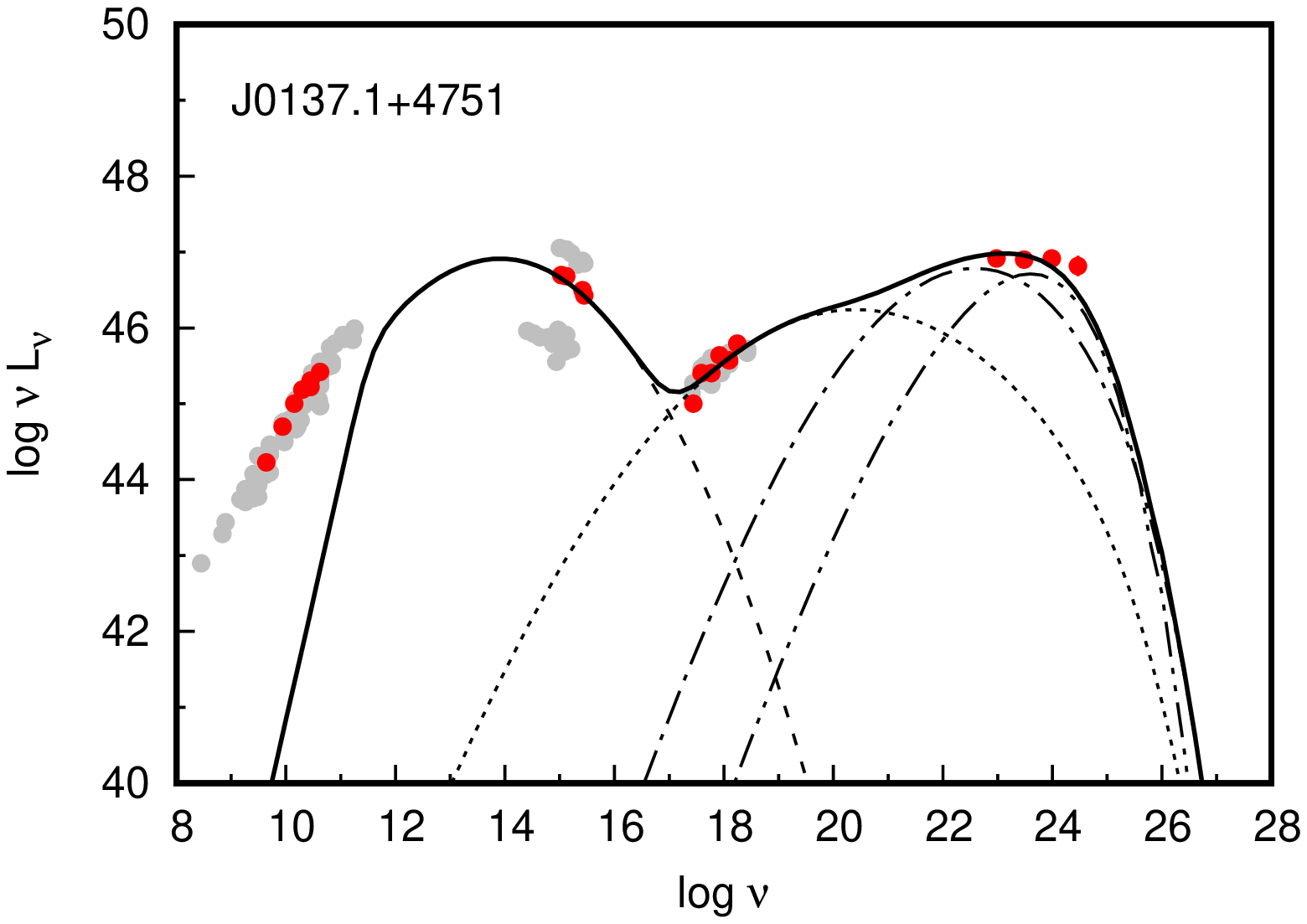}
\includegraphics[height=4cm, width=3cm, angle=-90]{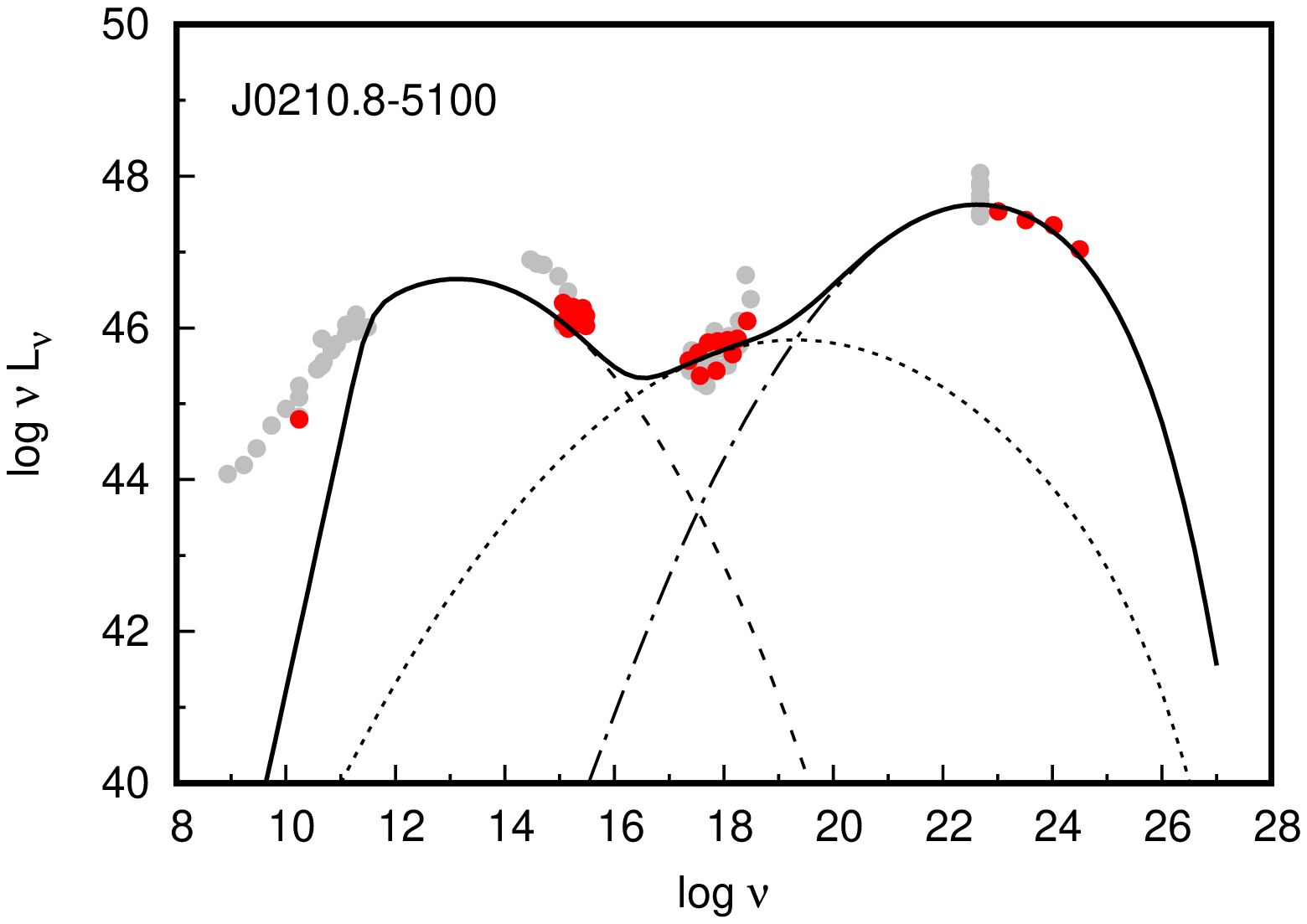}
\includegraphics[height=4cm, width=3cm, angle=-90]{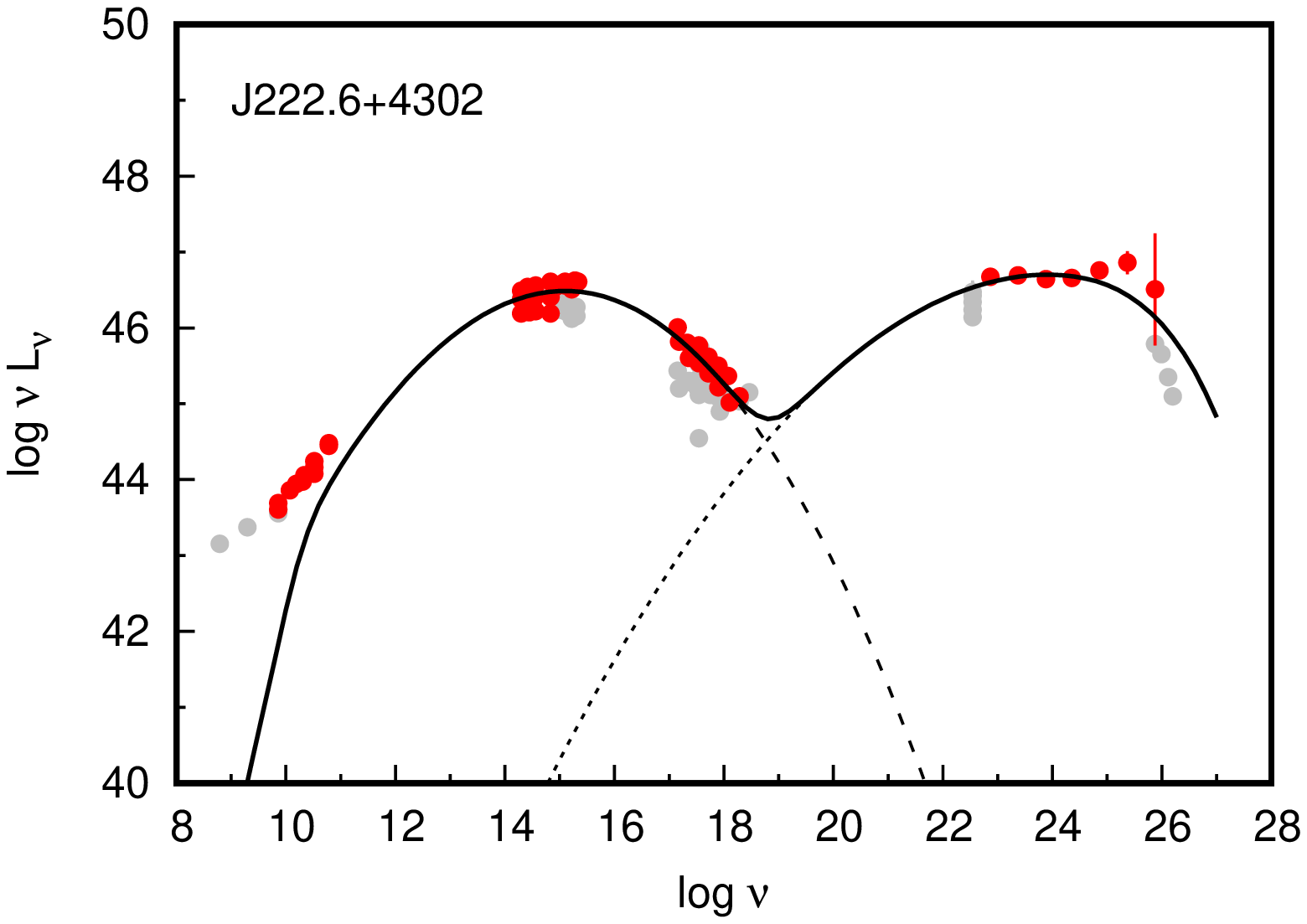}
\includegraphics[height=4cm, width=3cm, angle=-90]{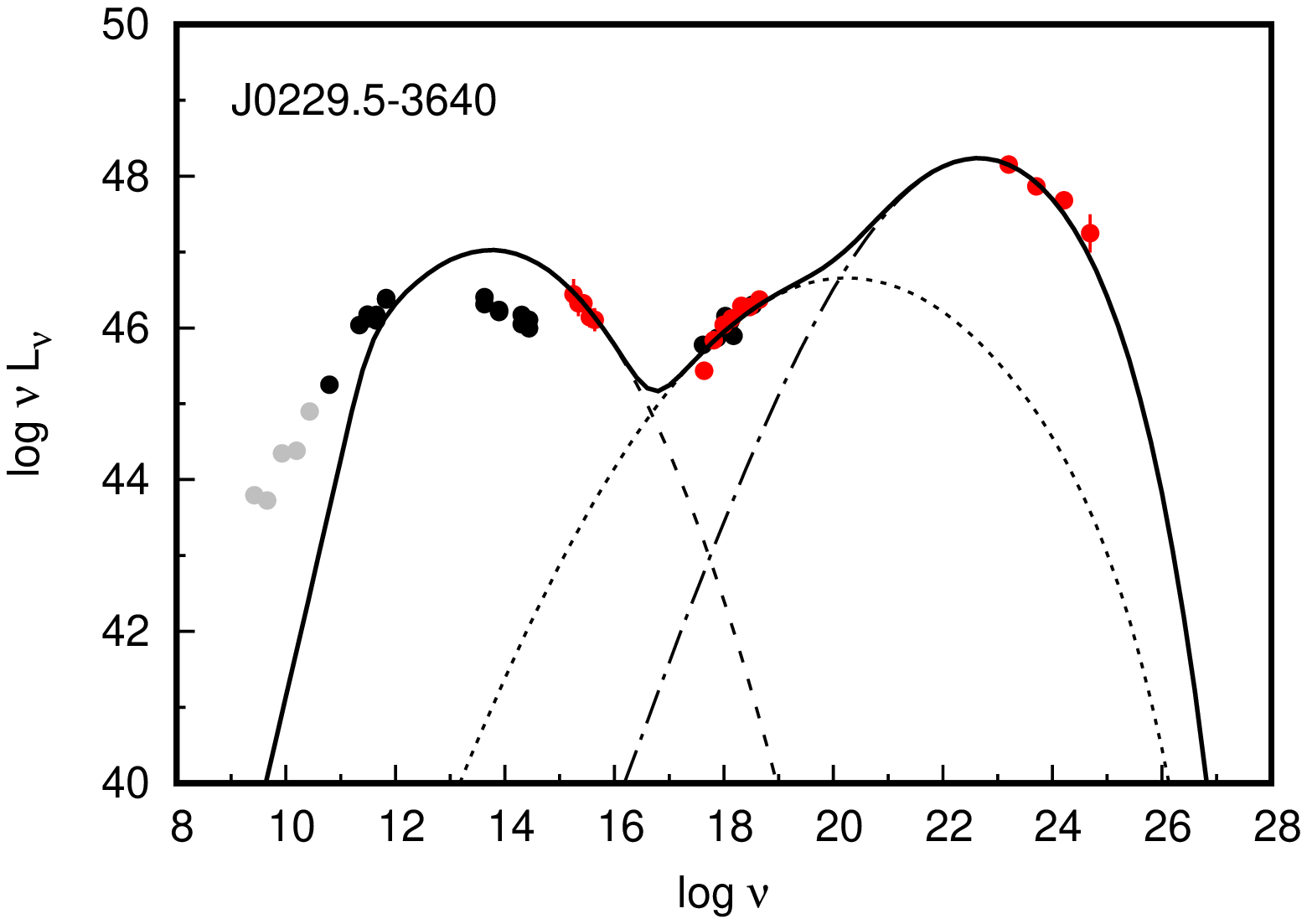}
\includegraphics[height=4cm, width=3cm, angle=-90]{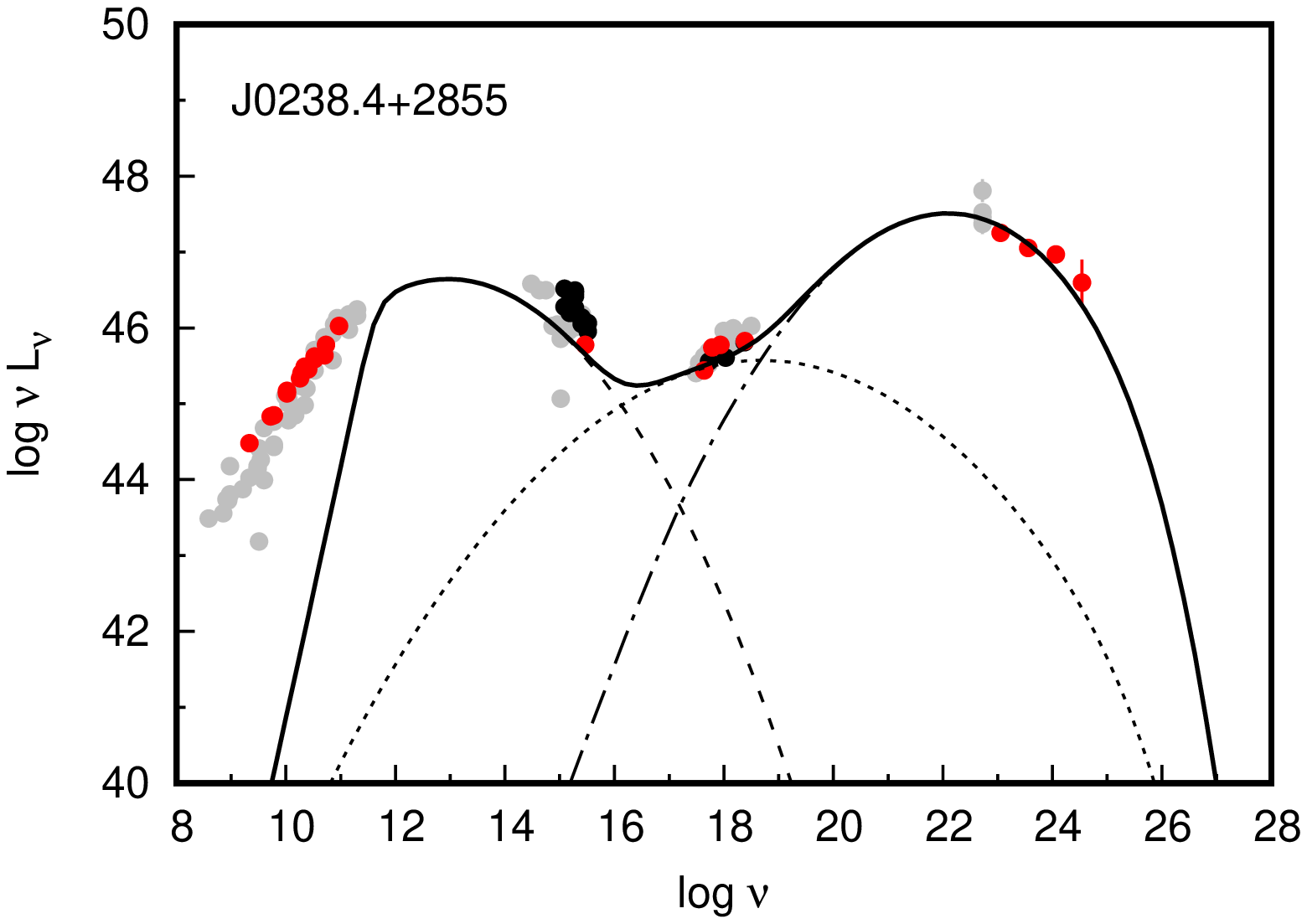}
\includegraphics[height=4cm, width=3cm, angle=-90]{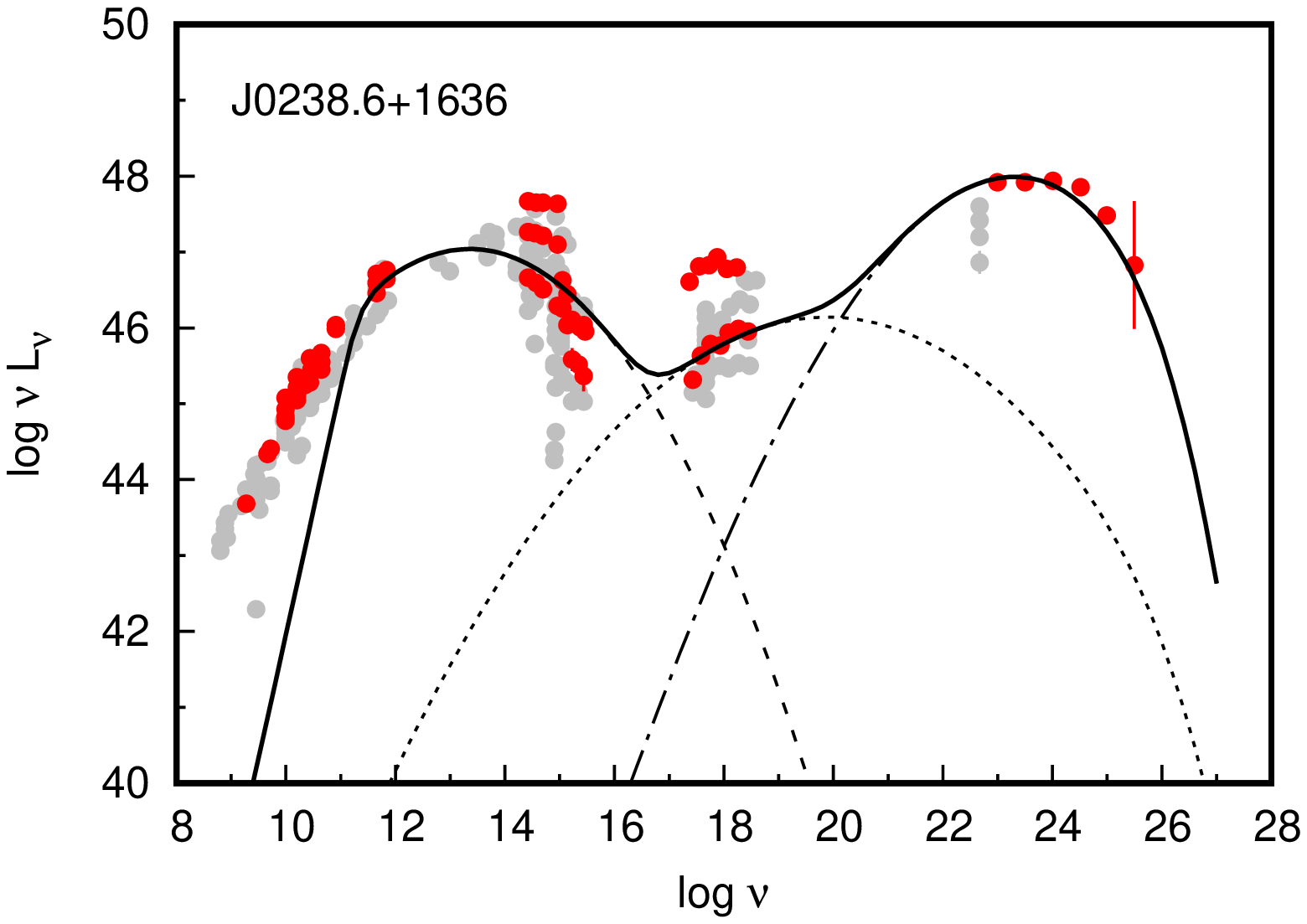}
\includegraphics[height=4cm, width=3cm, angle=-90]{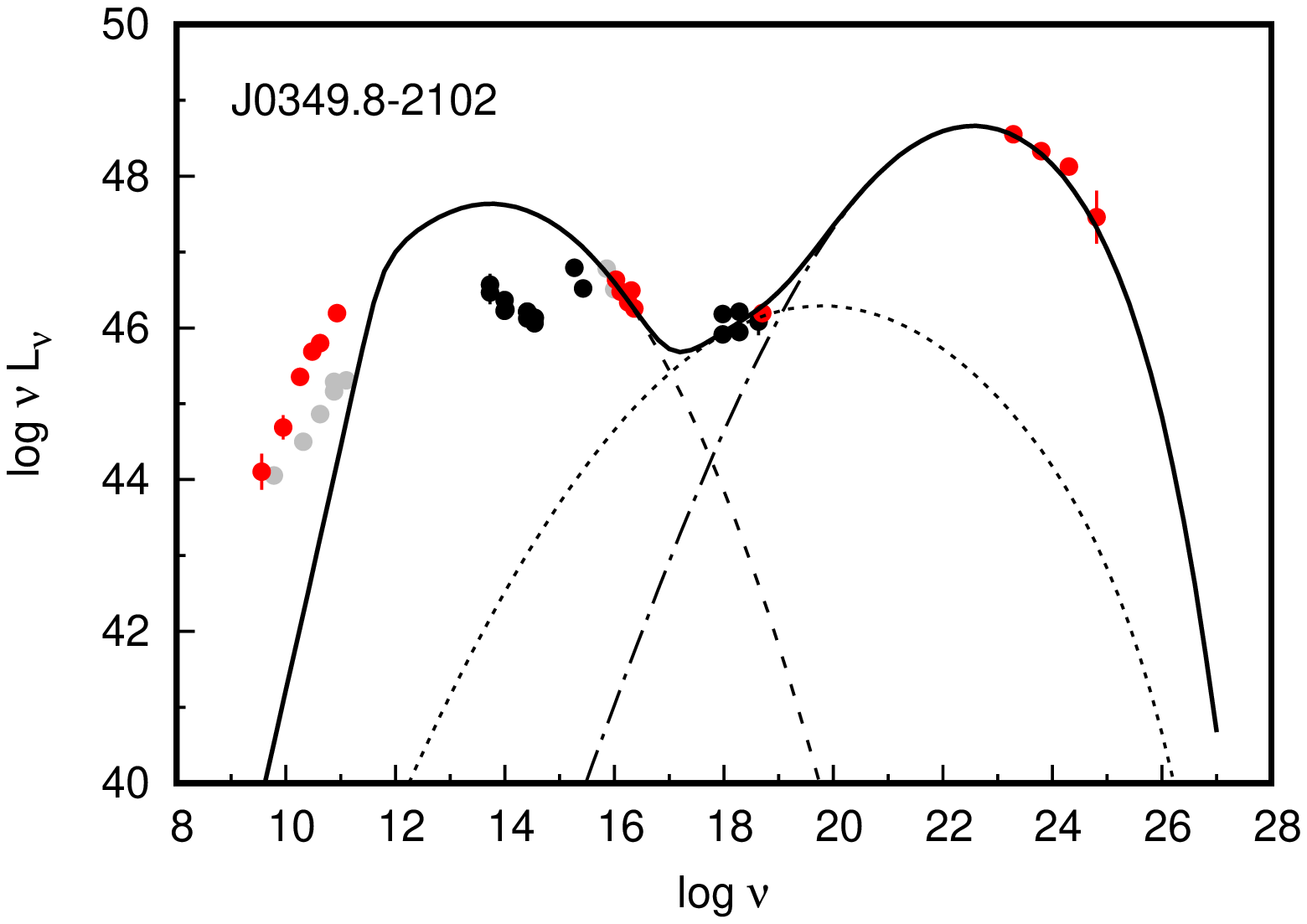}
\includegraphics[height=4cm, width=3cm, angle=-90]{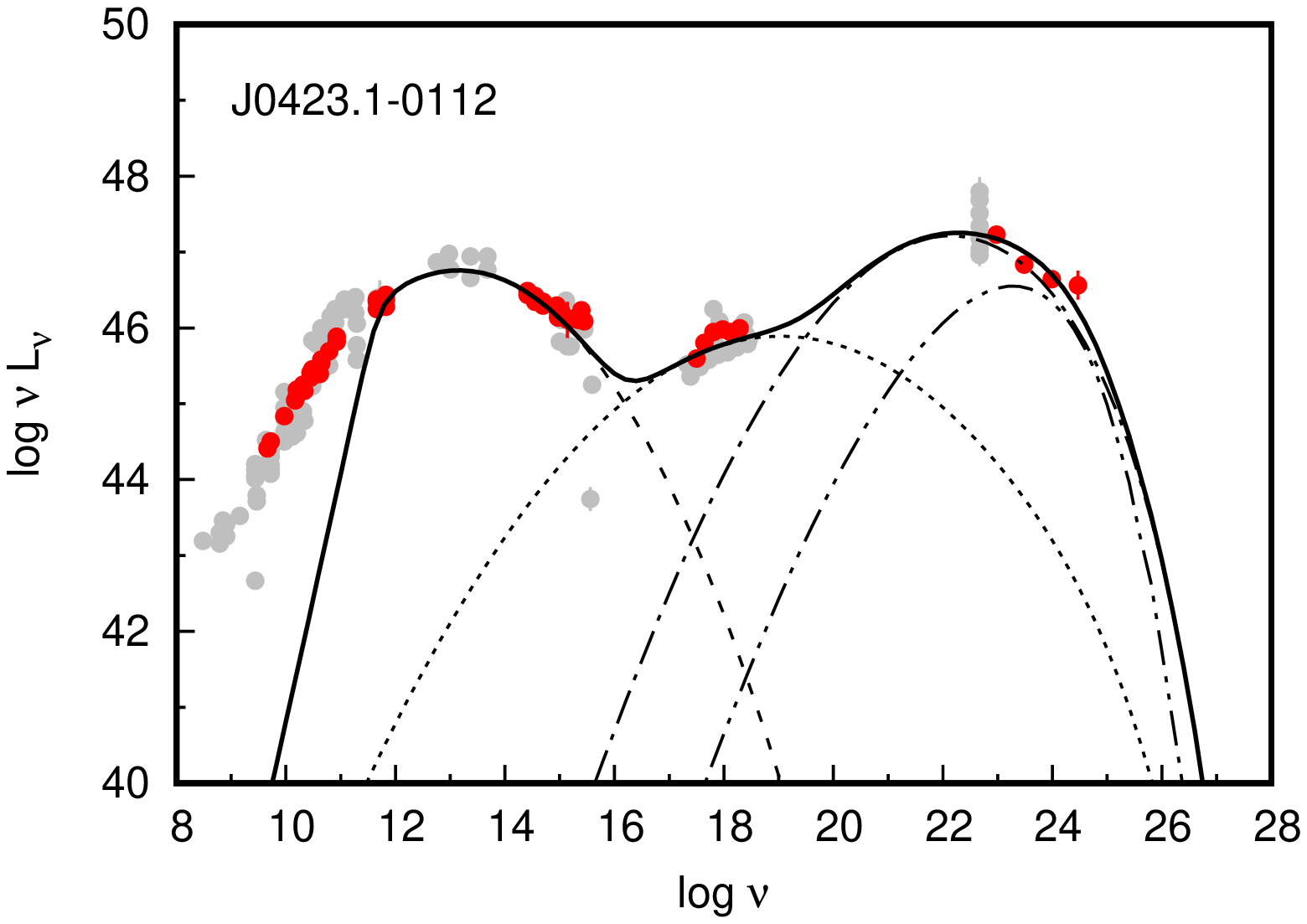}
\includegraphics[height=4cm, width=3cm, angle=-90]{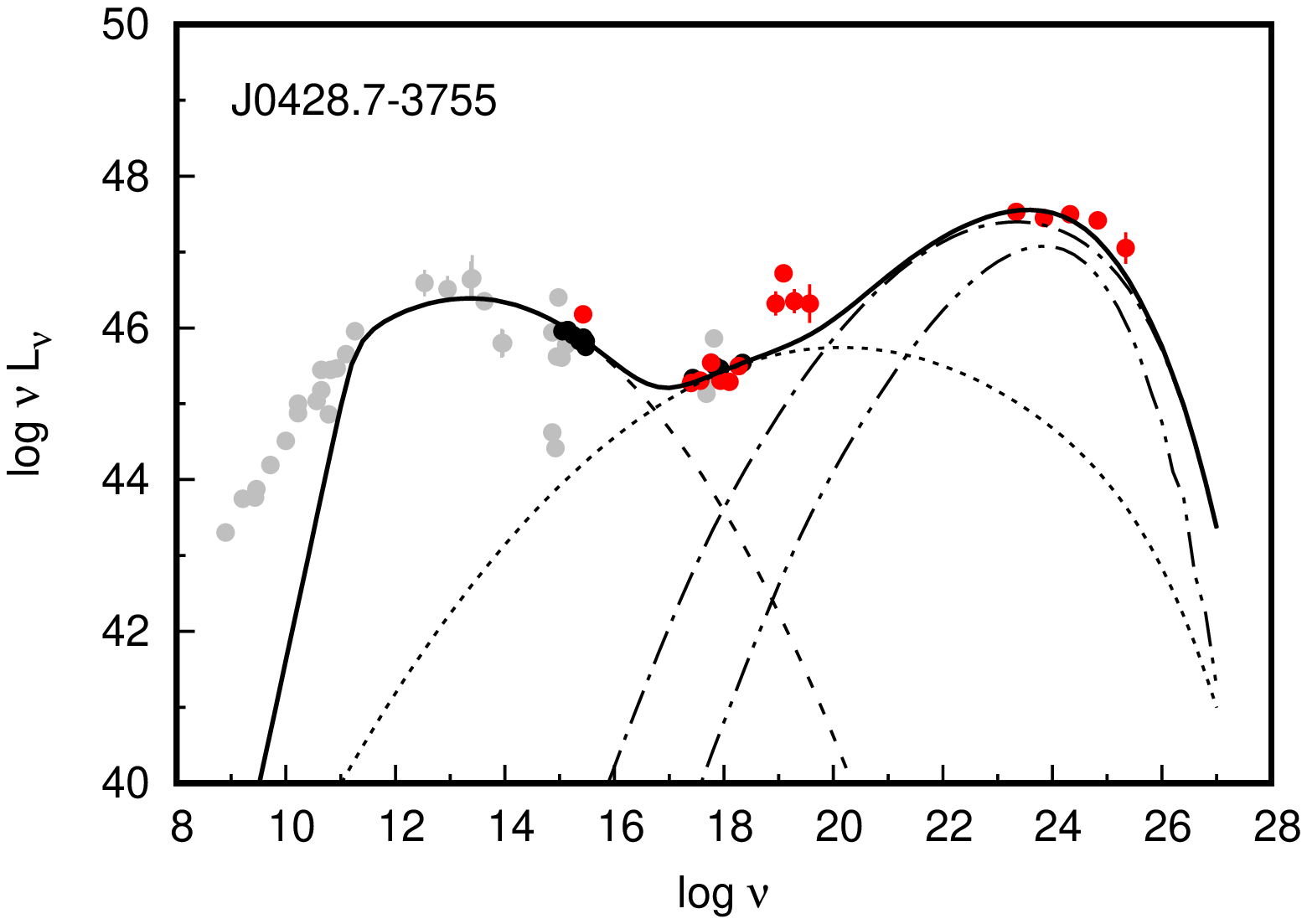}
\includegraphics[height=4cm, width=3cm, angle=-90]{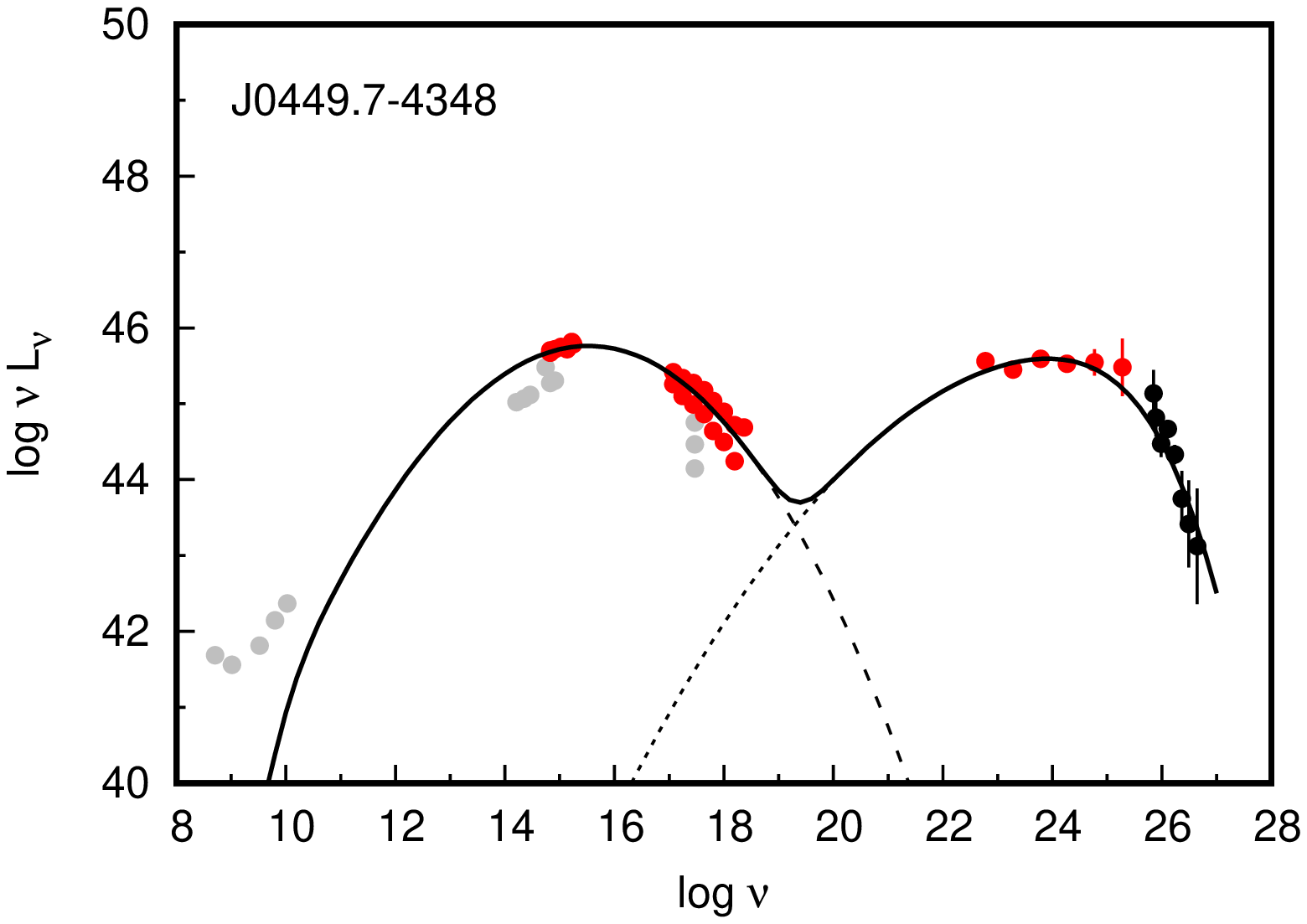}
\includegraphics[height=4cm, width=3cm, angle=-90]{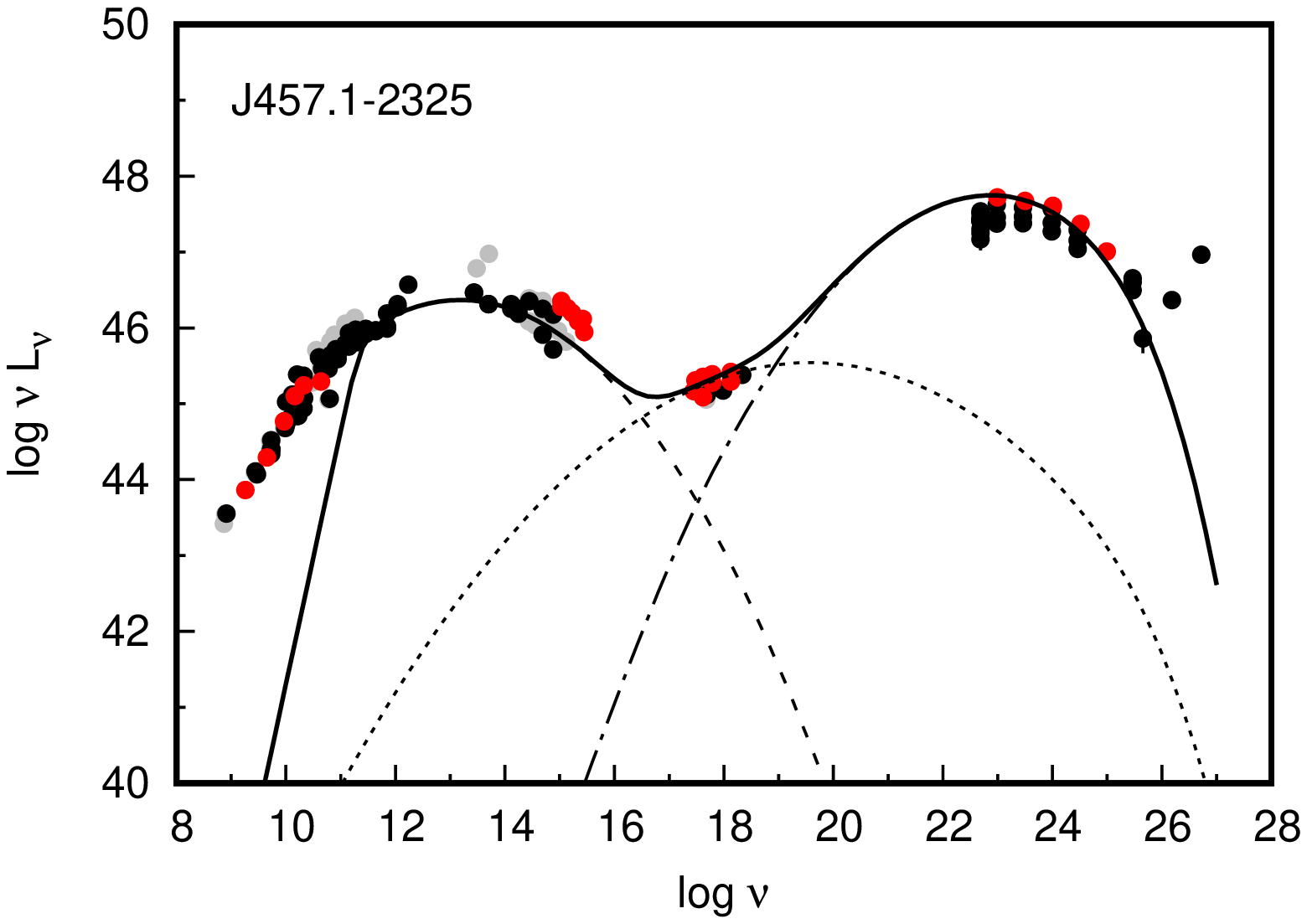}
\includegraphics[height=4cm, width=3cm, angle=-90]{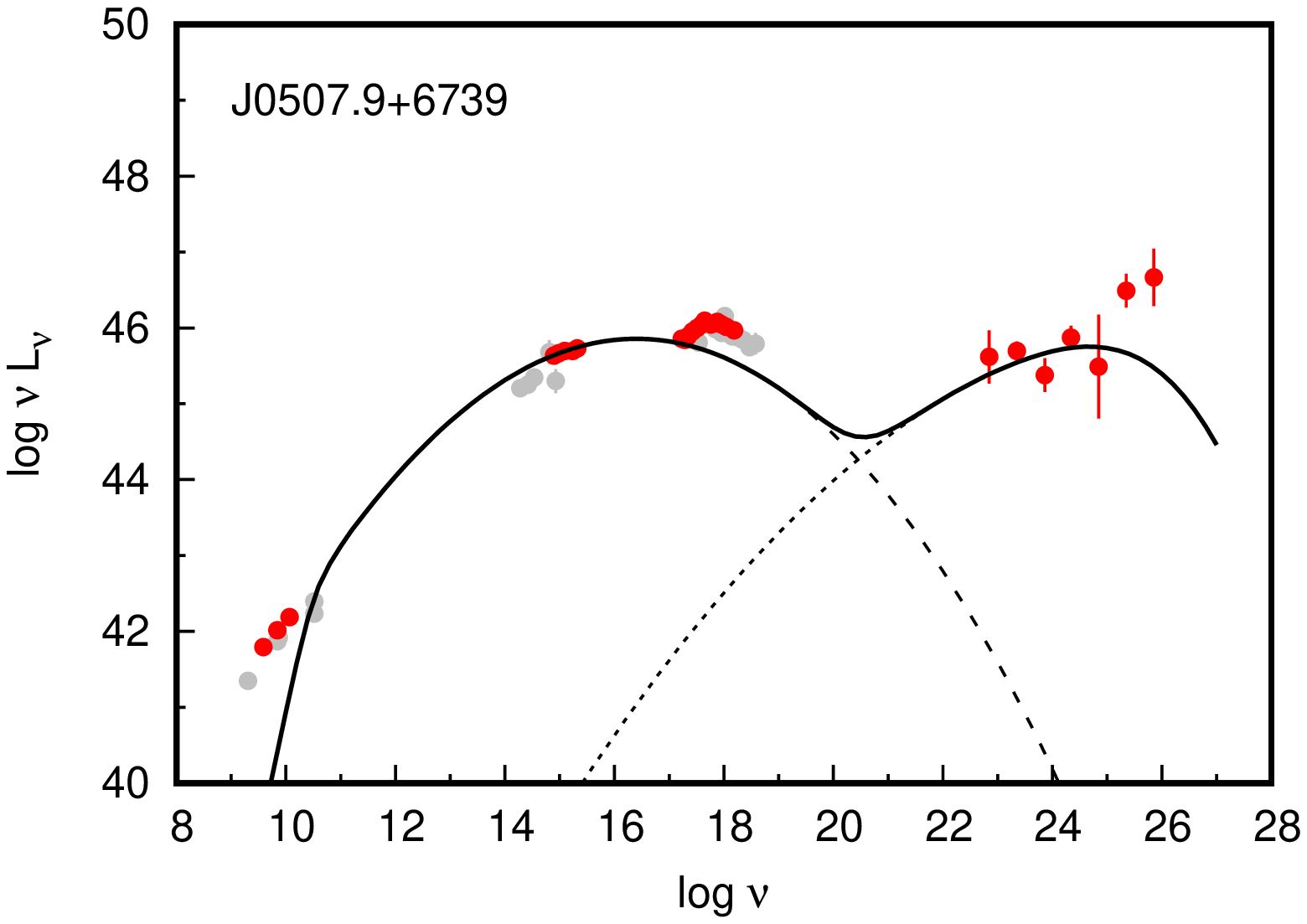}
\includegraphics[height=4cm, width=3cm, angle=-90]{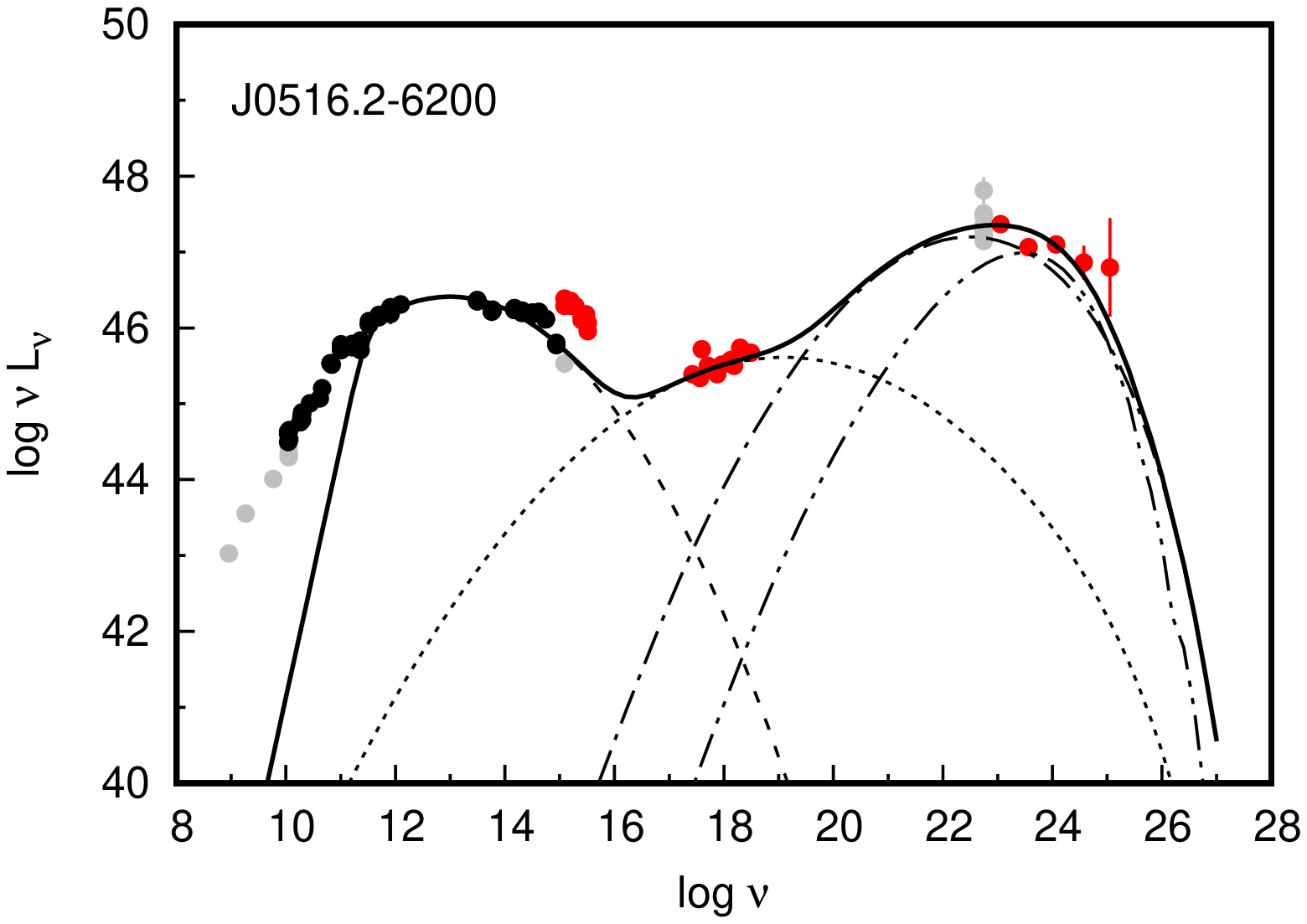}
\includegraphics[height=4cm, width=3cm, angle=-90]{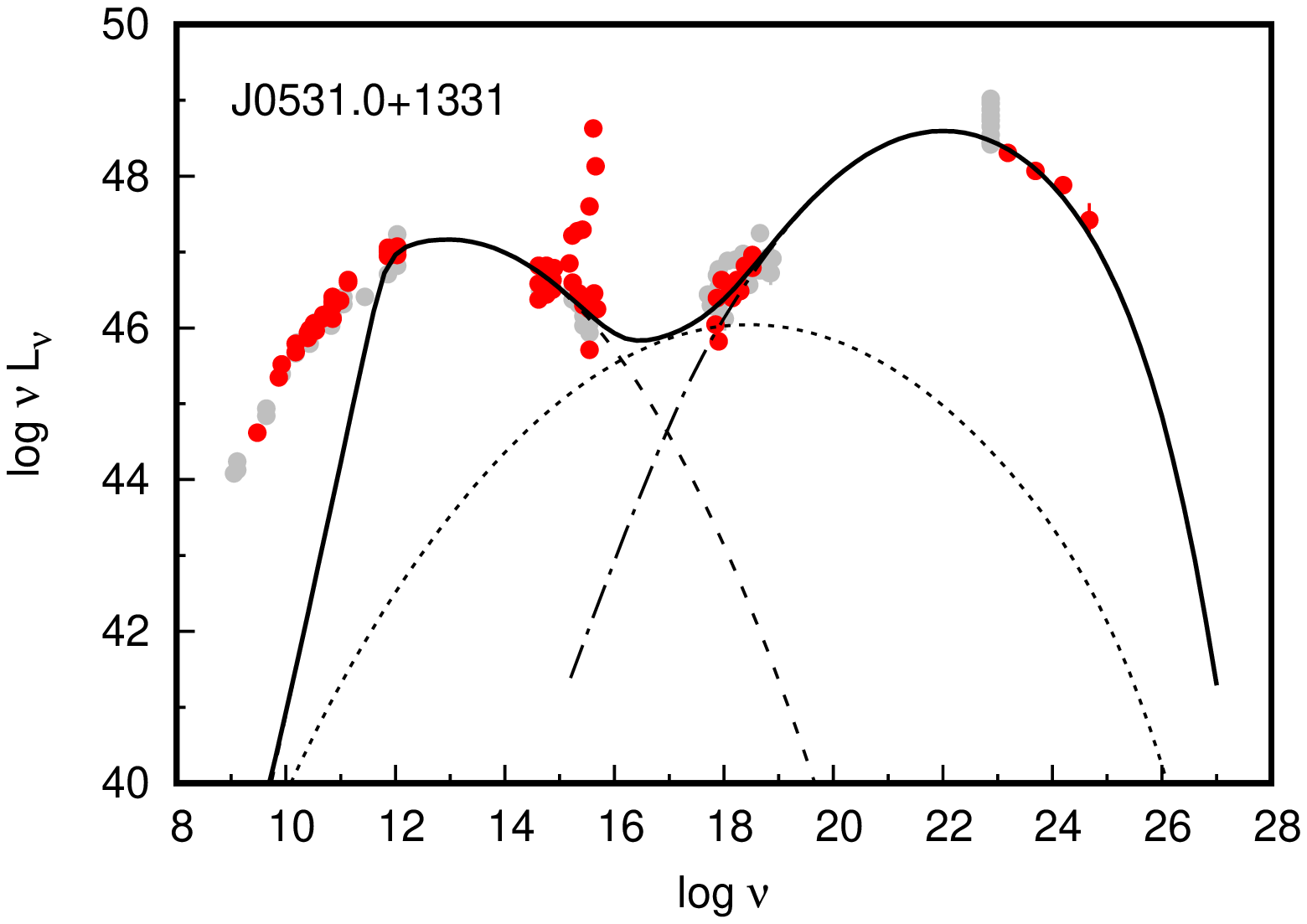}
\caption{The SED modeling of LBAS blazars. The simultaneous data are shown as red circles, while gray and black circles represent the historical and additional archival observations, respectively. The synchrotron, SSC, EC dust, EC BLR, and total model curves are represented by dashed, dotted, dashed-dotted, dashed-double dotted, and solid lines, respectively.}
\label{fig:fits}
\end{figure}

We constrain the global physical parameters and curvature by fitting the SEDs. The causality condition for variability timescale sets an upper limit on the size of the emitting region, $R\lesssim\delta c\Delta t/(1+z)$. The variability timescales may be different in various sources \citep[see][for a review,]{1997ARA&A..35..445U}, \citet{2011MNRAS.410..368B} and \citet{2011ApJ...736..131A} for the well-studied blazars 3C 454.3 and Mrk 421, respectively, and \citet{2013MNRAS.430.1324N} for a systematic study indicating a typical variability timescale in the source frame in the Fermi-LAT band of $\sim 1$ day; see also \citet{2010ApJ...718..279G, 2014MNRAS.443.2940H}, while for consistency, we employ the average value $\Delta t \approx 1$ day \citep[e.g.,][]{2014ApJS..215....5K, 2015ApJ...807...51Z, 2018ApJS..235...39C}. For the injected electron distribution, each SED for a particular class of blazars, e.g., FSRQ, can be considered a different state of evolution of electrons and its its spectral curvature. Similar to \citet{2014Natur.515..376G} and \citet{2015MNRAS.448.1060G}, we present the best model fits based on visual inspection for each source SED in Figure \ref{fig:fits} and the corresponding jet parameters are reported in Table \ref{tab:parameters}. The red points in Figure \ref{fig:fits} represent the quasi-simultaneous spectral data, while the gray and black ones represent historical observations \citep[see][for a detailed description]{2010ApJ...716...30A} to guide our fitting. It can be seen that, for all blazars, either the synchrotron or IC component has a good coverage of SED data, which makes the modeling less uncertain. While the sample is limited due to availability of simultaneous observations of blazars, the meaningful statistical trends can highlight the relationship between curvature and source physical parameters. The error bars of the physical parameters $b$ and $\gamma_p$ are obtained using the average of uncertainties in the SED peak frequency and an average SED curvature (see Table 1 of C14). 
 
\section{Results and Discussion} \label{sec:results}
\begin{figure*}
\centering
\includegraphics[height=4cm, width=3cm, angle=-90]{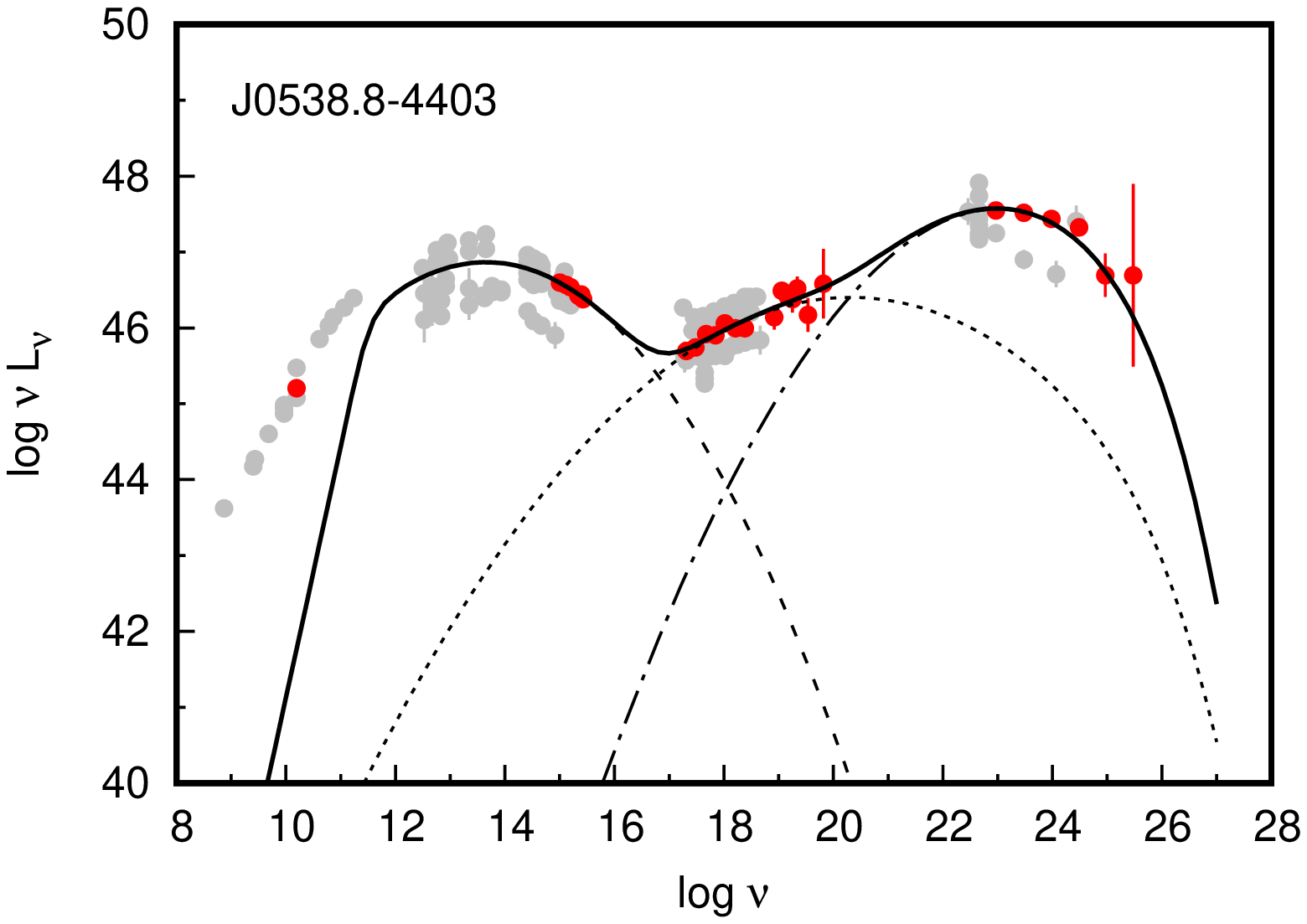}
\includegraphics[height=4cm, width=3cm, angle=-90]{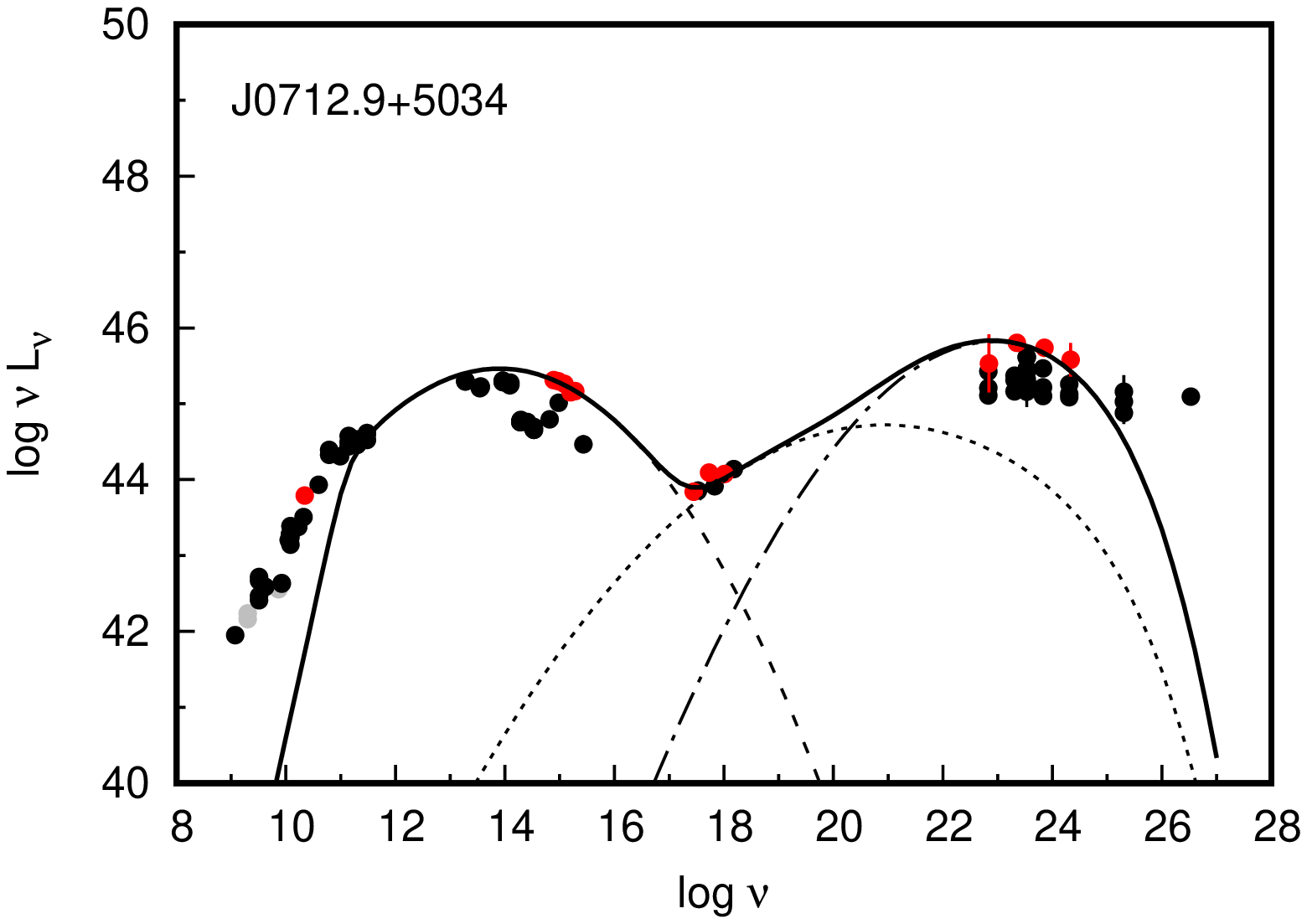}
\includegraphics[height=4cm, width=3cm, angle=-90]{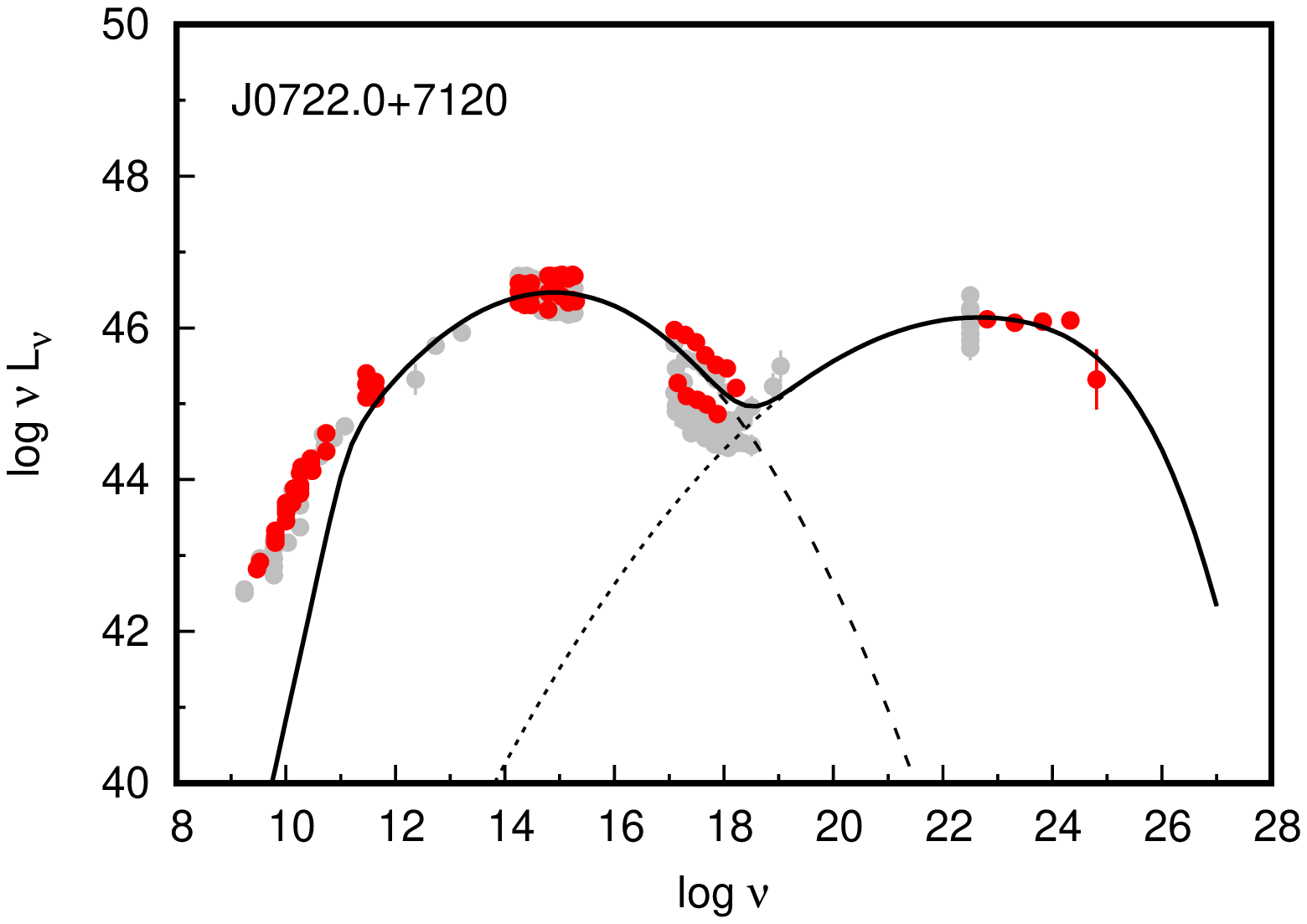}
\includegraphics[height=4cm, width=3cm, angle=-90]{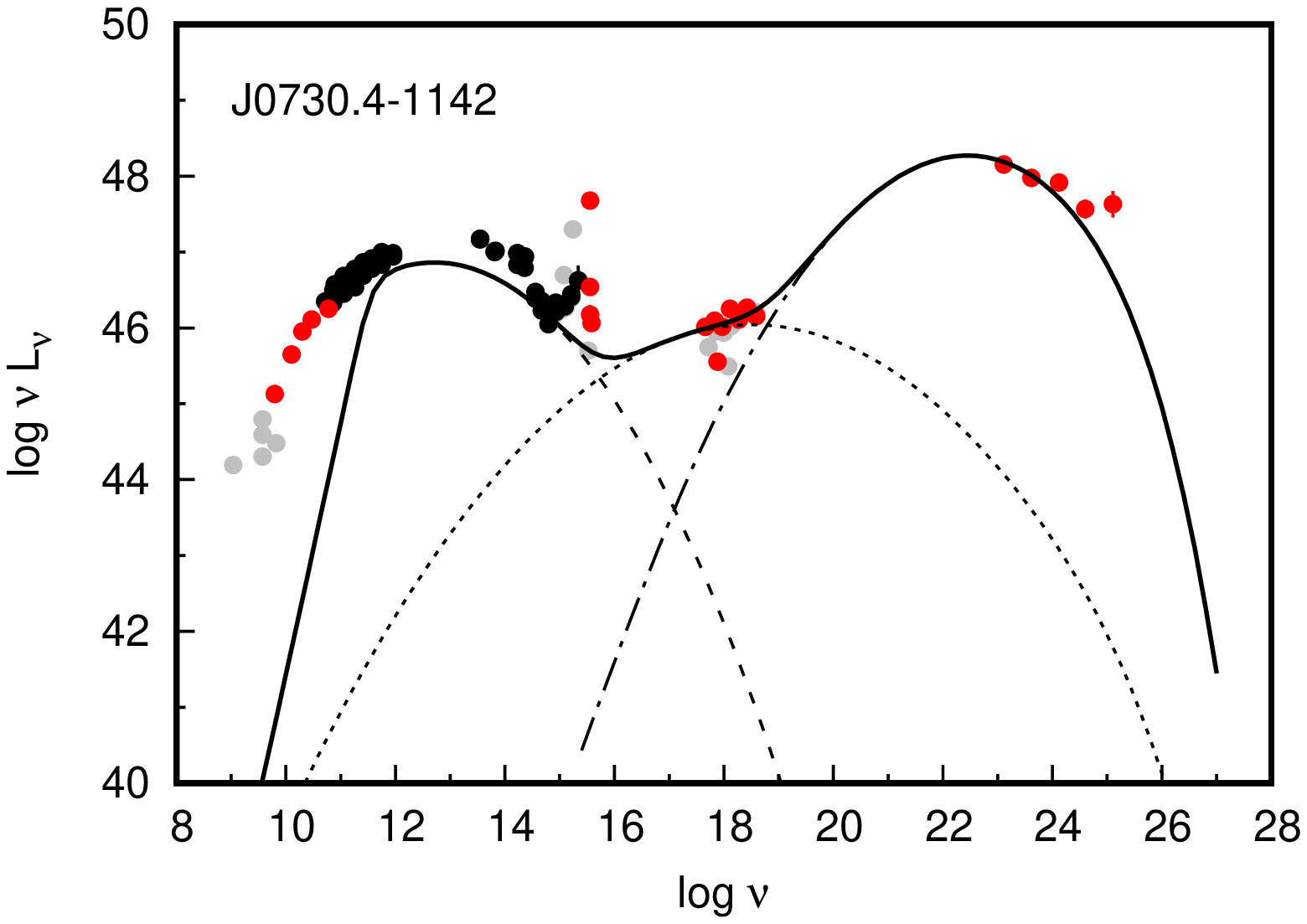}
\includegraphics[height=4cm, width=3cm, angle=-90]{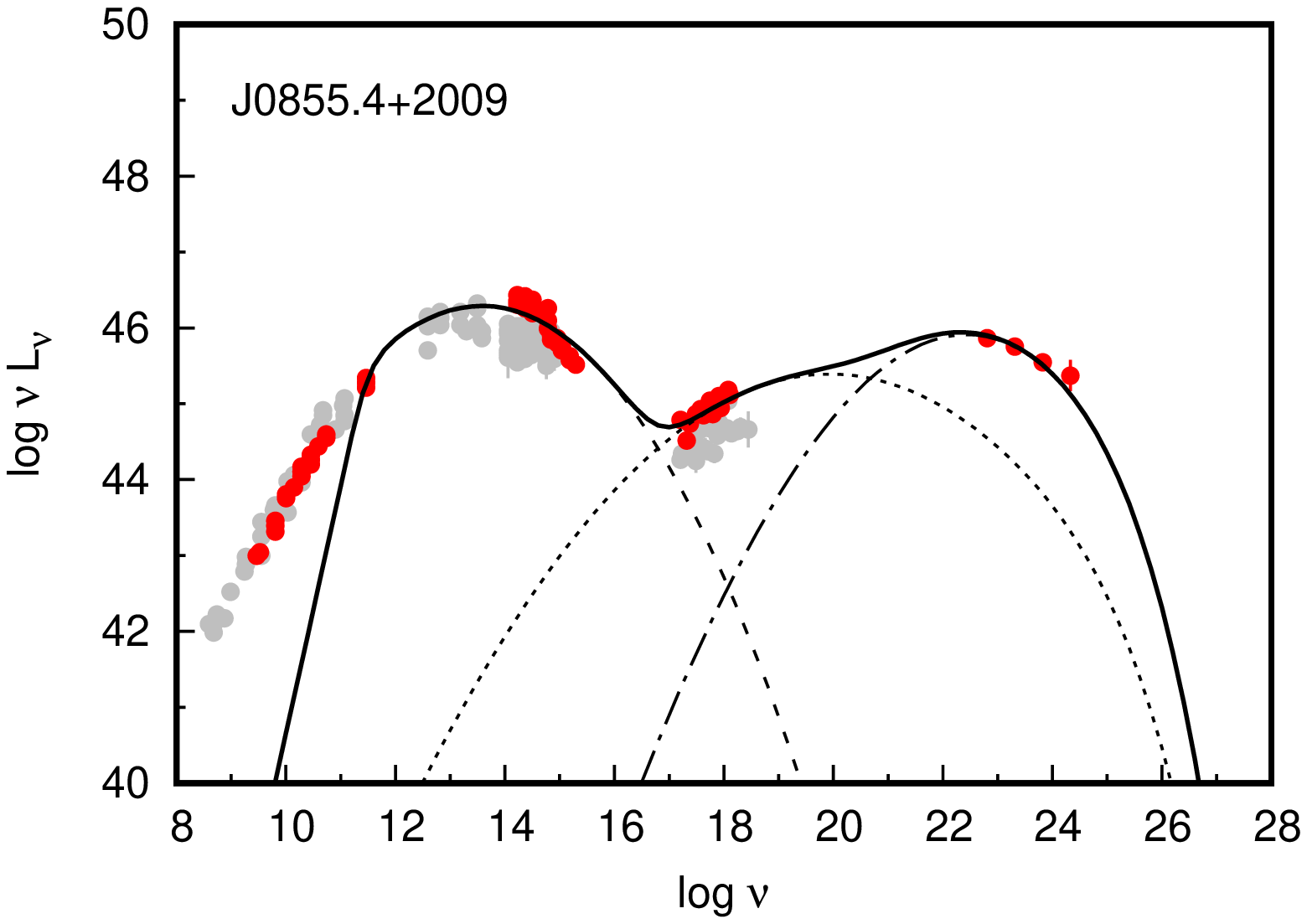}
\includegraphics[height=4cm, width=3cm, angle=-90]{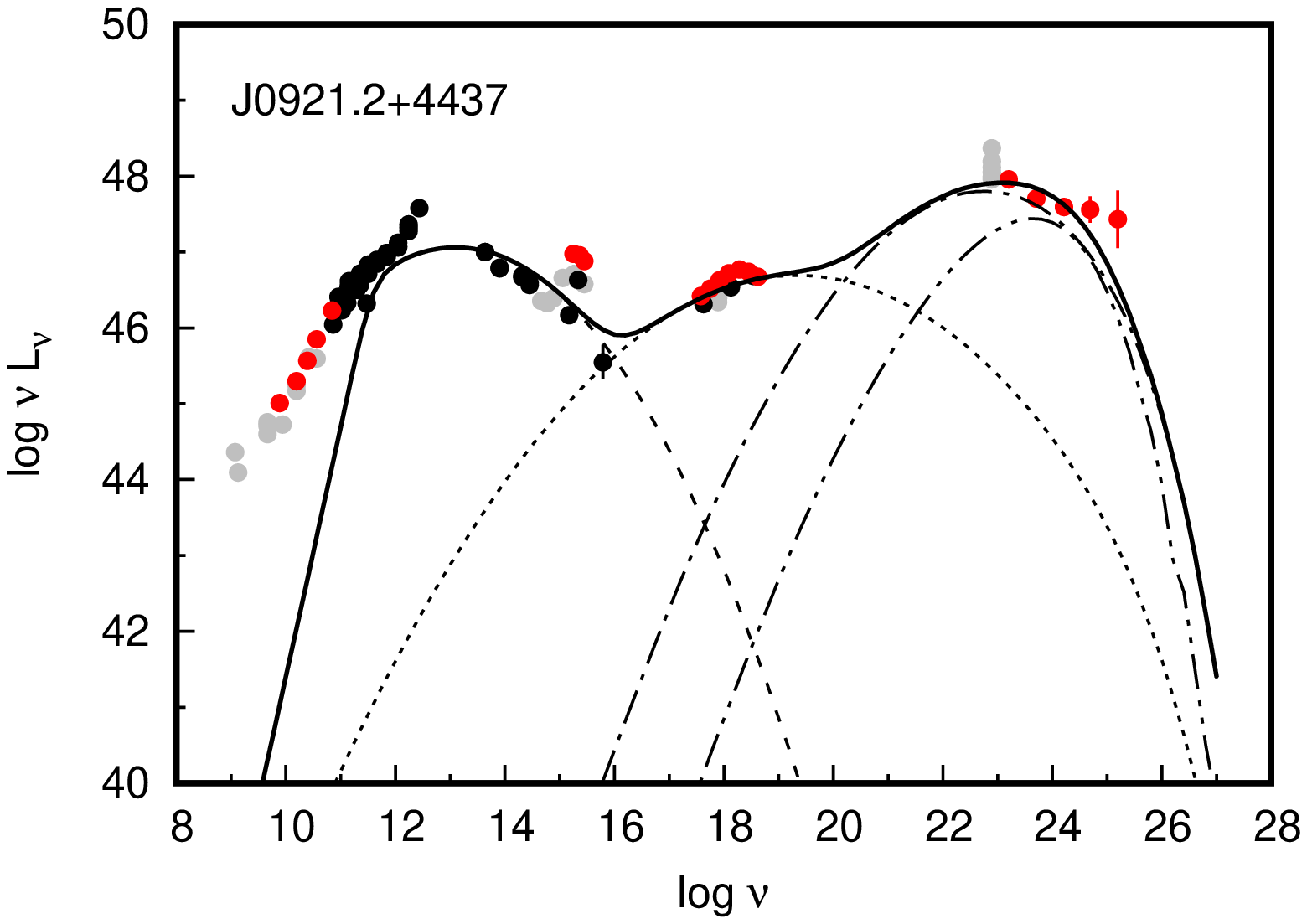}
\includegraphics[height=4cm, width=3cm, angle=-90]{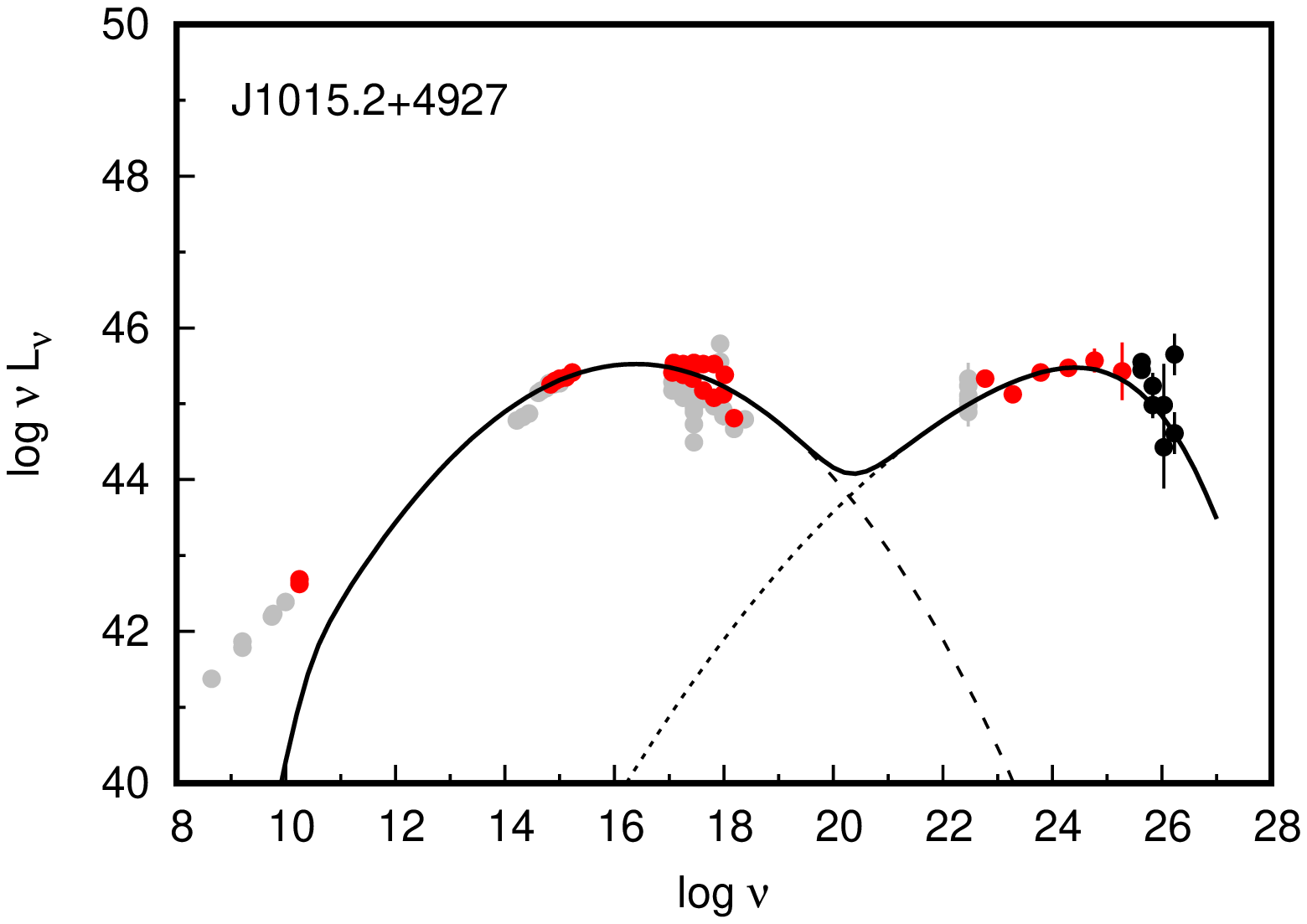}
\includegraphics[height=4cm, width=3cm, angle=-90]{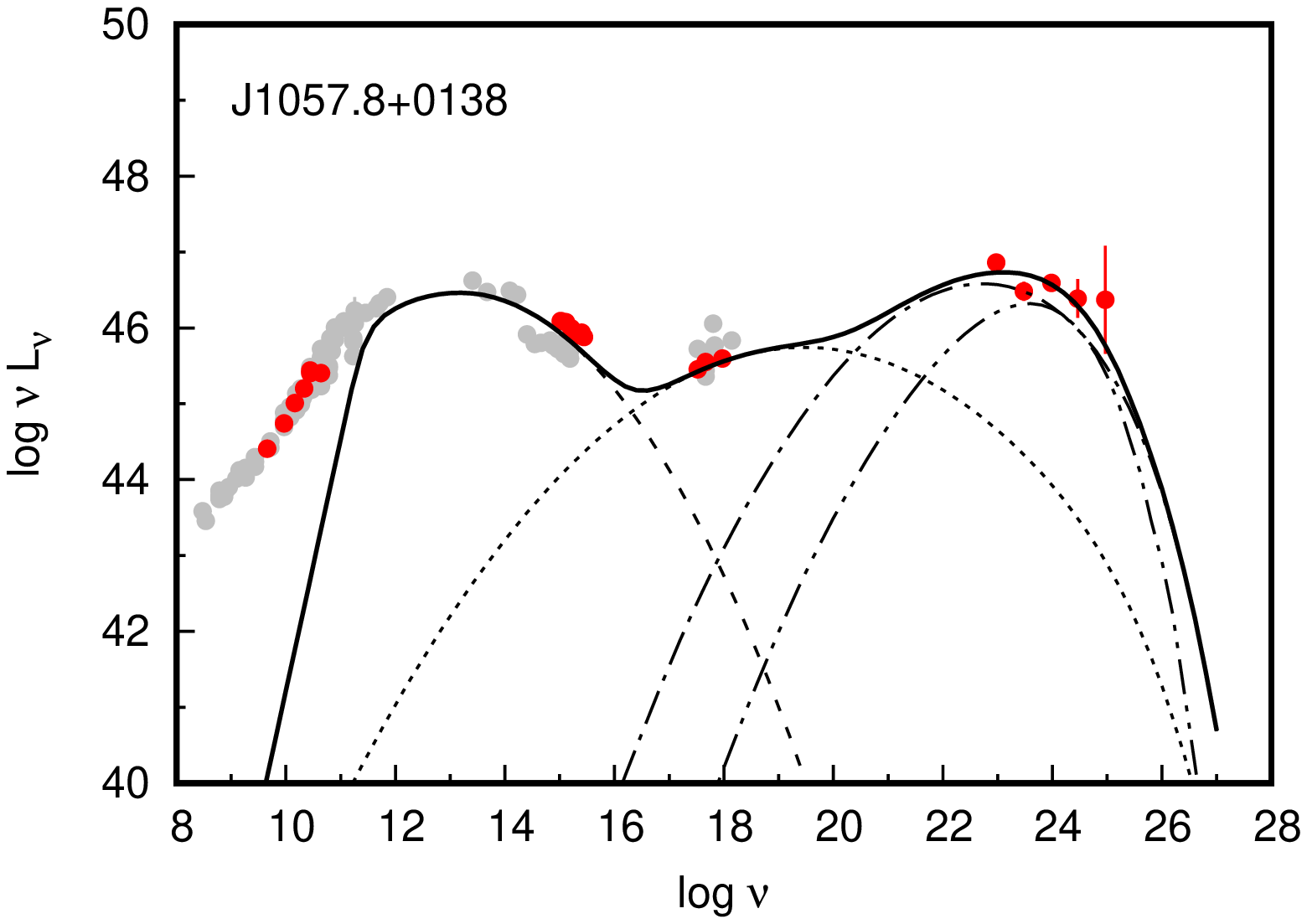}
\includegraphics[height=4cm, width=3cm, angle=-90]{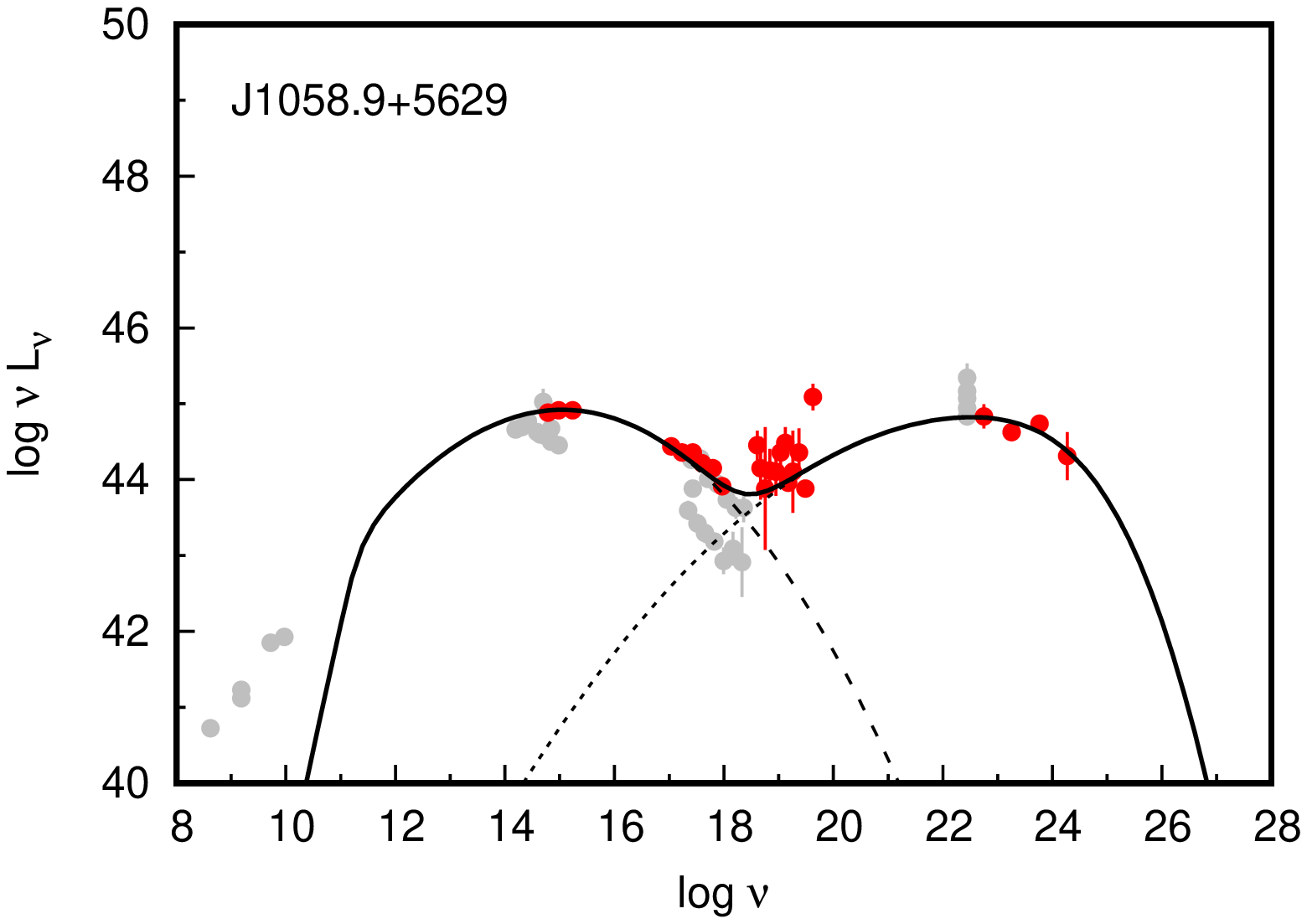}
\includegraphics[height=4cm, width=3cm, angle=-90]{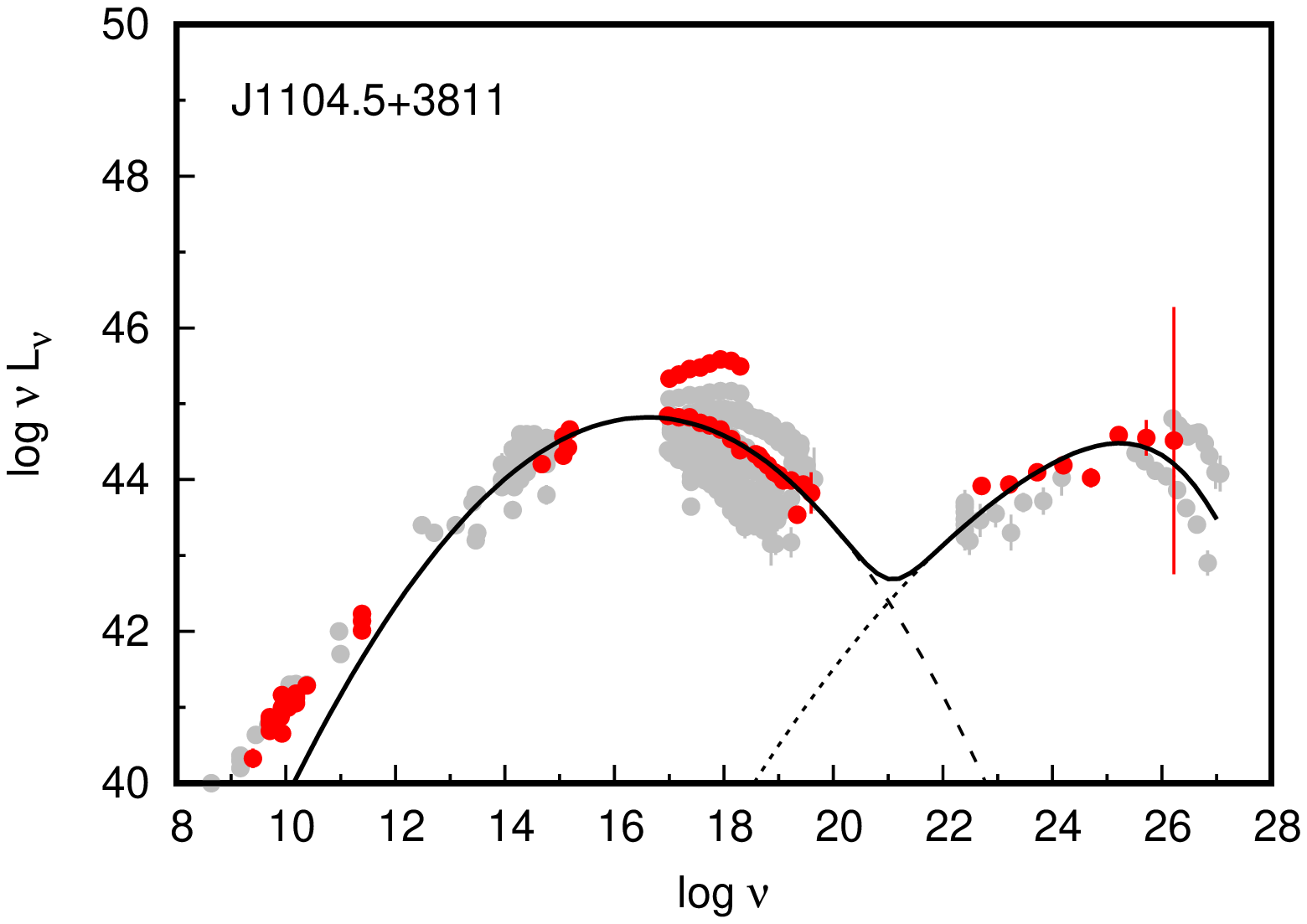}
\includegraphics[height=4cm, width=3cm, angle=-90]{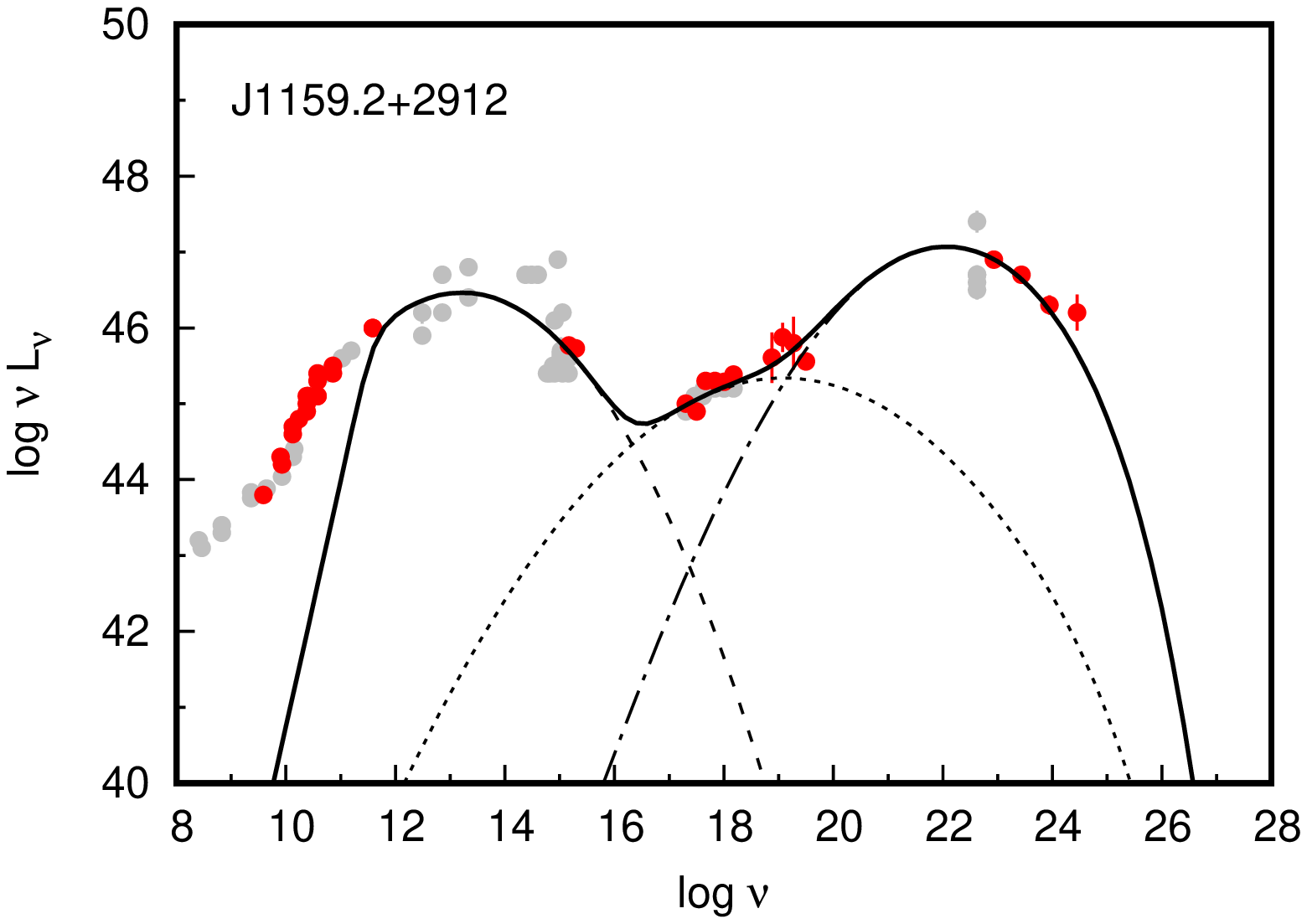}
\includegraphics[height=4cm, width=3cm, angle=-90]{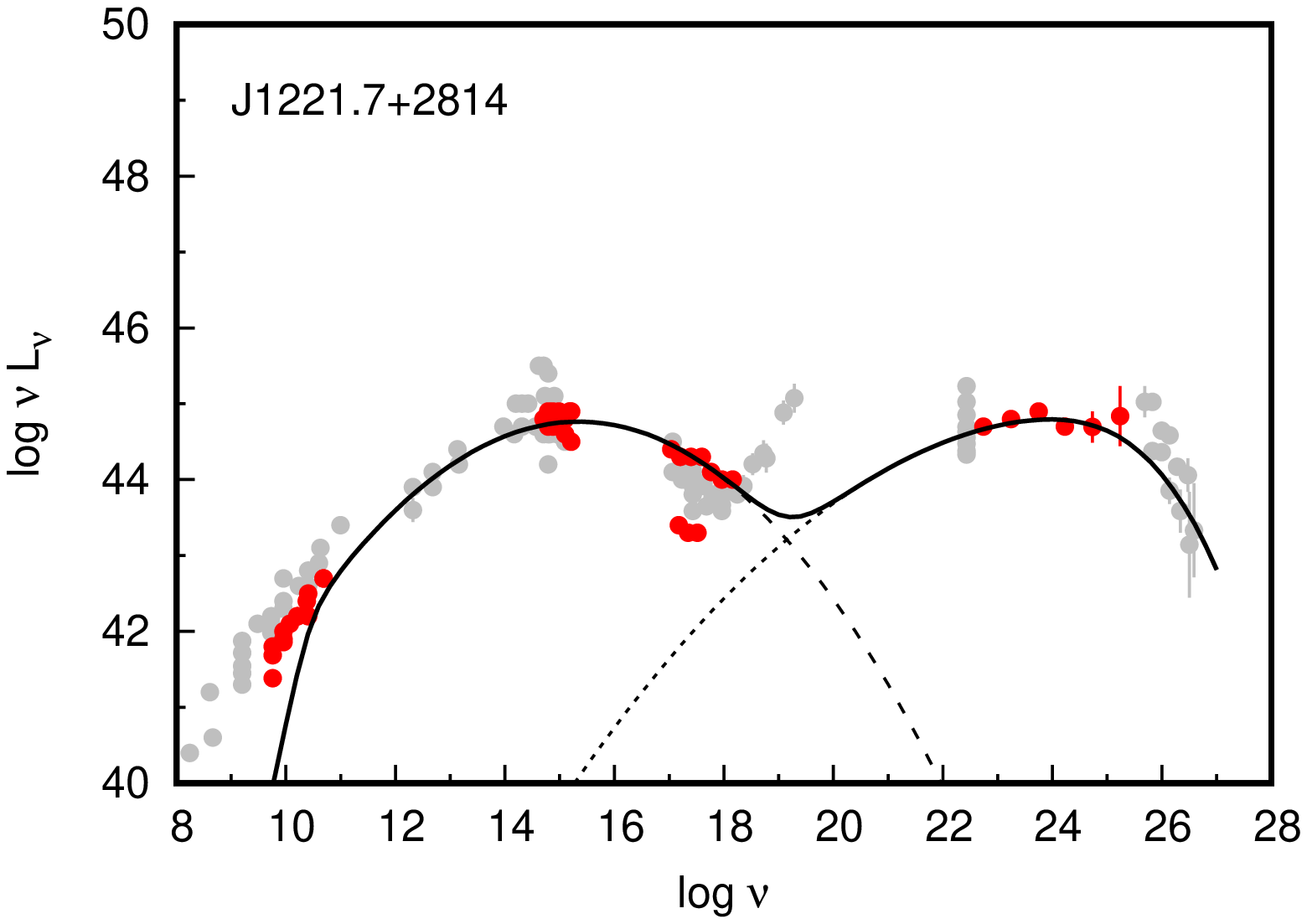}
\includegraphics[height=4cm, width=3cm, angle=-90]{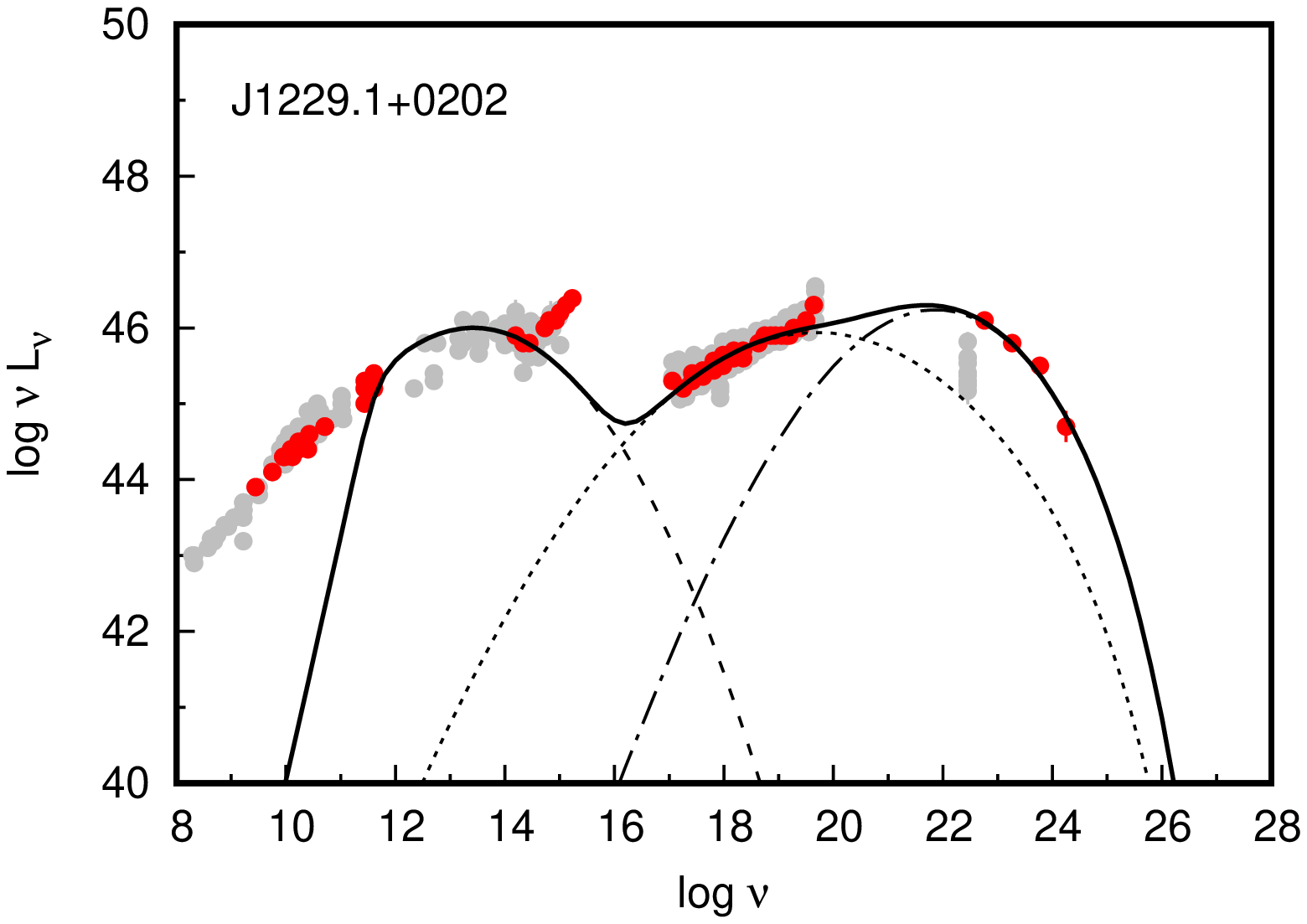}
\includegraphics[height=4cm, width=3cm, angle=-90]{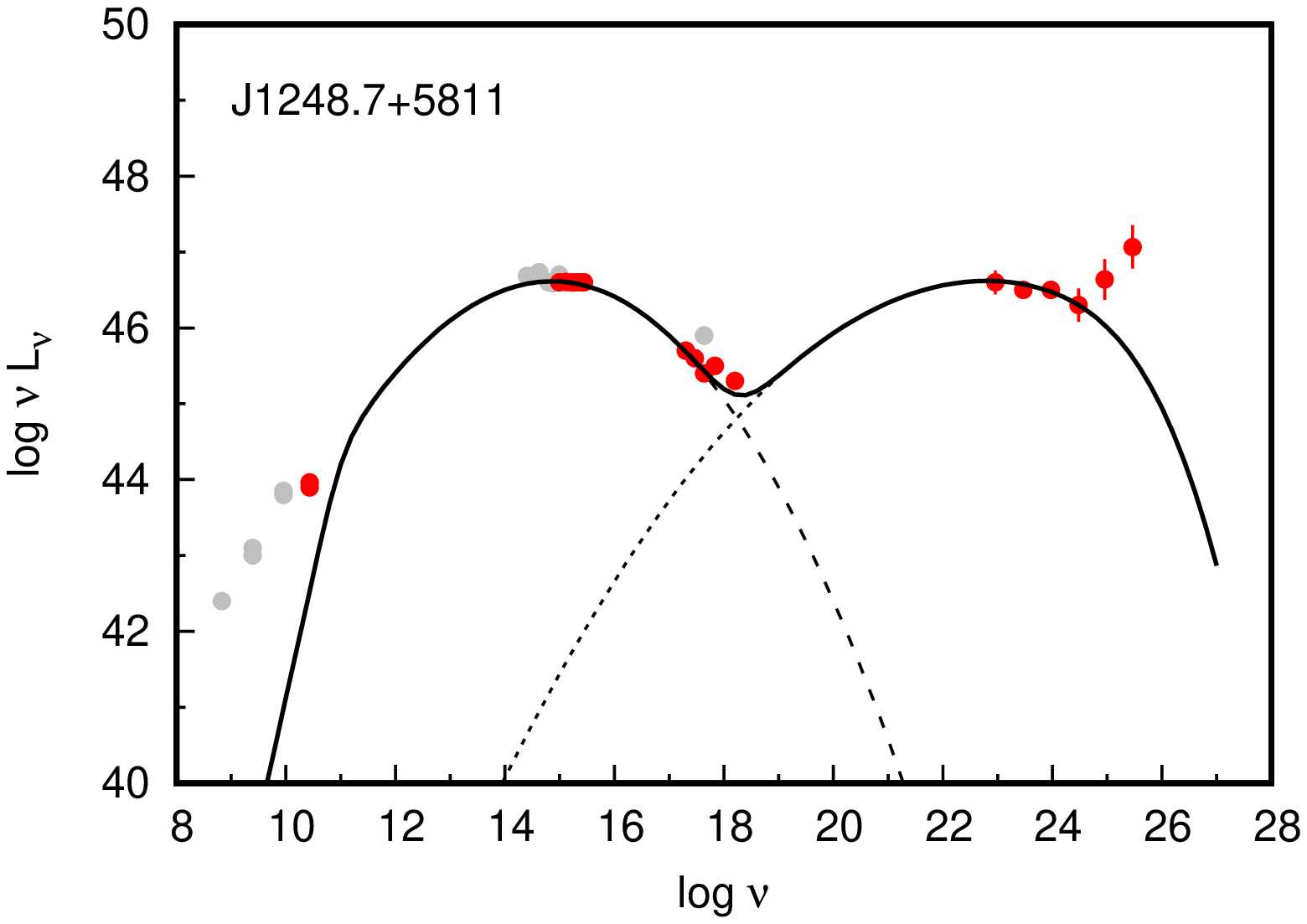}
\includegraphics[height=4cm, width=3cm, angle=-90]{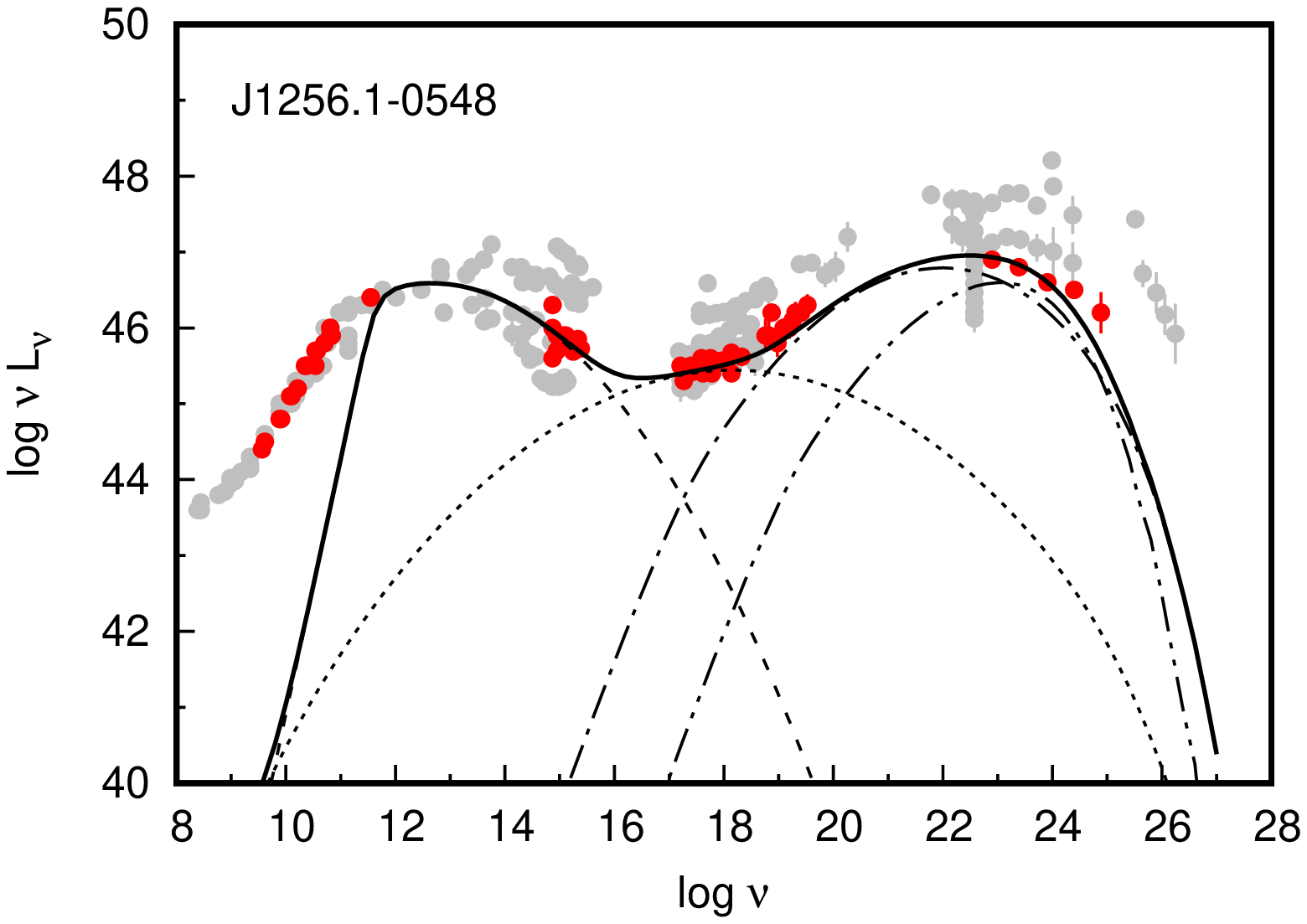}
\includegraphics[height=4cm, width=3cm, angle=-90]{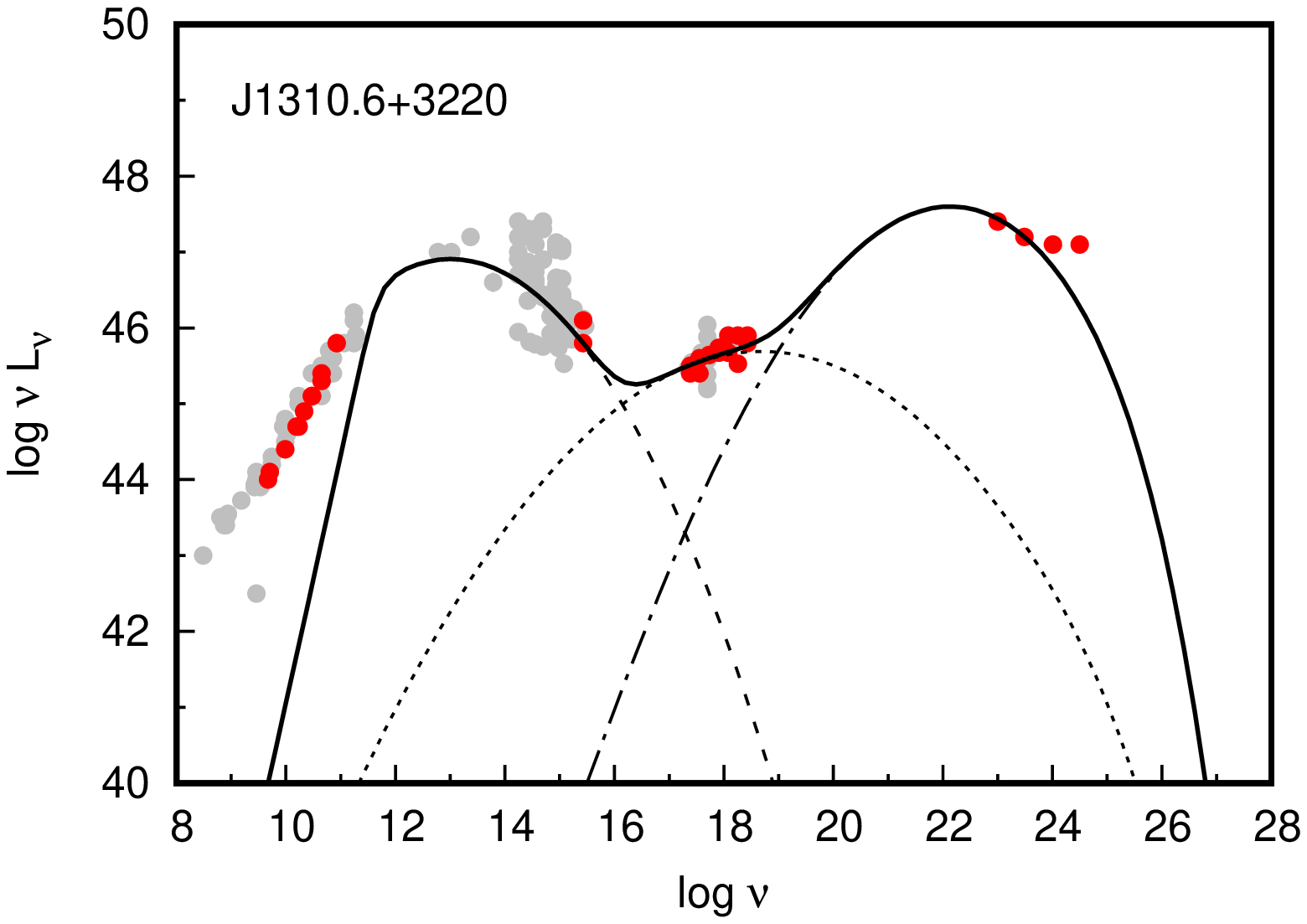}
\center{Figure \ref{fig:fits} --- continued.}
\end{figure*}
\begin{figure*}
\centering
\includegraphics[height=4cm, width=3cm, angle=-90]{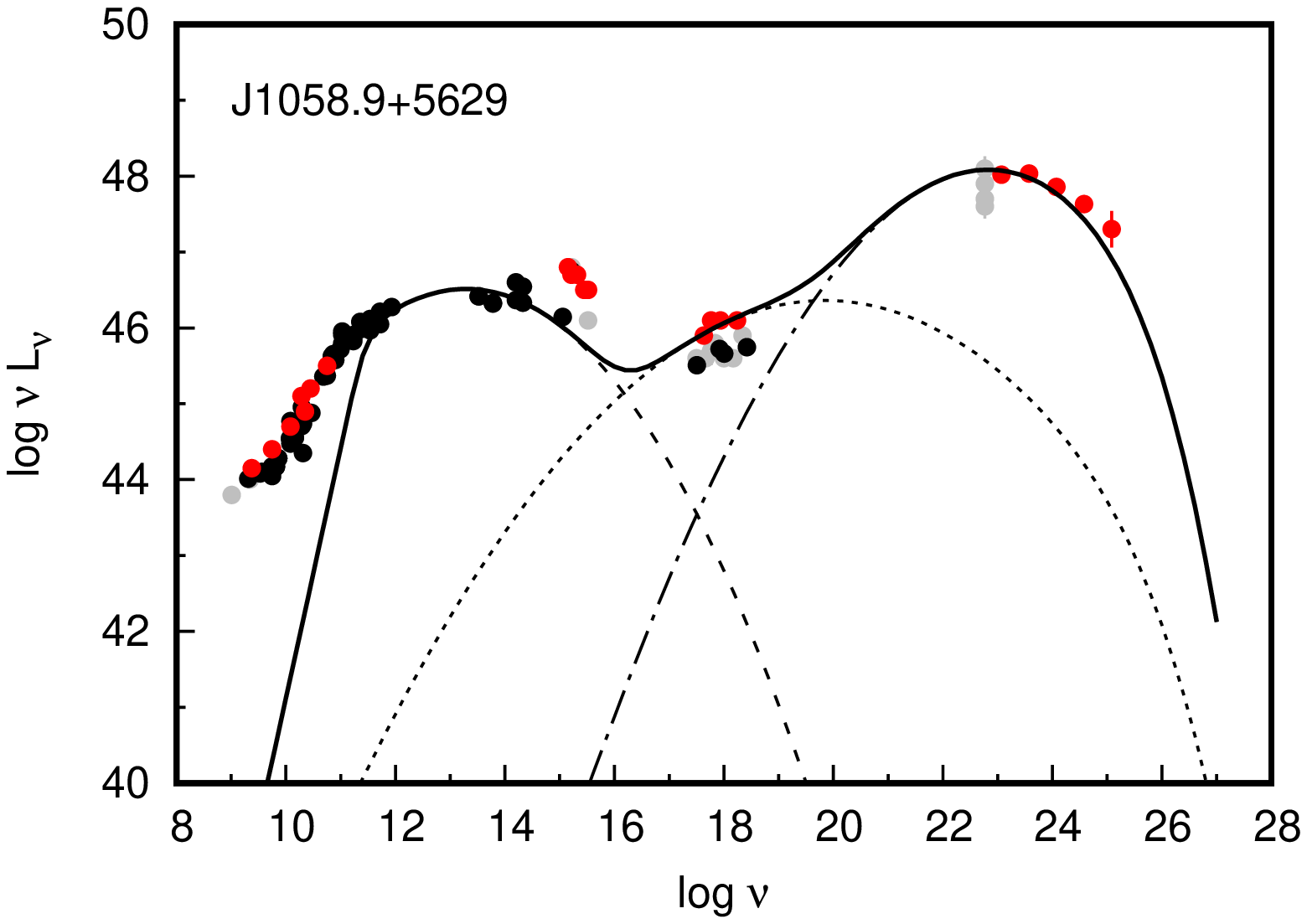}
\includegraphics[height=4cm, width=3cm, angle=-90]{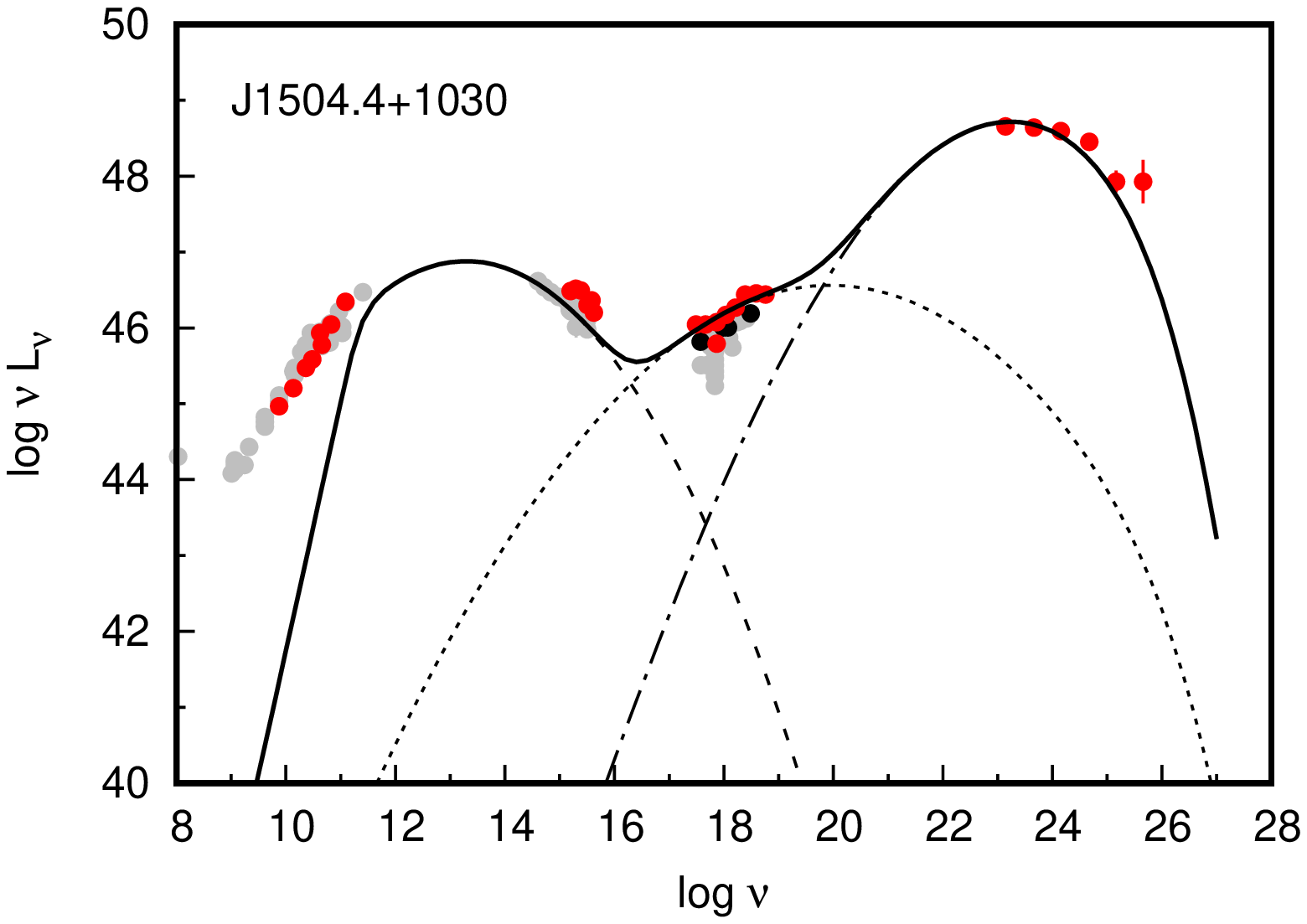}
\includegraphics[height=4cm, width=3cm, angle=-90]{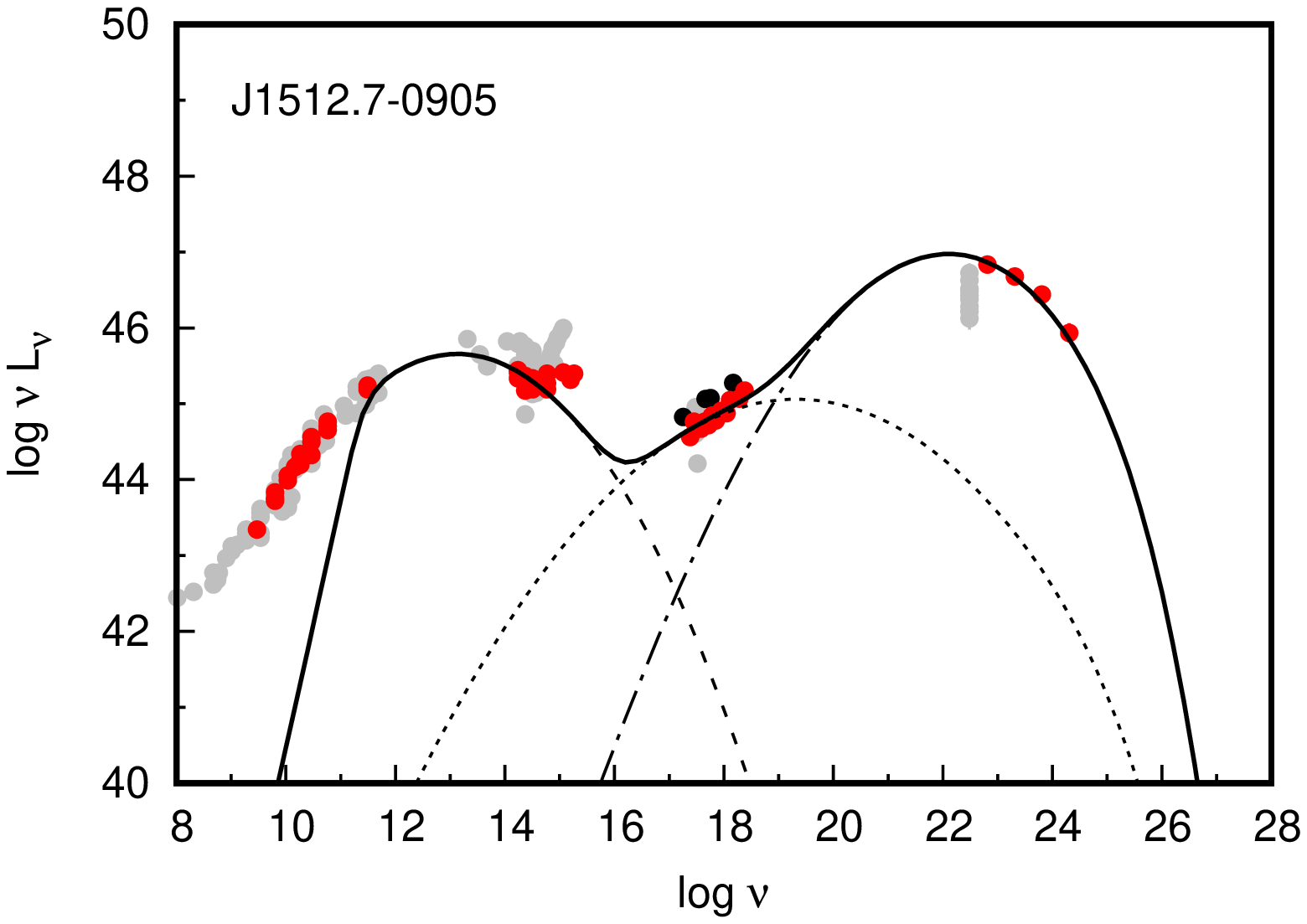}
\includegraphics[height=4cm, width=3cm, angle=-90]{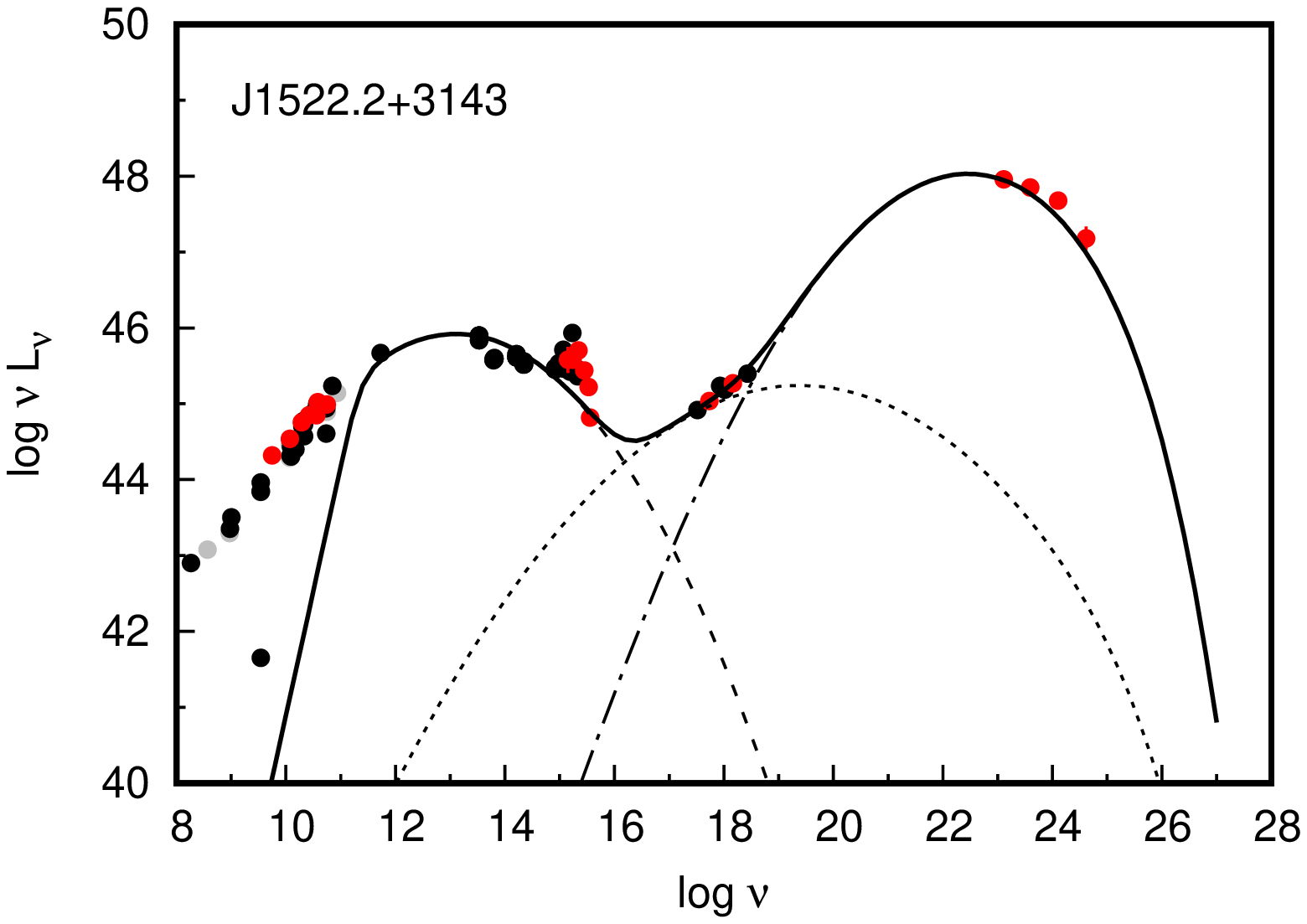}
\includegraphics[height=4cm, width=3cm, angle=-90]{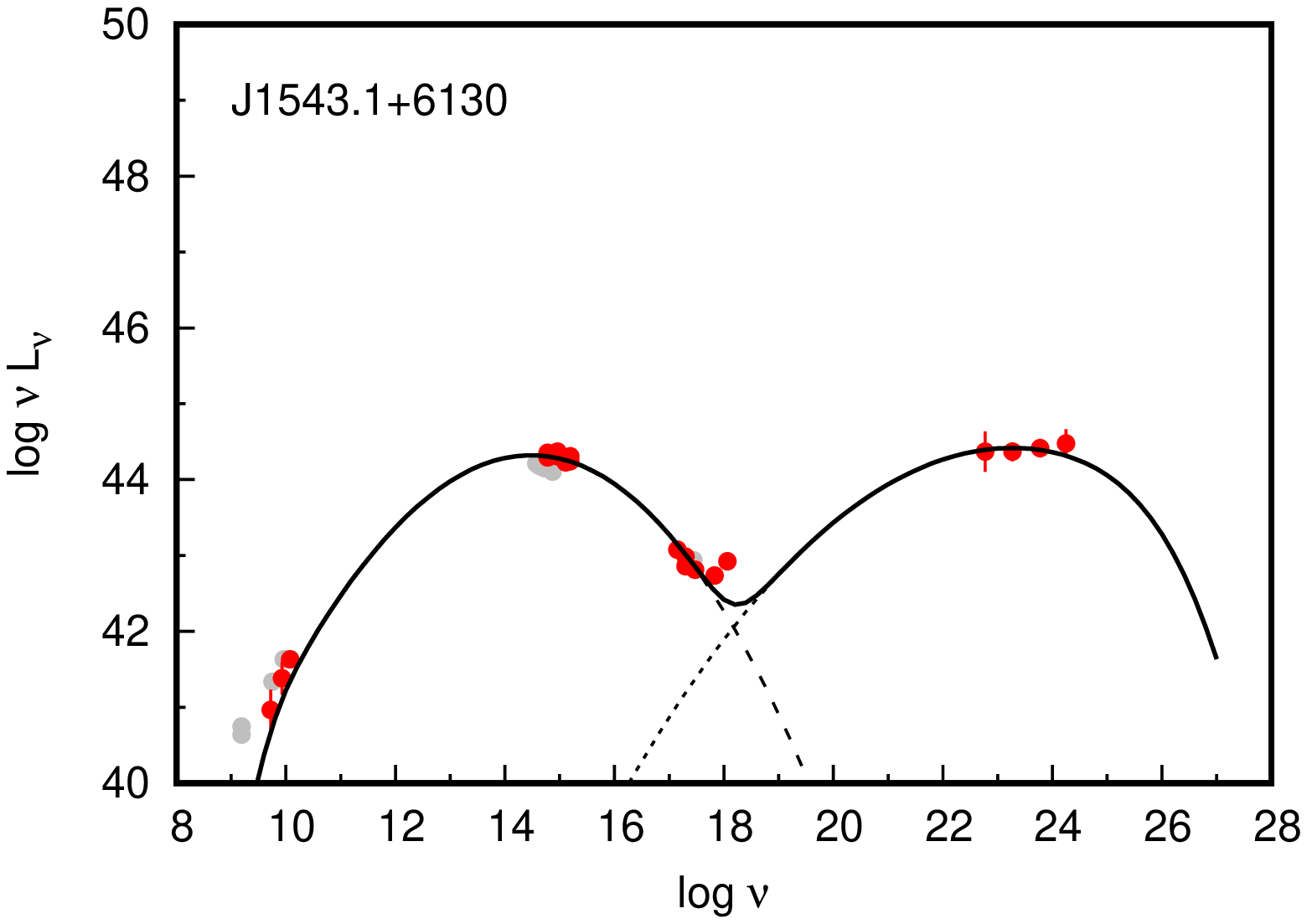}
\includegraphics[height=4cm, width=3cm, angle=-90]{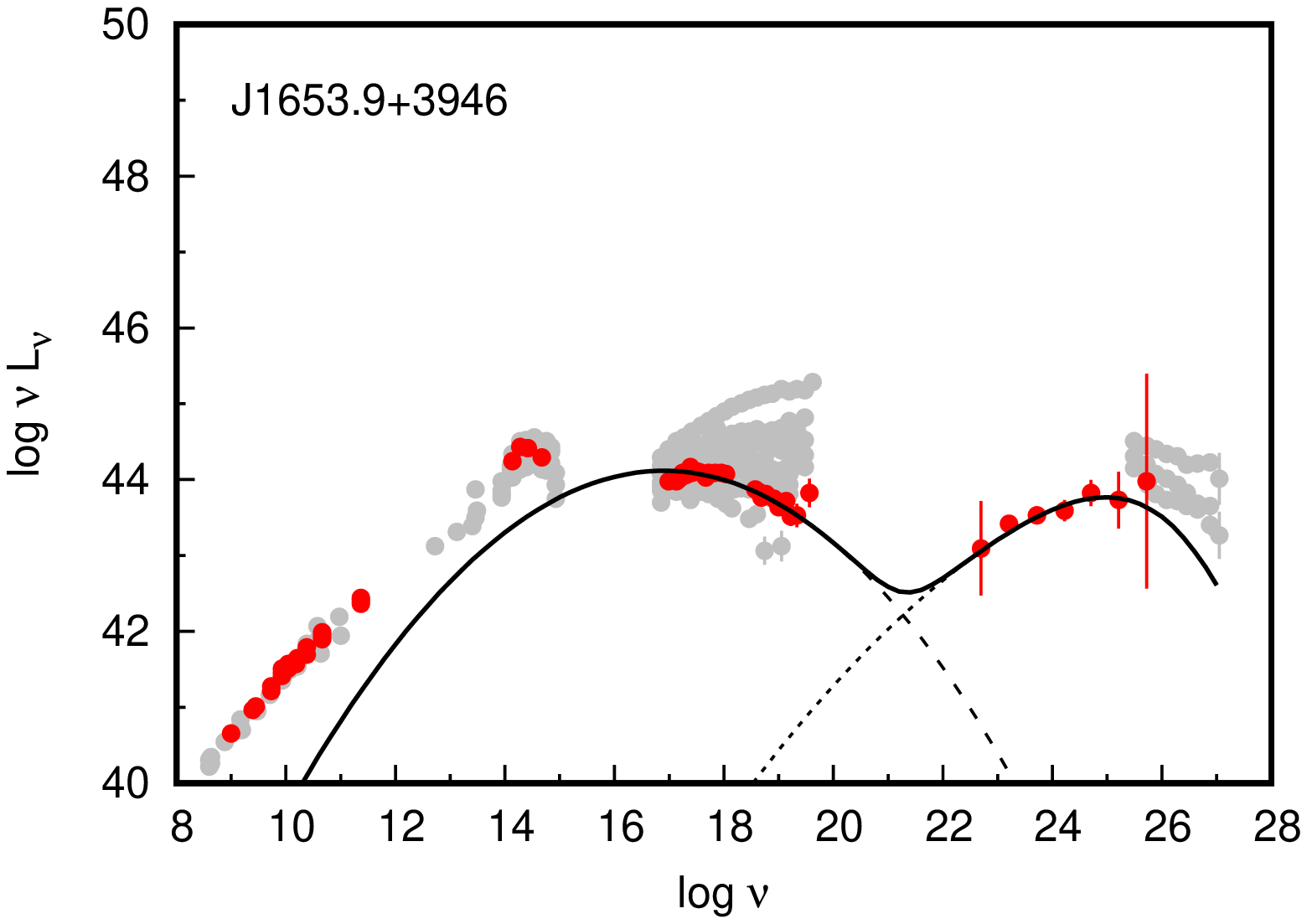}
\includegraphics[height=4cm, width=3cm, angle=-90]{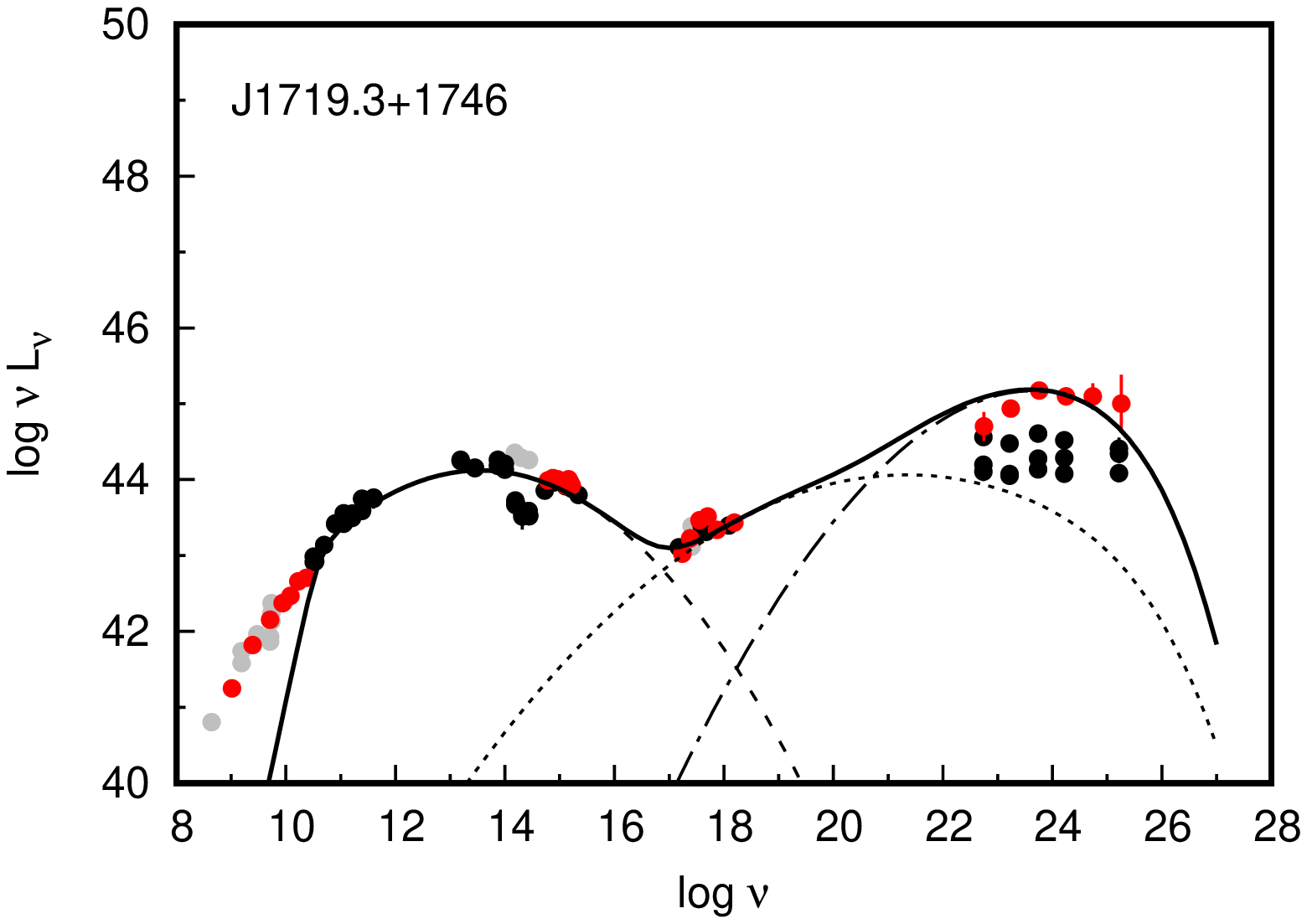}
\includegraphics[height=4cm, width=3cm, angle=-90]{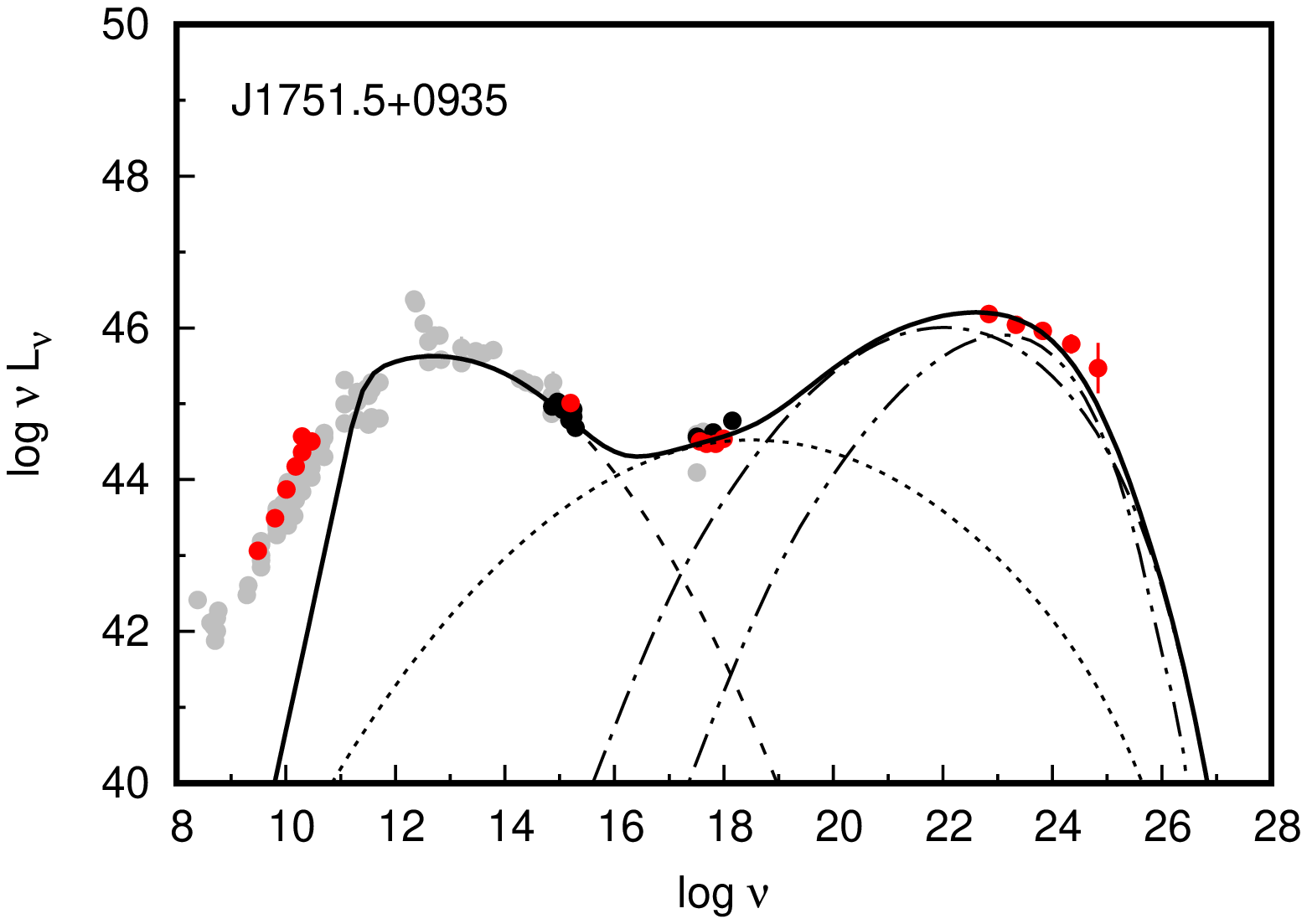}
\includegraphics[height=4cm, width=3cm, angle=-90]{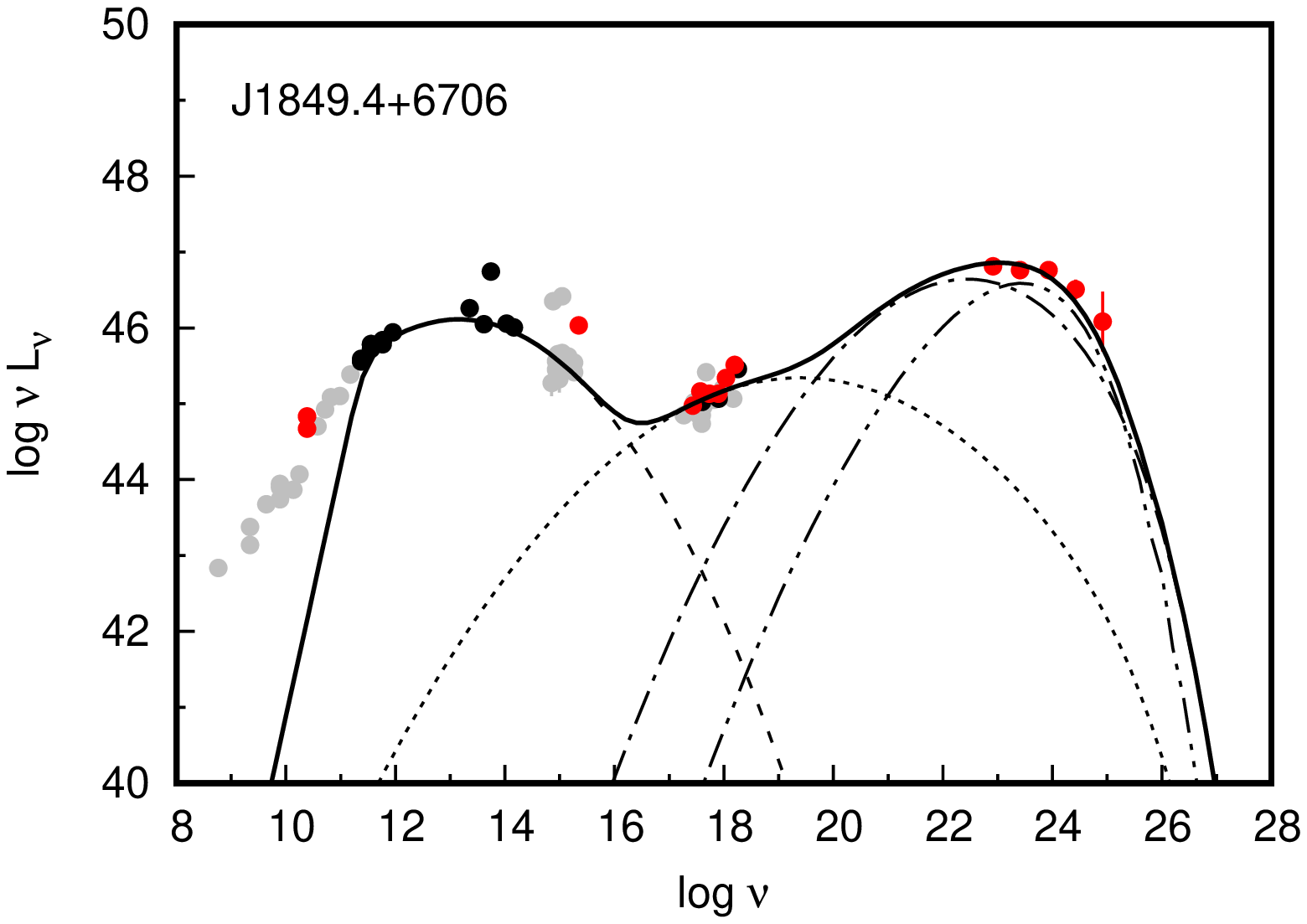}
\includegraphics[height=4cm, width=3cm, angle=-90]{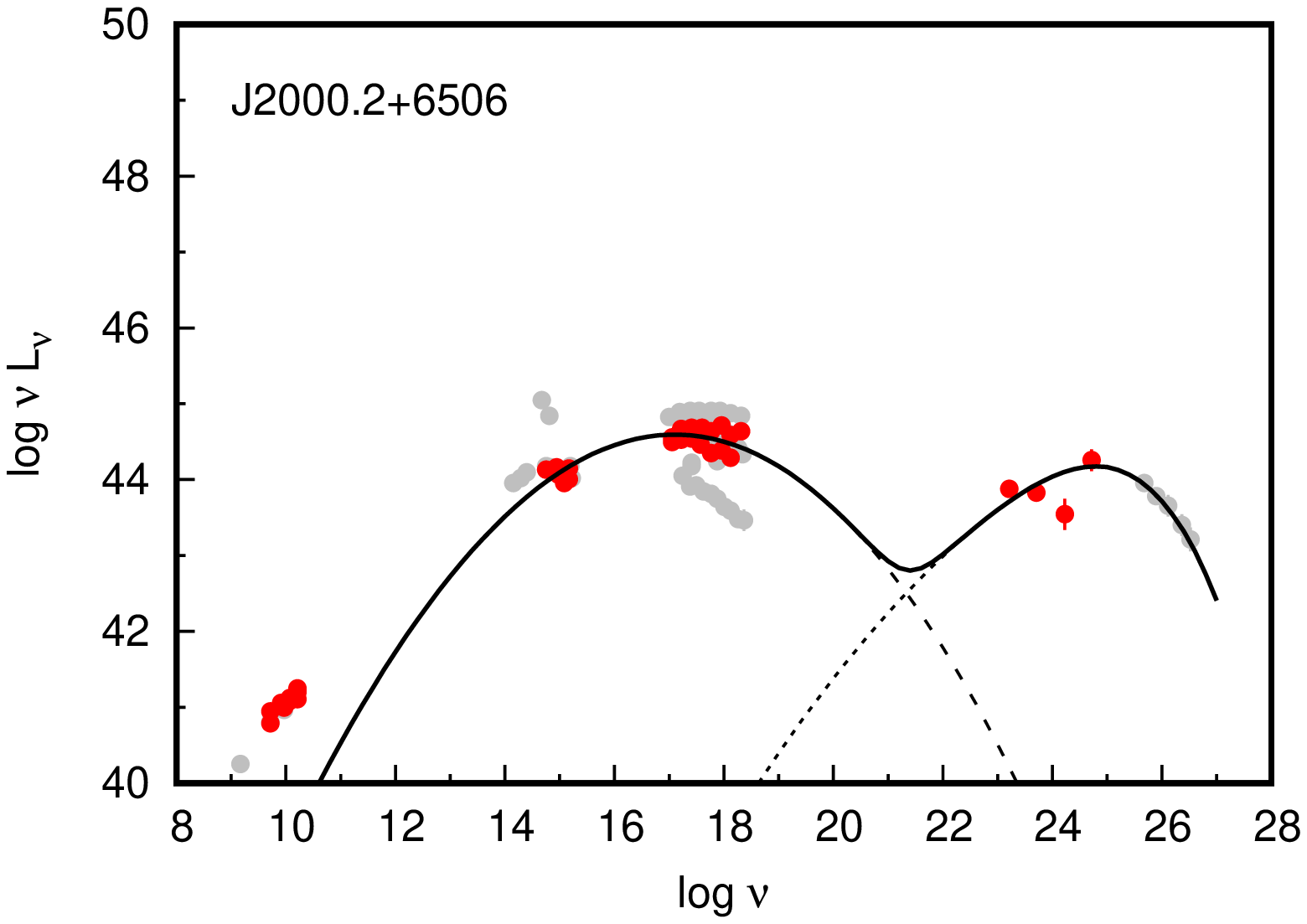}
\includegraphics[height=4cm, width=3cm, angle=-90]{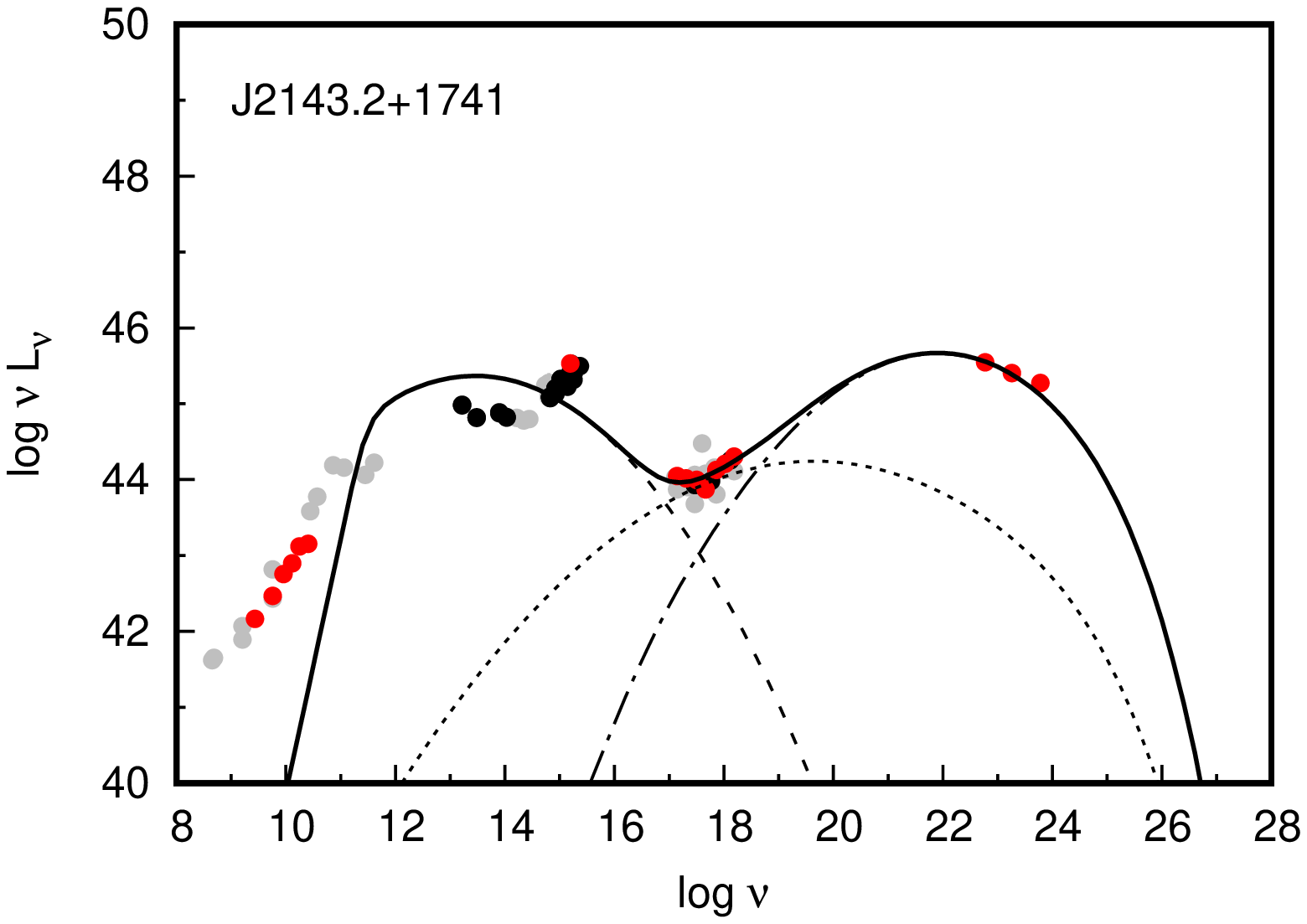}
\includegraphics[height=4cm, width=3cm, angle=-90]{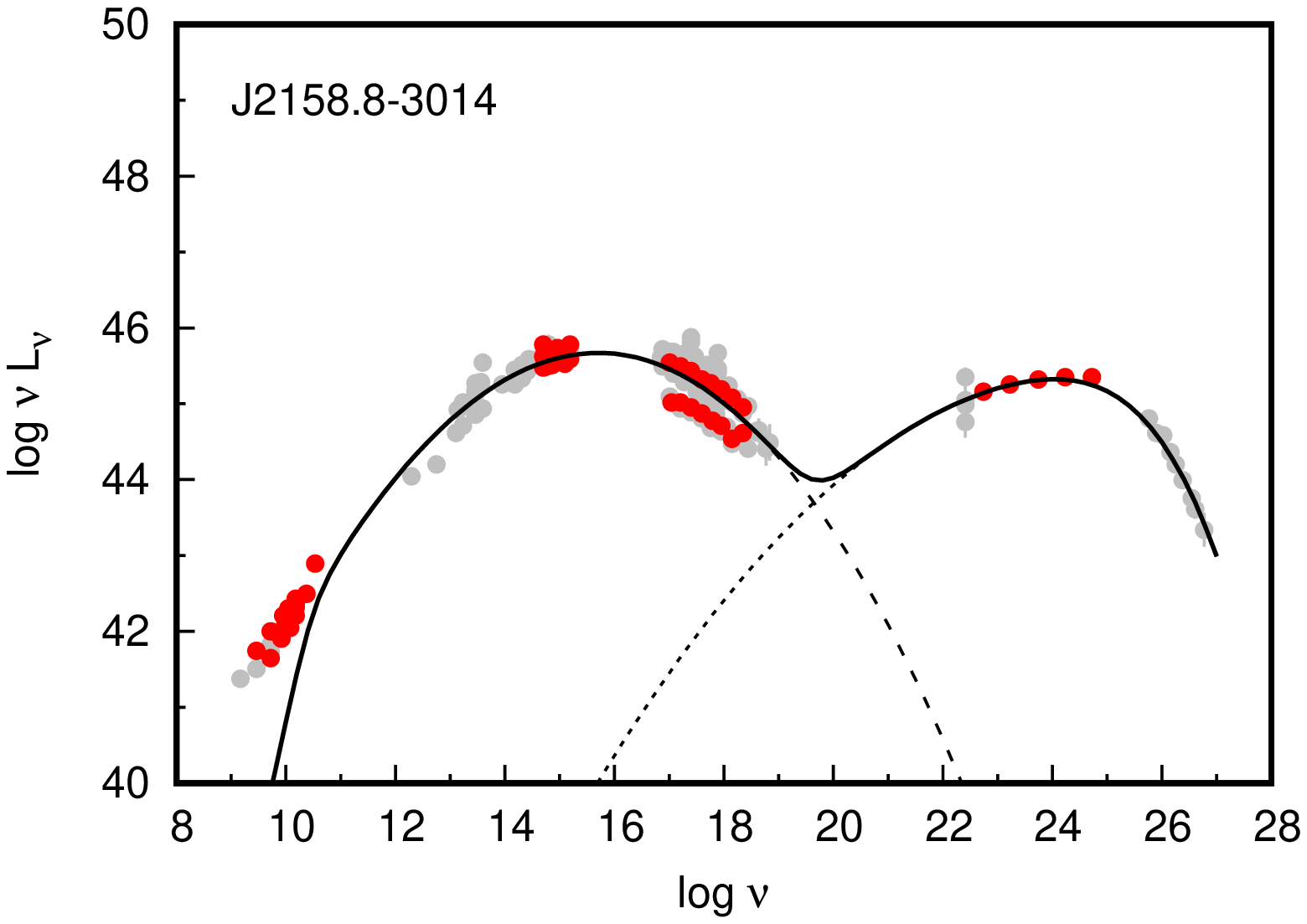}
\includegraphics[height=4cm, width=3cm, angle=-90]{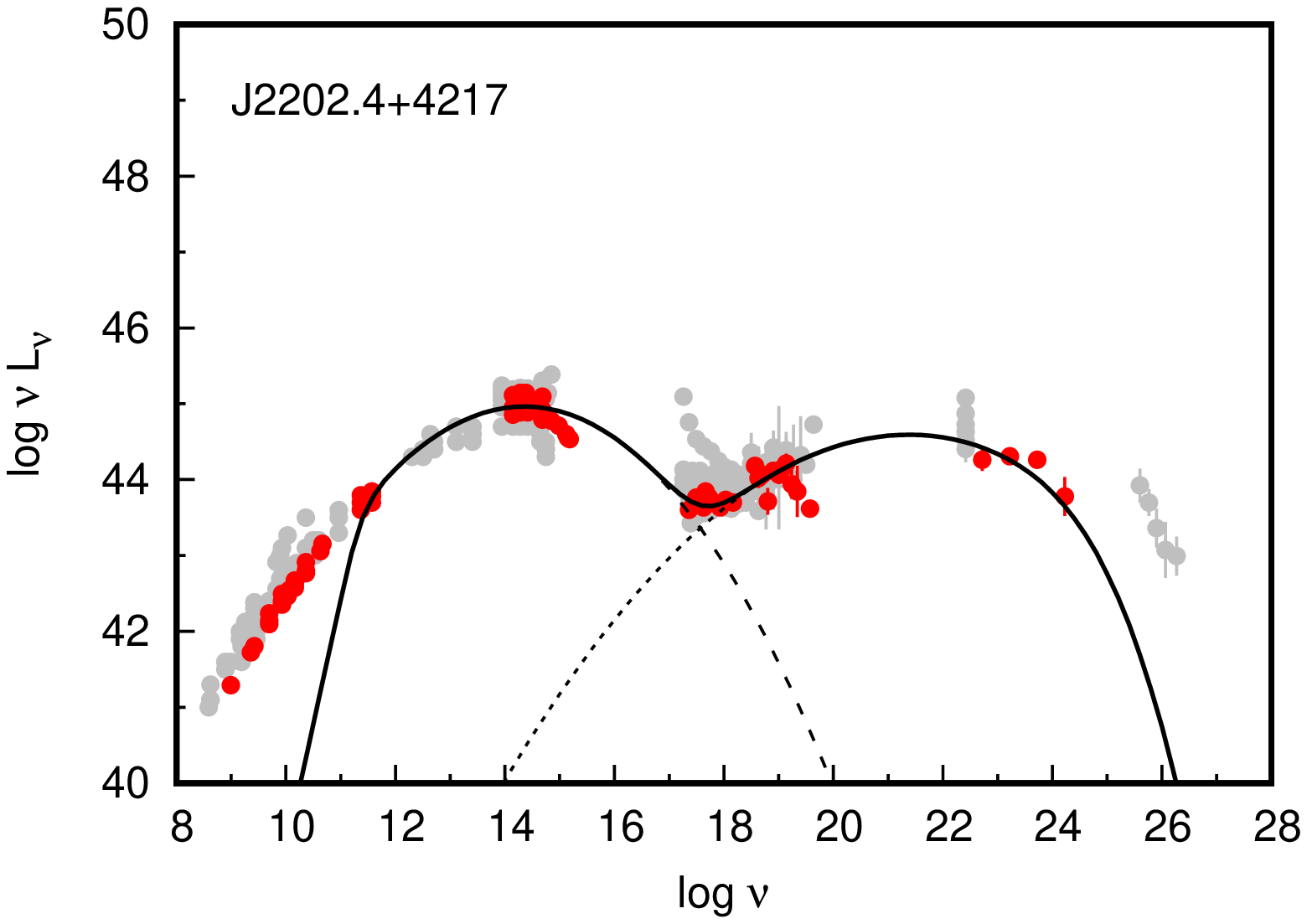}
\includegraphics[height=4cm, width=3cm, angle=-90]{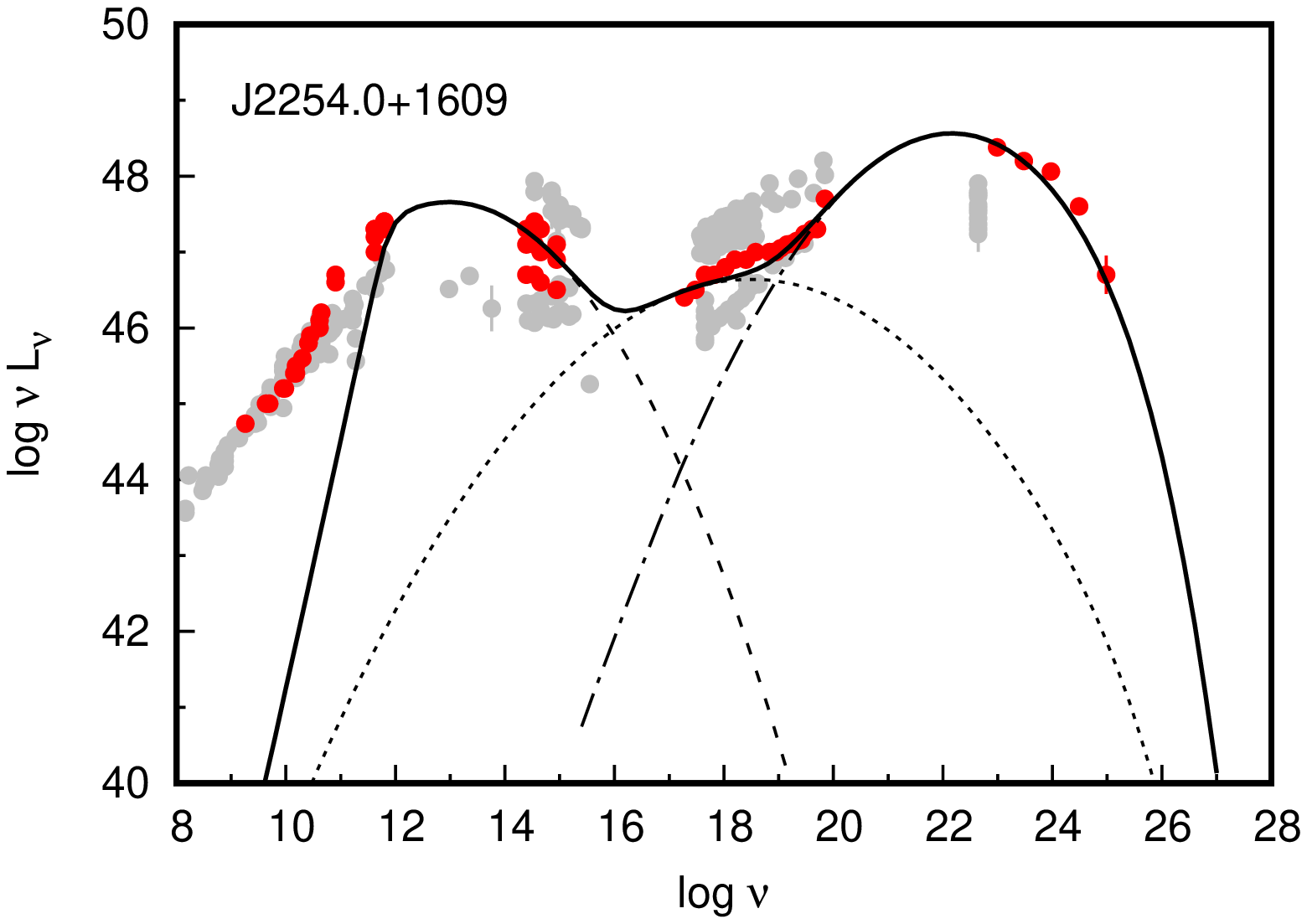}
\includegraphics[height=4cm, width=3cm, angle=-90]{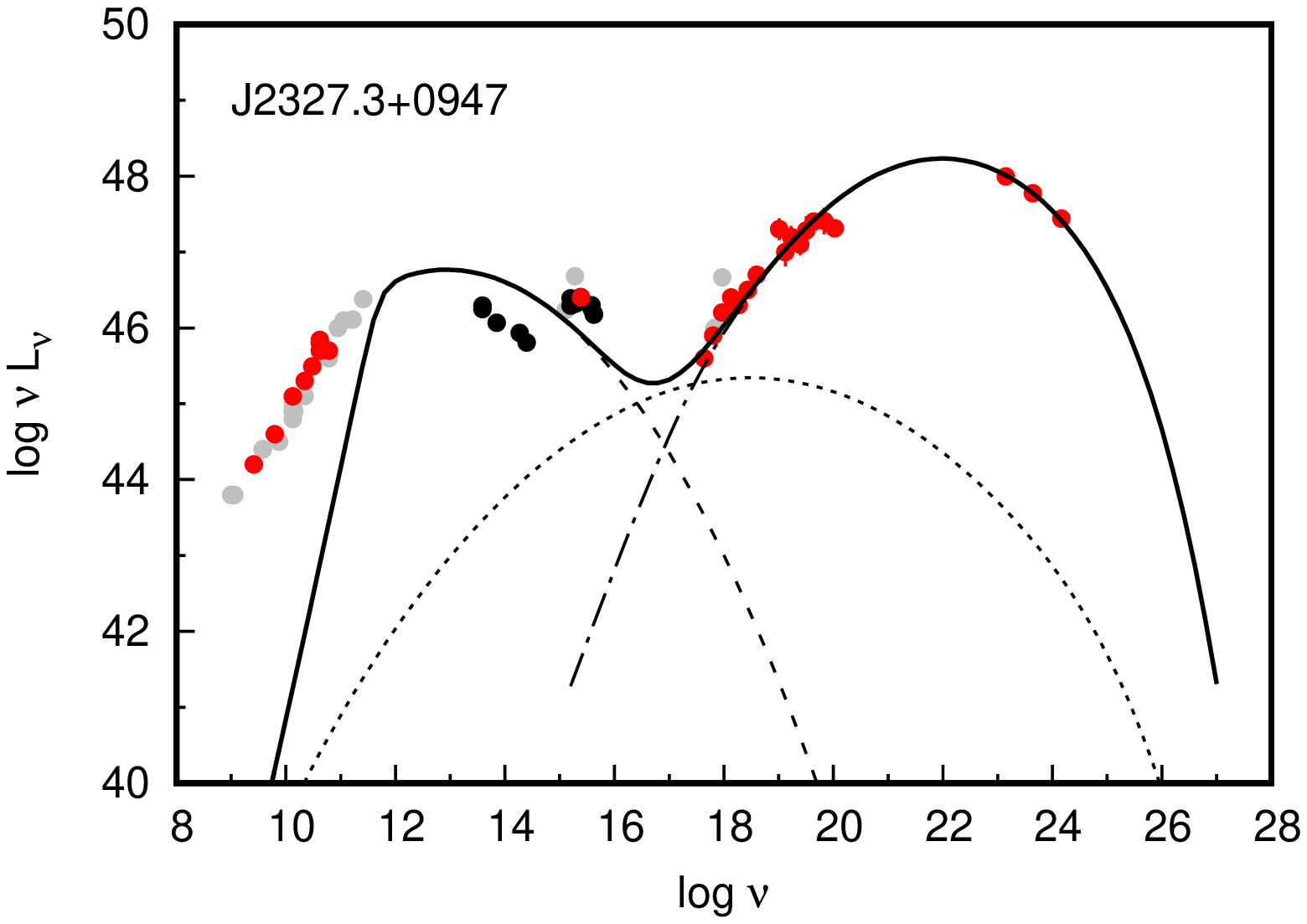}
\includegraphics[height=4cm, width=3cm, angle=-90]{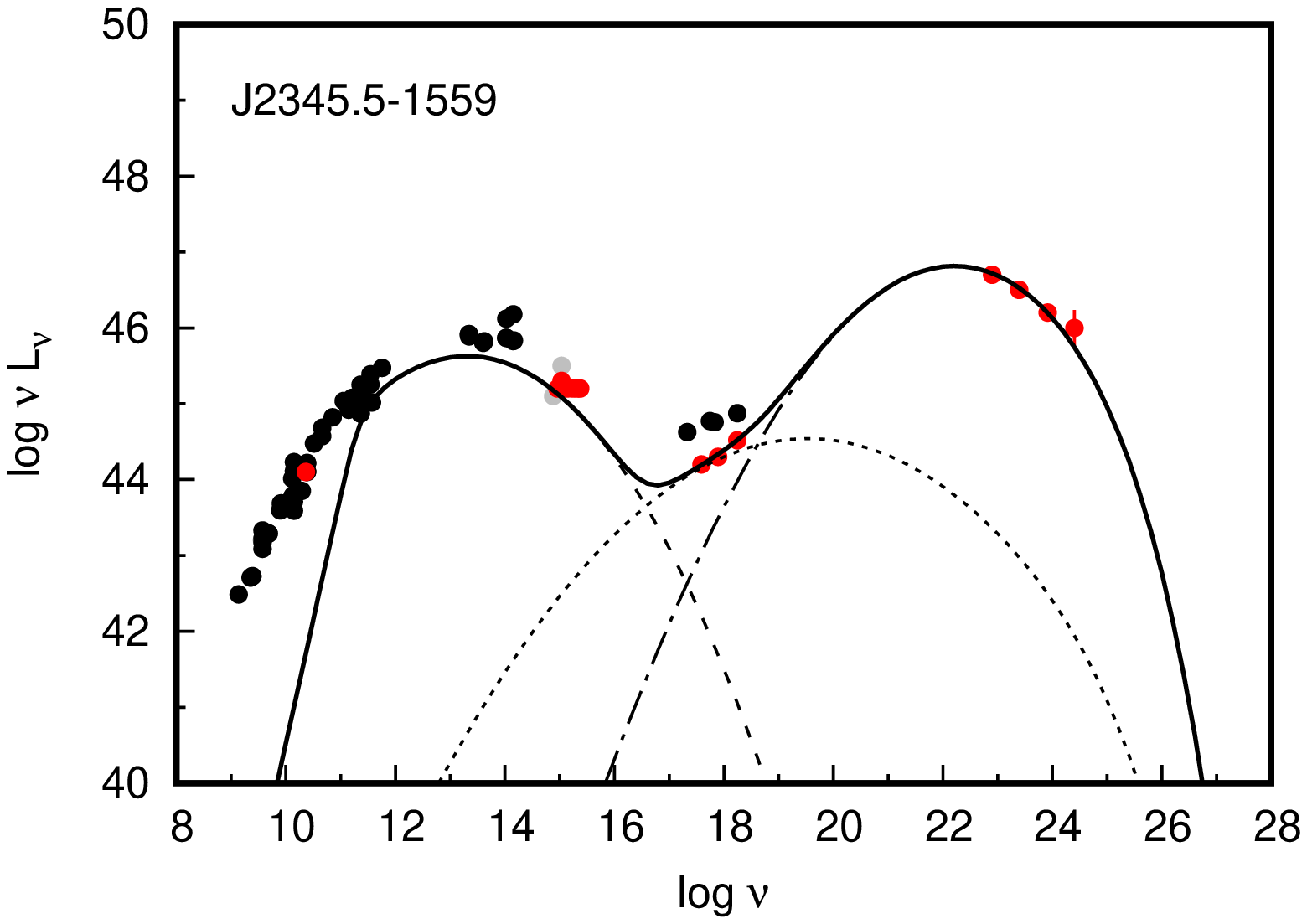}
\center{Figure \ref{fig:fits} --- continued.}
\end{figure*}
\begin{deluxetable}{cccccccccc}[t]
\tablecaption{SED modeling parameters of LBAS blazars. \label{tab:parameters}}
\tablecolumns{10}
\tabletypesize{\footnotesize}
\tablehead{
\colhead{Name (0FGL)} &
\colhead{$z$} &
\colhead{Type} &
\colhead{SED} &
\colhead{$B$} &
\colhead{$\log R$} &
\colhead{$\delta$} &
\colhead{$\gamma_p$} &
\colhead{$b$} &
\colhead{$u'$}\\
\colhead{(1)} &
\colhead{(2)} &
\colhead{(3)} &
\colhead{(4)} &
\colhead{(5)} &
\colhead{(6)} &
\colhead{(7)} &
\colhead{(8)} &
\colhead{(9)} &
\colhead{(10)}
}
\startdata
J0033.6-1921    &       0.610   &       BLLAC   &       HSP     &       0.06    &       6.55E+16        &       25      &       37803   &       0.51    &       3.86E-06        \\
J0050.5-0928    &       0.635   &       BLLAC   &       ISP     &       0.13    &       5.65E+16        &       22      &       4870    &       0.60    &       1.71E-05        \\
J0137.1+4751    &       0.859   &       FSRQ    &       LSP     &       0.02    &       1.08E+17        &       42      &       20775   &       0.66    &       4.31E-07        \\
J0210.8-5100    &       1.003   &       FSRQ    &       LSP     &       0.21    &       4.24E+16        &       16      &       16886   &       0.75    &       4.60E-05        \\
J0222.6+4302    &       0.444   &       BLLAC   &       ISP     &       0.11    &       5.12E+16        &       20      &       57880   &       0.42    &       1.32E-05        \\
J0229.5-3640    &       2.115   &       FSRQ    &       LSP     &       0.34    &       5.19E+16        &       20      &       5788    &       0.65    &       1.17E-04        \\
J0238.4+2855    &       1.213   &       FSRQ    &       LSP     &       0.33    &       3.02E+16        &       12      &       43404   &       0.50    &       1.09E-04        \\
J0238.6+1636    &       0.940   &       BLLAC   &       LSP     &       2.84    &       1.30E+16        &       5       &       4870    &       0.57    &       8.19E-03        \\
J0349.8-2102    &       2.944   &       FSRQ    &       LSP     &       0.04    &       5.03E+16        &       19      &       129578  &       0.55    &       1.36E-06        \\
J0423.1-0112    &       0.915   &       FSRQ    &       LSP     &       0.06    &       4.15E+16        &       16      &       25854   &       0.47    &       4.00E-06        \\
J0428.7-3755    &       1.112   &       BLLAC   &       LSP     &       0.20    &       5.70E+16        &       22      &       6879    &       0.71    &       4.24E-05        \\
J0449.7-4348    &       0.205   &       BLLAC   &       HSP     &       0.01    &       5.92E+16        &       23      &       18303   &       0.75    &       1.36E-07        \\
J0457.1-2325    &       1.003   &       FSRQ    &       LSP     &       0.08    &       2.80E+16        &       11      &       163129  &       0.43    &       6.96E-06        \\
J0507.9+6739    &       0.416   &       BLLAC   &       HSP     &       0.38    &       1.95E+16        &       8       &       115486  &       0.51    &       1.43E-04        \\
J0516.2-6200    &       1.300   &       BLLAC   &       LSP     &       0.21    &       4.48E+16        &       17      &       20537   &       0.56    &       4.33E-05        \\
J0531.0+1331    &       2.070   &       FSRQ    &       LSP     &       2.01    &       1.64E+16        &       6       &       2304    &       0.70    &       4.08E-03        \\
J0538.8-4403    &       0.892   &       BLLAC   &       LSP     &       0.75    &       5.27E+16        &       20      &       1252    &       1.00    &       1.97E-02        \\
J0712.9+5034    &       0.400   &       BLLAC   &       LSP     &       0.17    &       8.42E+16        &       33      &       861     &       0.71    &       4.01E-03        \\
J0722.0+7120    &       0.310   &       BLLAC   &       ISP     &       0.47    &       5.65E+16        &       22      &       1252    &       1.20    &       5.42E-02        \\
J0730.4-1142    &       1.589   &       FSRQ    &       LSP     &       0.36    &       7.16E+16        &       28      &       516     &       0.75    &       1.45E-02        \\
J0855.4+2009    &       0.306   &       BLLAC   &       LSP     &       0.08    &       1.28E+17        &       49      &       1296    &       0.82    &       9.29E-04        \\
J0921.2+4437    &       2.190   &       FSRQ    &       LSP     &       0.88    &       8.54E+16        &       33      &       746     &       0.96    &       1.24E-01        \\
J1015.2+4927    &       0.212   &       BLLAC   &       HSP     &       0.46    &       6.33E+16        &       24      &       613     &       0.88    &       1.29E-02        \\
J1057.8+0138    &       0.888   &       BLLAC   &       LSP     &       0.06    &       1.05E+17        &       41      &       1728    &       0.57    &       1.39E-03        \\
J1058.9+5629    &       0.143   &       BLLAC   &       ISP     &       0.11    &       8.85E+16        &       34      &       1103    &       0.63    &       4.43E-03        \\
J1104.5+3811    &       0.030   &       BLLAC   &       HSP     &       0.16    &       7.76E+16        &       30      &       781     &       0.75    &       5.77E-03        \\
J1159.2+2912    &       0.729   &       FSRQ    &       LSP     &       0.48    &       8.15E+16        &       31      &       410     &       0.70    &       9.40E-02        \\
J1221.7+2814    &       0.102   &       BLLAC   &       ISP     &       0.20    &       7.55E+16        &       29      &       1505    &       0.68    &       3.22E-03        \\
J1229.1+0202    &       0.158   &       FSRQ    &       LSP     &       0.26    &       4.52E+16        &       17      &       2304    &       0.71    &       2.32E-03        \\
J1248.7+5811    &       0.847   &       BLLAC   &       ISP     &       0.12    &       1.02E+17        &       39      &       559     &       0.76    &       5.84E-03        \\
J1256.1-0548    &       0.536   &       FSRQ    &       LSP     &       0.50    &       5.24E+16        &       20      &       1053    &       0.83    &       1.57E-03        \\
J1310.6+3220    &       0.997   &       FSRQ    &       LSP     &       0.13    &       9.05E+16        &       35      &       917     &       0.80    &       3.03E-03        \\
J1457.6-3538    &       1.424   &       FSRQ    &       LSP     &       0.14    &       8.14E+16        &       31      &       994     &       0.71    &       9.06E-04        \\
J1504.4+1030    &       1.839   &       FSRQ    &       LSP     &       0.57    &       5.81E+16        &       22      &       613     &       0.96    &       1.91E-02        \\
J1512.7-0905    &       0.360   &       FSRQ    &       LSP     &       0.76    &       2.93E+16        &       11      &       950     &       1.00    &       1.46E-02        \\
J1522.2+3143    &       1.487   &       FSRQ    &       LSP     &       0.27    &       8.14E+16        &       31      &       400     &       0.59    &       3.57E-03        \\
J1543.1+6130    &       0.117   &       BLLAC   &       ISP     &       0.41    &       7.95E+16        &       31      &       487     &       0.90    &       1.25E-02        \\
J1653.9+3946    &       0.033   &       BLLAC   &       HSP     &       0.12    &       6.88E+16        &       27      &       1326    &       0.75    &       8.57E-03        \\
J1719.3+1746    &       0.137   &       BLLAC   &       LSP     &       0.07    &       1.03E+17        &       40      &       1479    &       0.82    &       4.80E-03        \\
J1751.5+0935    &       0.322   &       BLLAC   &       LSP     &       0.32    &       4.24E+16        &       16      &       876     &       0.90    &       3.05E-02        \\
J1849.4+6706    &       0.657   &       FSRQ    &       LSP     &       0.18    &       5.93E+16        &       23      &       972     &       0.82    &       5.96E-02        \\
J2000.2+6506    &       0.047   &       BLLAC   &       HSP     &       0.02    &       5.16E+16        &       20      &       5464    &       0.52    &       6.02E-05        \\
J2143.2+1741    &       0.213   &       FSRQ    &       LSP     &       0.20    &       6.05E+16        &       23      &       566     &       0.63    &       3.51E-03        \\
J2158.8-3014    &       0.116   &       BLLAC   &       HSP     &       0.21    &       6.26E+16        &       24      &       917     &       0.76    &       6.62E-03        \\
J2202.4+4217    &       0.069   &       BLLAC   &       ISP     &       0.83    &       3.45E+16        &       13      &       876     &       0.62    &       2.02E-02        \\
J2254.0+1609    &       0.859   &       FSRQ    &       LSP     &       0.42    &       9.75E+16        &       38      &       410     &       0.88    &       2.15E-02        \\
J2327.3+0947    &       1.843   &       FSRQ    &       LSP     &       0.49    &       7.86E+16        &       30      &       415     &       0.65    &       1.05E-01        \\
J2345.5-1559    &       0.621   &       FSRQ    &       LSP     &       0.40    &       4.69E+16        &       18      &       907     &       0.85    &       3.73E-02        \\
\enddata
\tablecomments{The Column 1 gives the source name as in the 0FGL catalog. Column 2 provides the source redsift $z$. Column 3 gives the blazar type. Column 4 describes the synchrotron SED type. Columns 5-8 provides the physical jet parameters, including magnetic field $B$ (G), size $R$ (cm), Doppler factor $\delta$ and electron peak energy $\gamma_p$, respectively. Column 9 gives the curvature parameter $b$ and Column 10 provides total ambient energy density in jet frame $u'$ (erg cm$^{-3}$).}
\end{deluxetable}

We find that SSC emission can successfully reproduce the whole SED of IBLs and HBLs (SSC blazars), which is consistent with previous studies \citep[e.g.,][and references therein]{2012ApJ...752..157Z, 2014MNRAS.439.2933Y}, while the $\gamma$-rays in FSRQs and LBLs (EC blazars) necessarily need an EC component to explain their GeV spectra \citep [e.g., see][]{2013ApJ...767....8Z, 2015MNRAS.454.1310Y}. In some EC blazars the VHE $\gamma$-rays in the LAT band necessarily need a BLR component, suggesting that the $\gamma$-ray region in some blazars may be at the edge of BLR. In a few HSP sources, for example J1104.5+3811, when multiple X-ray observations are available, we fit the low state spectra conforming to our steady state model. 

\subsection{Physical Parameter Distributions}
\begin{figure}[htb!]
\centering
\includegraphics[height=5cm, width=7cm]{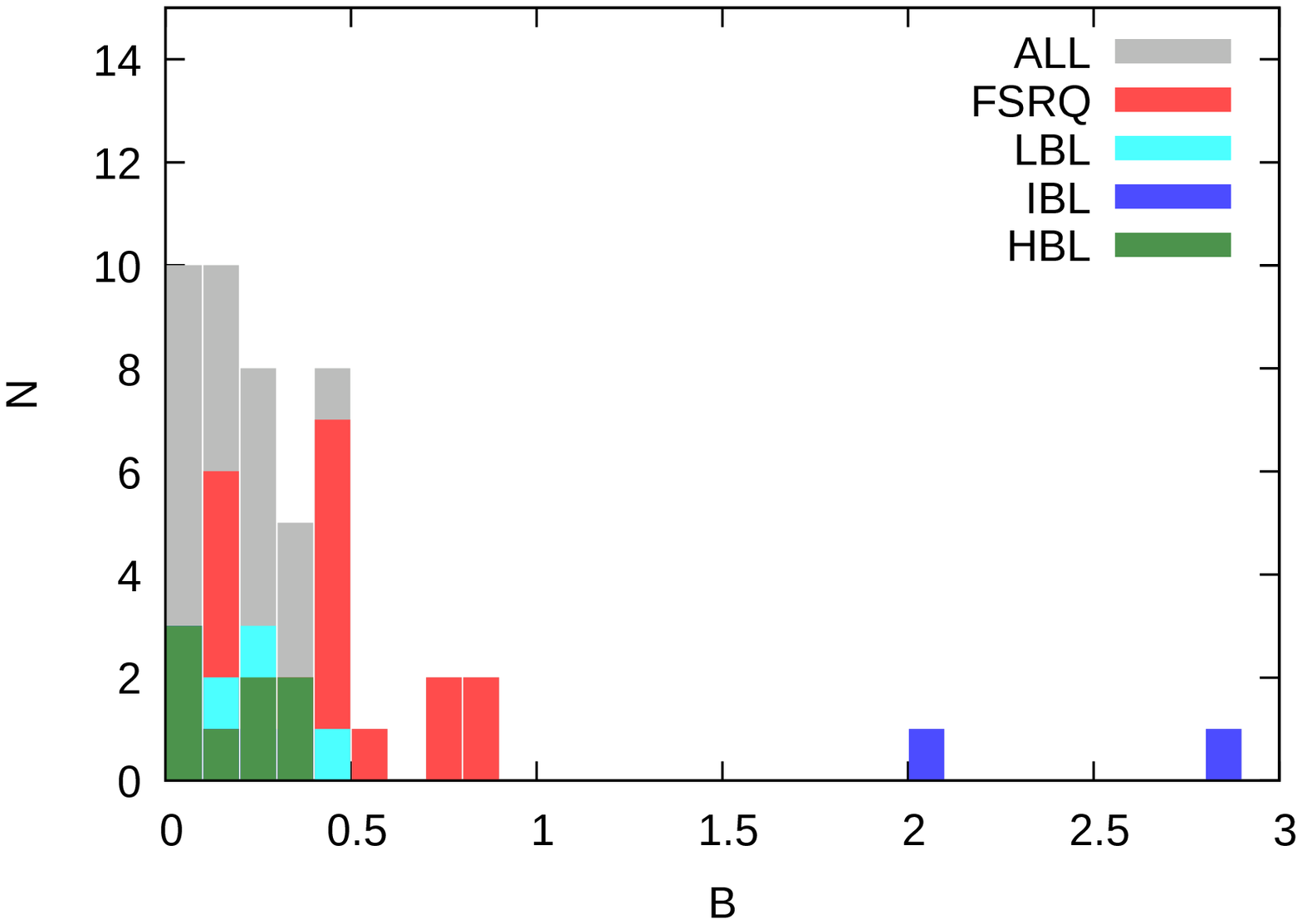}
\includegraphics[height=5cm, width=7cm]{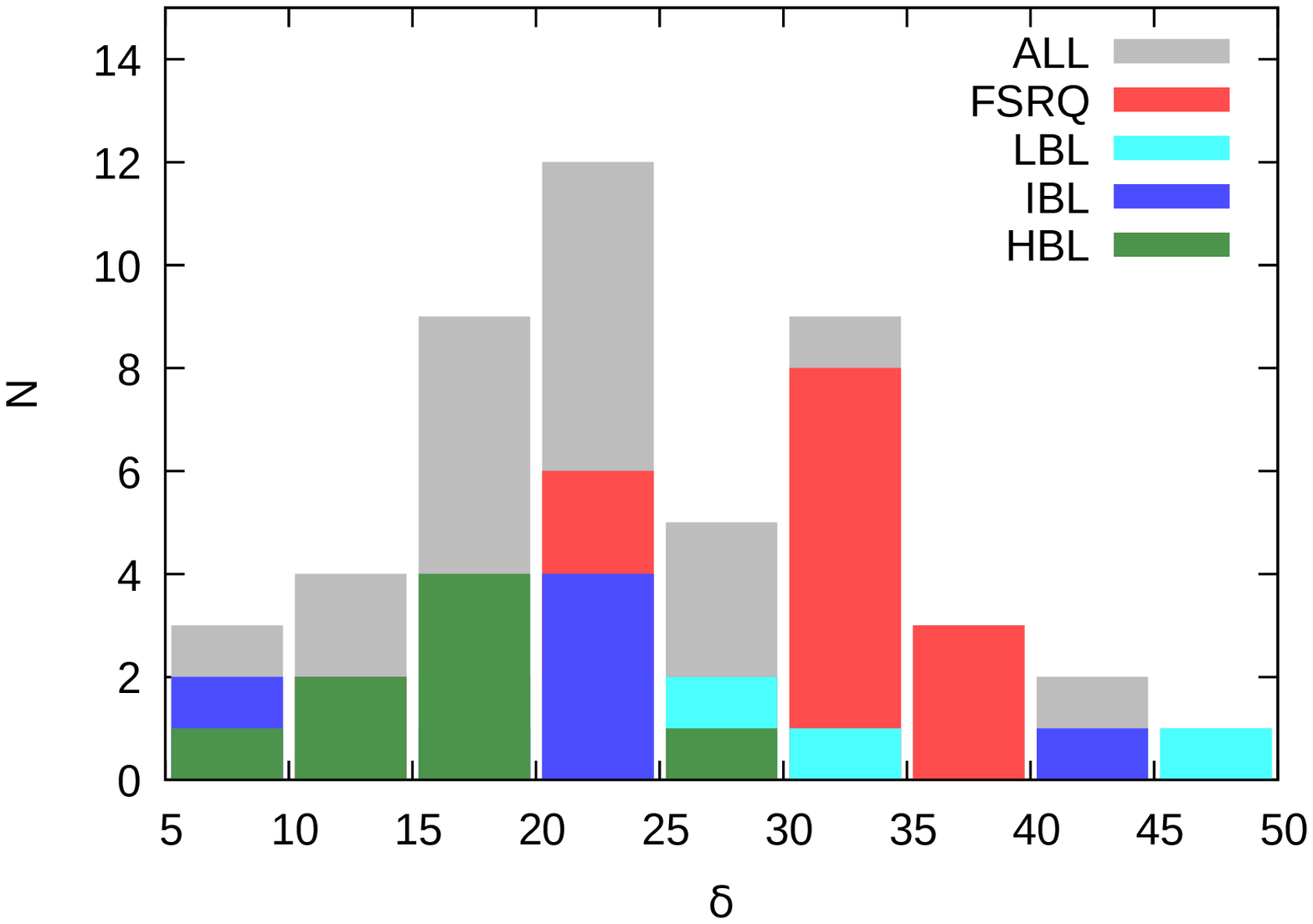}\\
\includegraphics[height=5cm, width=7cm]{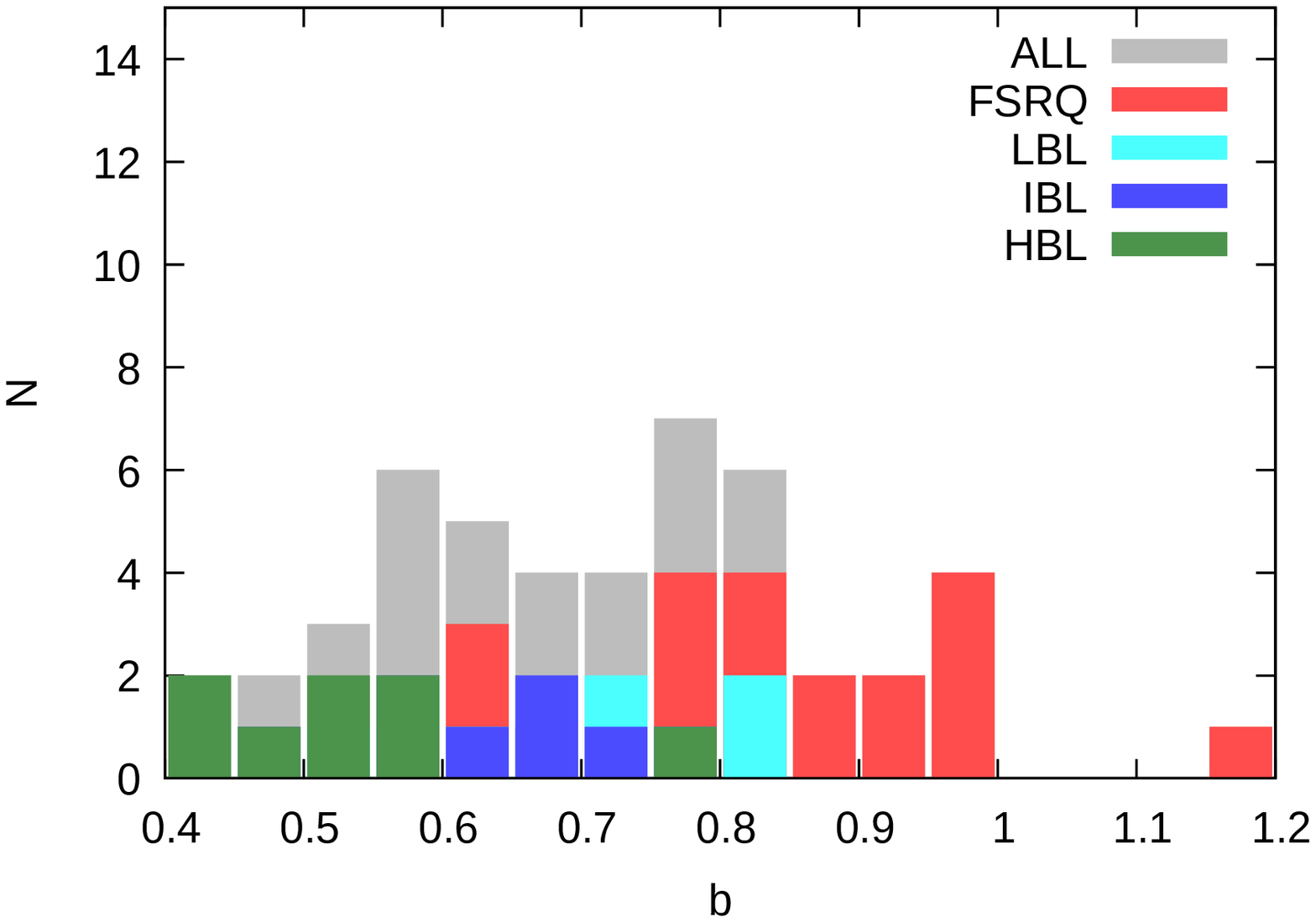}
\includegraphics[height=5cm, width=7cm]{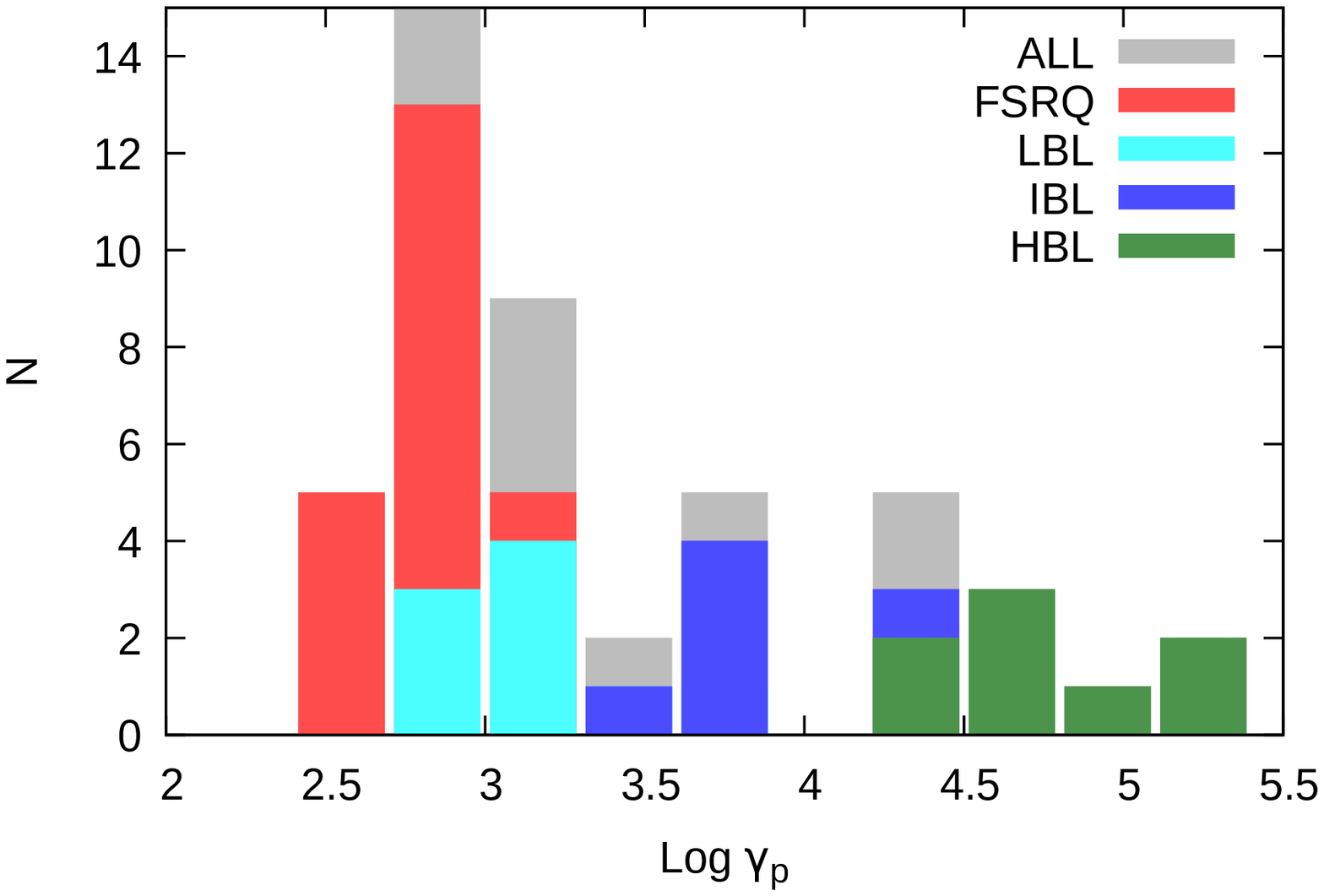}
\caption{Distributions of physical parameters of the blazar sample.}
\label{fig:hist} 
\end{figure}     
Figure \ref{fig:hist} presents the distribution of the modeling parameters corresponding to best model fits presented in Figure \ref{fig:fits}. It can be seen that the parameters corresponding to FSRQs and BL Lac classes occupy different parameter space, especially for Doppler factor $\delta$, electron peak energy $\gamma_p$ and the electron curvature $b$. On average, the $\delta$ for FSRQs is the largest among all classes. While magnetic field $B$ varies in the range $\sim0.02-1$ G, the EC blazars, on average, have higher magnetic field ($\leq0.2$ G) as compared to SSC one ($\geq0.2$ G). The peak energy $\gamma_p$ systematically increases from low to high synchrotron peaked blazars. While $\gamma_p$ for EC blazars remains in the range $\sim 400-5000$, its value for SSC dominated blazars varies in a significantly large range ($\sim 5000-160,000$). This shows that the synchrotron peak frequency shifts in EC blazars are mainly dominated by cooling due to a higher $B$ and $\delta$, which increases the total ambient energy density in the jet $u'$, whereas the peak shifts in SSC sources are caused by $\gamma_p$. Interestingly, the curvature $b$ decreases systematically from FSRQs toward HBLs, suggesting that the curvature may be related to other parameters. Our model parameter distributions and their average values are consistent with others studies \citep{2013ApJ...767....8Z, 2014Natur.515..376G, 2014MNRAS.439.2933Y, 2015MNRAS.448.1060G, 2018ApJS..235...39C} and, therefore, justifies the physical relationship of $b$ and $\gamma_p$, which changes drastically from EC to SSC cases.

\subsection{Curvature $b$ verses Peak Energy $\gamma_p$}
\begin{figure}[htb!]
\centering
\includegraphics[height=7cm, width=9cm]{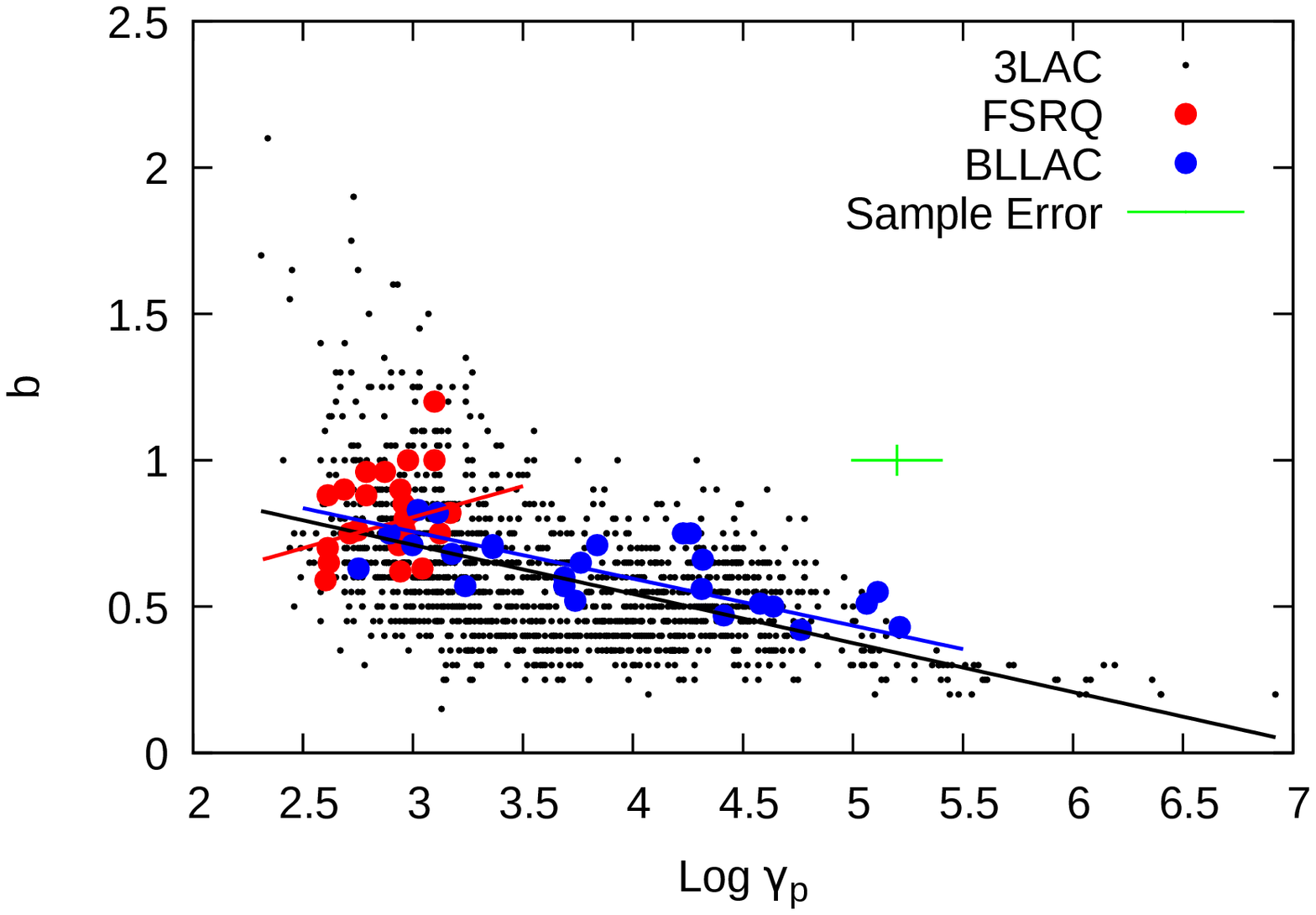}
\caption{Relationship between curvature parameter $b$ and electron break energy $\gamma_p$ for blazars. The colored points and lines show the FSRQs and BL Lac objects in the LBAS sample and the best linear fit, while the green line shows the sample mean error bar. The black dots and line show the parameters and corresponding best linear fit to the 3LAC blazar sample from \citet{2018ApJS..235...39C}.}
\label{fig:x0_b}
\end{figure}
We study the relationship between spectral curvature $b$ and peak energy of EED $\gamma_p$, shown in Figure \ref{fig:x0_b}, suggesting that the curvature in FSRQs and BL Lac objects evolves differently. The LBAS blazars show an anti-correlation between $\gamma_p$ and $b$ for of BL Lac objects. The best linear fit to BL Lac objects (blue line in Figure \ref{fig:x0_b}) yields a relationship $b\simeq -0.16\log \gamma_p + 1.24$ and the Pearson test suggests a significant anticorrelation with coefficient $R_p\approx -0.7$ and a chance probability $P\sim 10^{-4}$. The FSRQs dominate at low $\gamma_p$ and deviate from the BL Lac objects. FSRQs show a positive linear relationship $b\simeq 0.2\log \gamma_p + 0.17$ (see the red line in Figure \ref{fig:x0_b}); however the correlation is not strong ($R_p\approx 0.3$ and $P\approx 0.14$). The 3LAC sample also shows an inverse relationship between curvature and peak energy with a similar slope ($=-0.17$) as in case of LBAS BL Lac objects; however the correlation is not strong due to large scatter at low $\gamma_p$. This may be due to significant overlap at low energies between FSRQs and BL Lac objects, as revealed in our exact modeling of LBAS blazars, suggesting that the curvature evolves differently against $\gamma_p$ in FSRQs as compared to BL Lac objects. We also compare our results with \citet{2017MNRAS.464..599D}, who fitted the sample of 29 TeV BL Lac objects having low and high states based on 1 TeV observations with a log-parabolic SSC model and found negative correlation between curvature and synchrotron peak frequency. We find that the curvature and peak energy of EED derived from their model parameters \citep[see Table 2 in][]{2017MNRAS.464..599D} follows a strong inverse relationship $r\simeq -0.6\log\gamma_p +3.43$ with $R_p\approx -0.54$ and $P\sim 10^{-4}$, where $r=b$ and $\gamma_p= \gamma_0 \exp[(3-s)/2r]$. Since FSRQs in our sample do not show the signature of acceleration, we suggest that the anticorrelation between synchrotron peak frequency and SED curvature found in C14 does not translate to an intrinsic inverse relationship between $\gamma_p$ and $b$ for all types of blazars, but only for BL Lac objects.\\
\\
There are two scenarios that can explain both the log-parabolic nature of EED and the negative correlation found in BL Lac objects: a "Statistical" and a "Stochastic" acceleration mechanism (see C14 for a detailed discussion). In the statistical acceleration scenario, the injected particles undergo a constant gain $\varepsilon$ acceleration such that $\gamma_i=\varepsilon \gamma_{i-1}$. Then if acceleration probability ($p_i$) is inversely proportional to electron energy itself ($p_i= g/\gamma^{q}$), \citet{2004A&A...413..489M} showed that the integrated EED can be approximated by a log-parabolic law,
\begin{equation}
\centering
N(\gamma)\propto \left(\frac{\gamma}{\gamma_{0}}\right)^{-s-r\log(\gamma/\gamma_{0})},
\end{equation}
where $\gamma_{0}$ is injected electron energy, $r=q/2\log\varepsilon$ is the curvature parameter, and $s$ is the spectral index that is related to injected energy and energy dependence. This EED resembles our assumed electron population given by Equation \ref{EED}, with peak,
\begin{equation}
\centering
\gamma_{p}=\gamma_0 10^{\frac{3-s}{2r}},
\end{equation}
and the curvature parameter $b=r$. The second scenario involves diffusive shock acceleration in which momentum diffusion in the injected electron population is described as a stochastic process. Solving kinetic equation for a monoenergetic injection and diffusion in momentum space, \citet{2011ApJ...739...66T} and C14 showed that the EED at time $t$ can be approximated by log-parabolic form and its curvature $r$ is related to the diffusion coefficient $D_p (\gamma,t)= D_{p0} (\gamma/\gamma_0)^q$, as
\begin{equation}
\centering
r \propto 1/D_p(\gamma) t
\end{equation}
where $q$ is the index of magnetic turbulence spectrum. The curvature decreases continuously during the acceleration process, especially in case of $q=2$ (hard-sphere approximation). Therefore, in both acceleration mechanisms, one expects an anti-correlation between electron peak energy and curvature of EED as shown by  Figure \ref{fig:x0_b}. \citet{2011ApJ...739...66T} did the detailed Monte Carlo simulations to study the combined effect of both stochastic acceleration and radiative cooling and found that initially the acceleration dominates the spectral evolution and induces a curvature in the EED. At a later stage of evolution, the cooling dominates in a short lived transition region where $b$ grows rapidly with $\gamma_p$, leading to final stage where $b$ retains a constant value \citep[see Figure 4 in][for details]{2011ApJ...739...66T}.\\
\\ 
A mild positive relationship between $b$ and $\gamma_p$ for FSRQs can be explained by strong cooling during the spectral evolution of injected electron population. Figure \ref{fig:x0_b} shows that curvature of log-parabolic EED, in typical conditions of an FSRQ, would evolve in a regime where a dominant radiative cooling completely compensates the acceleration and the curvature either rises mildly and reaches nearly a steady value, as predicted by \citet{2011ApJ...739...66T}. This happens as the cooling timescale at the peak energy $t_c(\gamma_p)= 3mc/4\sigma_T \gamma_p u'$ becomes nearly equal to the typical acceleration timescale $t_A(\gamma_p)$ in the evolution of injected particle spectra. When cooling becomes relevant, it should restrict the growth of $\gamma_p$ as seed photon energy density would increase \citep{1998MNRAS.301..451G}. The $\gamma_p$ for FSRQs in our sample indeed vary within a very narrow range. The strong cooling in FSRQs is manifested by the fact that total ambient energy density in the jet $u'=u_B'+u_s'+u_{ext}'$ is larger in FSRQs than that in BL Lac objects, due to higher magnetic field $B$ and an additional EC component with $u_{ext}'$. \citet{2016MNRAS.456.2173Y} also studied the relationship between the peak energy of log-parabolic EED and its curvature by modeling 14 SEDs of 3C 279 representing different epochs, and found a positive correlation between them. However, their results might be ambiguous due to poorly constrained $\gamma_p$ due to a lack of data around synchrotron peak (see their Figure 1). Nevertheless, our modeling confirms that the log-parabolic EED in FSRQs evolves within a cooling dominated regime where its curvature grows mildly with $\gamma_p$ or remains almost steady.\\
\\
\begin{figure}
\centering
\includegraphics[height=7cm, width=9cm]{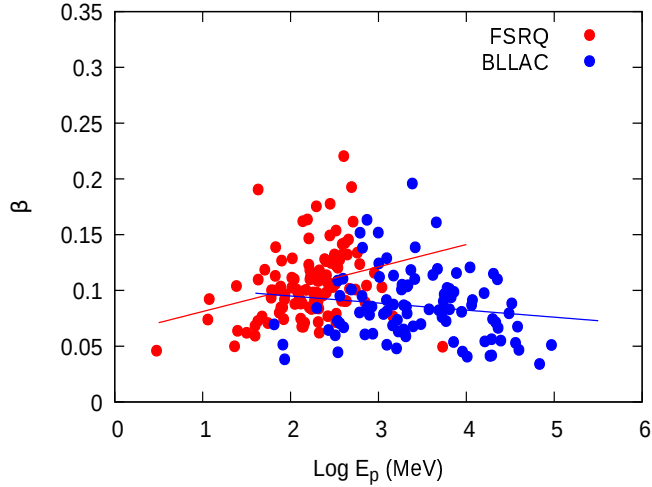}
\caption{Relationship between and peak energy $E_p$ (MeV) and curvature parameter $\beta$ of log-parabola for 4LAC blazars \citep{2020ApJ...892..105A}. The colored points show BL Lac and FSRQ populations and the solid lines represent the best linear fitting.}
\label{fig:4LAC}
\end{figure}
We also compare our modeling results with 4LAC blazars, detected in the wide energy band from 50 MeV to 1 TeV \citep{2020ApJ...892..105A}. \citet{2020ApJS..247...33A} showed that the LAT spectra of nearly 600 blazars can be described by a log-parabola. The log-parabolic photon spectrum is described as
\begin{equation}
\frac{dN}{dE}=k\left(\frac{E}{E_0} \right)^{-\alpha-\beta\log(E/E_0)}, 
\end{equation} 
with $\alpha$ being the spectral slope at reference energy $E_0$ and $\beta$ is the curvature parameter of log-parabola. The peak of the spectrum $E_p$ becomes
\begin{equation}
E_p=E_0 \exp\left( \frac{2-\alpha}{2\beta} \right).
\end{equation}
We selected 275 confirmed blazars with significant curvature ($>4\sigma$), including 96 BL Lac objects and 125 FSRQs, and studied the relation between peak energy $E_p$ and the curvature $\beta$ of log-parabolic fit, as shown in Figure \ref{fig:4LAC}. The BL Lac objects show a mild inverse relationship with slope$\approx -0.015$, whereas the FSRQs show a positive linear relationship with slope $\approx 0.1$, similar to our broadband modeling. These results are in agreement with our SED modeling as shown in Figure \ref{fig:x0_b}, suggesting that curvature of injected EED in FSRQs and BL Lac objects should evolve differently due to different ambient conditions.
 
\section{Summary} \label{sec:summary}
We study the evolution of spectral curvature of the log-parabolic EED in a combined cooling and acceleration scenario by IC modeling of quasi-simultaneous broadband SEDs of a complete sample of LBAS blazars. We find that the SSC model can explain the high energy emission in IBLs and HBLs, whereas an EC component is important to explain the emission and observed curvature at GeV energies in FSRQs and LBLs, as in previous studies \citep[e.g.,][]{2013ApJ...767....8Z, 2014ApJ...782...82D}. While the dust seed photons can explain the $\gamma$-ray spectra in most EC blazars, the contribution from BLR is important in some sources, suggesting that the $\gamma$-ray location in the jets can vary from subparsec to parsec scales. The magnetic field and Doppler factor in blazars decrease continuously from FSRQs toward HBLs, suggesting that the total jet energy density decreases along the blazar sequence.

We find that curvature in the EED, which originates due to a stochastic component in the underlying acceleration mechanism, evolves against peak energy $\gamma_p$ in BL Lac objects and FSRQs differently due to different ambient conditions. In BL Lacs, the curvature evolves in a purely stochastic acceleration dominated regime and cooling is of secondary importance or practically irrelevant due to lower jet energy density. The curvature in typical FSRQs conditions evolves in a transition regime where the cooling overtakes the underlying acceleration. Our results are consistent with theoretical predictions of \citet{2011ApJ...739...66T}, i.e, the curvature decreases in the acceleration regime and grows mildly or remains steady in cooling dominant conditions. This happens in FSRQs as the cooling timescale at peak energy $\gamma_p$ gets shorter than the typical acceleration timescale due to an additional EC component. A higher jet energy density in FSRQs restricts the growth of $\gamma_p$ and provides the extra curvature at the high energy tail of EED. This explains why, on the average, FSRQs have higher curvature compared to BL Lacs and their peak energy varies in a small range. We see a similar trend in 4LAC BL Lac objects and FSRQs even for a single band from 50 MeV to 1 TeV, suggesting that a positive evolution of curvature may be a consistent feature of FSRQs. Although blazars detected in 4LAC with a log-parabolic spectra are $<20\%$, these sources are bright with average significance of detection $\sigma>40$. This suggests that all blazars may have significantly curved spectra \citep{2018RAA....18...56K}. Since the Fermi-LAT spectra usually do not show the signature of internal absorption \citep{2018MNRAS.477.4749C}, the curvature is likely to be intrinsic to emitting electrons. As the curvature decreases from FSRQs toward HBLs opposing the peak energy, it may be considered as a third parameter of blazar sequence as it reveals the underlying conditions of acceleration and cooling. It would be important to study the evolution of curvature in blazar jets in a more realistic scenario in which processes including injection, particle acceleration, radiative cooling, and escape of particles compete against each other.

\acknowledgments
We thank the anonymous referee for their valuable comments and suggestions that helped improve the manuscript. We also thank Professor Shuangliang Li, Haritma Gaur, Minhua Zhou, and Jiawen Li for the fruitful discussions. This research has made use of the NASA/IPAC Extragalactic Database (NED), which is operated by the Jet Propulsion Laboratory, California Institute of Technology, under contract with the National Aeronautics and Space Administration. Part of this work is based on archival data, software, or online services provided by the Space Science Data Center - ASI. This work is supported by the National Natural Science Foundation of China (grant no: U1831138 and 11873073).


\bibliographystyle{aasjournal}
\bibliography{ms}{}
\end{document}